\documentclass{article}

\usepackage{booktabs}
\usepackage{graphicx}
\usepackage[table,xcdraw]{xcolor}
\usepackage{enumitem}
\usepackage{multirow}
\usepackage{makecell}
\setlist[itemize]{itemsep=0.2em, topsep=0pt, partopsep=0pt, parsep=0pt} 
\usepackage{lineno}

\usepackage{blindtext}
\usepackage{algorithm}                                                
\usepackage{algpseudocode}                                            
\usepackage{amsmath}
\usepackage{subcaption}
\usepackage[font=bf]{caption}
\usepackage[abbrvbib, preprint]{jmlr2e}

\usepackage{lastpage}

\ShortHeadings{Earth system foundation model  - Heterogeneous data integration and forecasting}{} %
\firstpageno{1}

\begin{document}

\title{Earth System Foundation Model (ESFM):  A unified framework for heterogeneous data integration and forecasting}

\begin{center}

\author{%
\name Firat Ozdemir$^{1}$ 
\name Yun Cheng$^{1}$ 
\name Salman Mohebi$^{1}$ 
\name Fanny Lehmann$^{2,5}$ 
\\%
\name Simon Adamov$^{3,6}$ 
\name Zhenyi Zhang$^{4}$ 
\name Leonardo Trentini$^{4}$ 
\name Dana Grund$^{5,6}$
\\%
\name Oliver Fuhrer$^{3,6}$
\name Torsten Hoefler$^{7}$
\name Siddhartha Mishra$^{5}$
\\%
\name Sebastian Schemm$^{8}$
\name Benedikt Soja$^{4}$
\name Mathieu Salzmann$^{1}$
\\
\vspace{1em}
\addr $^{1}$ Swiss Data Science Center (SDSC), ETH Zurich \& EPFL, Zurich \& Lausanne, Switzerland \\
\addr $^{2}$ ETH AI Center, ETH Zurich, Zurich, Switzerland \\
\addr $^{3}$ Federal Office for Meteorology and Climatology MeteoSwiss, Zurich, Switzerland \\
\addr $^{4}$ Institute of Geodesy and Photogrammetry, ETH Zurich, Zurich, Switzerland \\
\addr $^{5}$ Computational and Applied Mathematics Laboratory, ETH Zurich, Zurich, Switzerland \\
\addr $^{6}$ Institute for Atmospheric and Climate Science (IAC), ETH Zurich, Zurich, Switzerland \\
\addr $^{7}$ Scalable Parallel Computing Lab, ETH Zurich, Zurich, Switzerland \\
\addr $^{8}$ Dep.\ of Applied Mathematics and Theoretical Physics, University of Cambridge, United Kingdom \\
}

\maketitle

\begin{abstract}%
Following their success in other areas, such as language processing, the idea of creating a foundation model for natural sciences has gained traction. 
A foundation model for the Earth system is one that has learned the statistical relationships between physical variables and can predict them. 
It is a versatile model that excels when finetuned to specific tasks, and not a task-specific weather forecasting model. 
To achieve this, foundational models are typically pre-trained on very large datasets. %
Here, we introduce Earth System Foundation Model (ESFM), a fully open model that builds on the 3D Swin UNet transformer backbone of the pioneering Aurora model.
ESFM introduces a series of extensions that significantly increase its functionality, thereby fostering the adoption of foundation models in the climate sciences.
Firstly, the encoding scheme and training protocols have been extended to handle diverse datasets, including those containing missing values across all spatio-temporal dimensions such as satellite data, as well as station data, all under one backbone. Axial attention is introduced to capture inter-variable dependencies.
As a result, ESFM skillfully predicts variables in regions or on pressure levels where no data is present at the initial time, while preserving inter-variable relationships, for example between temperature, pressure, and humidity. 
Individual variable tokenization enables different sets of variables to be shuffled during training and simplifies the process of building extensions for new downstream tasks. 
Adaptive layer norm-based ensembles allow for a simple yet effective way to transform deterministic ESFM to a probabilistic FM.
We present our findings using dense gridded data, such as ERA5 and CMIP6, as well as regionally masking of dense data, sparse gridded MODIS satellite data, and station data. 
Our results demonstrate competitive or superior performance relative to state-of-the-art benchmarks in each domain. 
Additional case studies of Super Typhoon Doksuri in July 2023 and the 2024 sudden stratospheric warming events show clear indications of accurate positional and magnitude estimations of extreme weather.  
The results show that ESFM retains all the strengths of previous foundation models, such as long term stability, but facilitates the application of foundation models on a variety of downstream tasks.%
\footnote{ESFM is available on \href{https://github.com/swiss-ai/ESFM}{github.com/swiss-ai/ESFM}.}

\end{abstract}
\end{center}

\clearpage
\tableofcontents
\clearpage

\begin{figure}[t]
    \centering
    \includegraphics[width=0.8\linewidth]{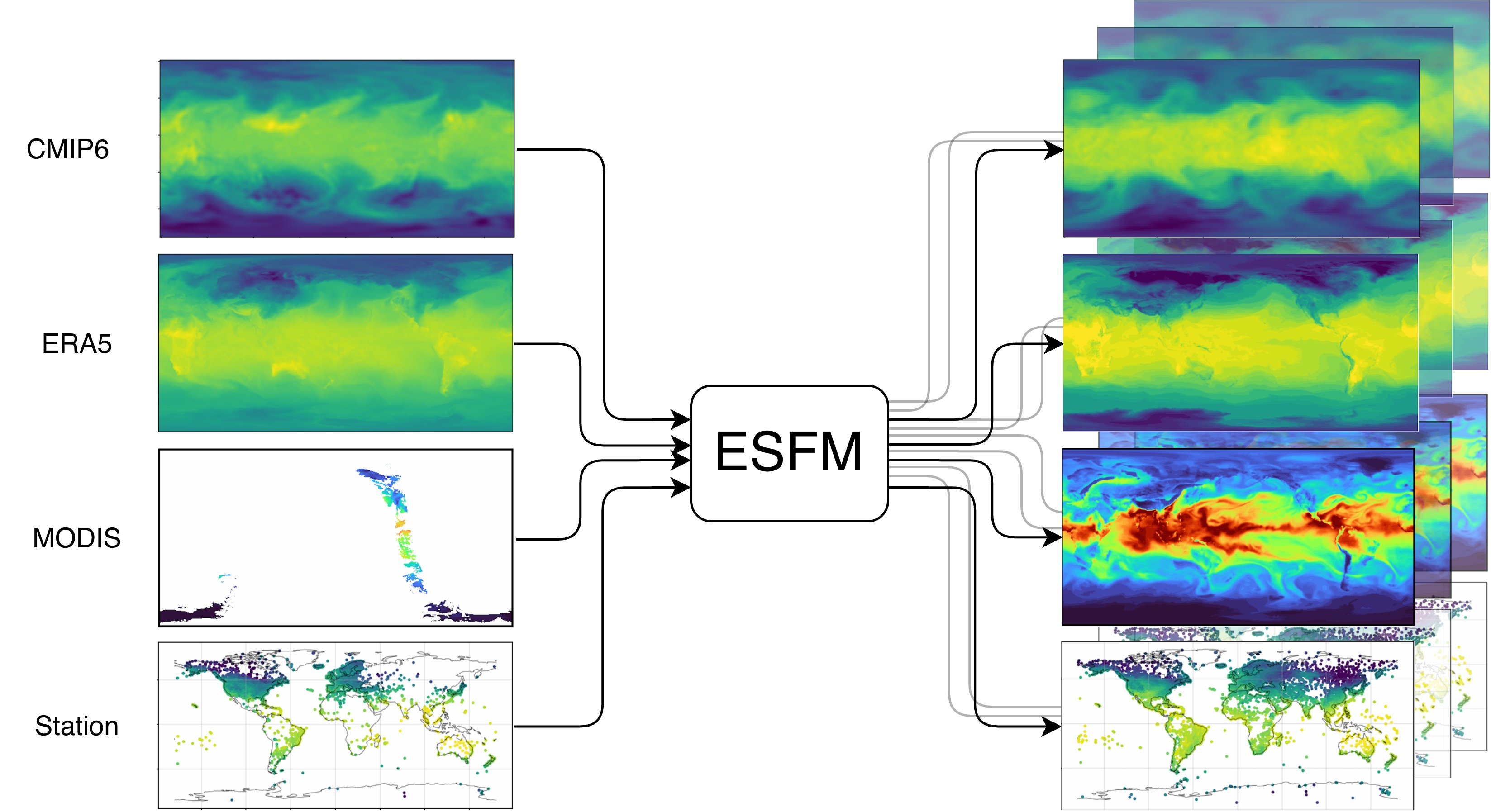}
    \caption{\textbf{ESFM unified framework:}
    ESFM is a flexible foundation model, ingesting multi-modal heterogeneous datasets with missing, sparse, or point data, and predict forecasts with ensembles; all using the same backbone.
    }
    \label{fig:ESFM-teaser}
\end{figure}

\section{Introduction}\label{sec:intro}
\subsection{Motivation}
Accurate weather forecasting is crucial for numerous applications in societal resilience, such as disaster mitigation and food security, and industrial sectors, including aviation, logistics, and renewable energy planning. 
Numerical weather prediction (NWP) models such as IFS from ECMWF and GFS from NOAA are widely used, and will continue to be used in particular because their physical groundings make them trustworthy when it comes to forecasting of extremely rare events. 
However, they are computationally intensive, and their compute footprints scale exponentially with the needs for increased spatial resolution, making them expensive to run frequently, for longer horizons, or with more ensemble members. 
In recent years, data-driven neural networks, some of which are considered to be foundation models (FMs), started to be competitive with NWPs, surpassing them at certain forecasting tasks up to several weeks of lead time. 
This development has been enabled by the availability of large datasets generated by the climate community, such as CMIP6~\citep{eyring2016cmip6}, and even more so through reanalysis data, such as ERA5~\citep{hersbachERA5GlobalReanalysis2020}, which assimilate observations from global weather station networks, satellite data and balloon soundings.
In particular, the ERA5-based WeatherBench2 dataset~\citep{rasp2024weatherbench} made weather data more accessible to the machine learning (ML) community and accelerated the exploration of various data-driven methods, which ultimately allowed machine learning weather prediction (MLWP) models to surpass NWP models for certain aggregate quality measures.

\subsection{Weather and climate data}
Data availability fueled the ML revolution in weather forecasting. 
In particular, the availability of high-quality spatially and temporally gridded datasets, such as the ERA5 reanalysis. 
Therefore, most existing models assume that the set of input variables is consistent, gridded in space and time, and complete. 
However, this is rarely the case when moving beyond reanalysis data and towards direct observation. 
An FM for the Earth system should be able to handle multiple modalities and be flexible to handle sparse, point, volume and missing data. 
In the Earth sciences, raw observations are naturally sparse in space and time, for example, satellite data or measurement stations, which are unequally distributed over the globe. 
Sensor failures or acquisition errors might be present in the raw data. 
Even within the CMIP6 archive, the stored data is rarely on the exact same grid, contains missing values, variables are accumulated over different time periods in different models and various other issues can arise. %
To date, these modalities challenge most FMs, since they tend to concatenate all input variables at the same grid resolutions into a single multi-channel tensor and then create patch embeddings. 
Those FMs that do not concatenate tend to aggregate variables soon after; however, without proper training, such approaches are unlikely to exploit partially missing input at test time. 
Ideally, an FM handles these issues within a unified backbone, eliminating the need for simple interpolation or missing-data filling techniques.

\subsection{Deep learning approaches for weather and climate modeling}
Following their success in the natural image domain, vision transformer (ViT)-based methods have also shown significant performance in the weather domain, such as ClimaX~\citep{nguyen2023climax} and Stormer~\citep{nguyen2024scaling}.
Additional works utilize windowed attention, such as Pangu-Weather~\citep{bi2023accurate}, FuXi~\citep{chen2023fuxi}, and Aurora~\citep{bodnar2025foundation}, which allows ingesting observations with a significantly longer sequence length by limiting self-attention to a window of neighboring tokens within a 3D structure that represent atmospheric variables across different pressure levels.
In the meantime, a separate line of work, such as FourCastNet~\citep{pathak2022fourcastnet} and ACE \citep{watt2023ace}, focuses on the frequency domain to reduce costs of long sequence self-attention tokens with AFNOs~\citep{guibas2021adaptive}.
This work is extended to better represent Earth geometry using spherical harmonics in SFNOs~\citep{bonev2023spherical}.
Another line of work considers each data to exist in its own space (referred to as data-grid) and learn encoder and decoder modules for each such grid to map it to the mesh-grid of the network. %
Such a mesh-grid allows for a processor backbone that is independent of the input grid and resolution.
Some of the most notable works using graph neural networks (GNNs) are GraphCast~\citep{lam2023graphcast} and GraphDOP~\citep{lean2025learning}.
AIFS~\citep{lang2024aifs} and FGN~\citep{alet2025skillful} use a similar encoder for data-to-mesh grid mapping, but run a sliding window attention on the mesh-grid (graph-transformer).
GenCast~\citep{price2023gencast} is another novel work that, after mapping the data-grid onto a mesh-grid, infers the change in the Earth state using a denoising diffusion model.
This allows GenCast to be naturally probabilistic.
Authors of ERDM~\citep{cachay2025elucidated} use denoising diffusion models with careful design and training considerations to achieve improved stability over increased lead time forecasts.
Finally, there are also works focusing on direct station data, such as MetNet3~\citep{andrychowicz2023deep} and Aardvark~\citep{vaughan2024aardvark}, which so far received less attention due to the extremely sparse nature of station data, but progress is rapid and direct observation-based ML weather forecasting seems to be a logical next step.

\subsection{ESFM overview} 
This work further develops the foundation model line of research in the weather and climate domain. 
We introduce the open Earth System Foundation Model (ESFM) framework, which can successfully exploit heterogeneous data with sparsity across all dimensions and variables for different forecasting tasks across very different grids and station data (Fig.~\ref{fig:ESFM-teaser}). 
The backbone of ESFM is derived from the 3D Swin UNet transformers~\citep{tang2022self} of Aurora, as described in \cite{bodnar2025foundation}. 
The 3D Swin UNet transformer architecture has recently been shown to exhibit the strongest data-scaling behavior compared to other options such as graph NNs or FNOs \citep{yu2026scaling}. 
The long-term stability of Aurora is an inherent advantage of ESFM. \\
\clearpage
\textbf{Our contributions} can be summarized as follows:
\begin{enumerate}
    \item \textbf{Missing data.}\ ESFM learns patch embeddings for each variable independently, allowing natural support for partially missing data across any dimension when coupled with masked training. As a result, ESFM predicts skillfully in regions with locally no data available at initial time.
    \item \textbf{Multimodality under one backbone.}\ ESFM uses multi-resolution tokenizers. 
    It groups observations of different but relatively similar horizontal resolutions and maps the data to shared sets of variable tokenizers. 
    This allows to control the number of increasing (de-)tokenizers when scaling input data to any pixel resolution, such as coarser ones from CMIP6, finer 0.25$^\mathrm{o}$ ERA5 data, very fine satellite data, or even point data from measurement stations; all while using the same backbone.
    As a result, ESFM can be trained on various datasets and thereby has improved transferability. 
    \item \textbf{Axial-attention.}\ ESFM employs axial-attention~\citep{ho2019axial} across variable tokens, which enables capturing variable inter-dependencies while being highly efficient with increasing number of variables. 
    \item \textbf{Positional embeddings.}\ ESFM adds 2D sine-cosine positional embeddings to latent embeddings in the decoder to reinforce positional information for detokenizers~\citep{he2021masked}.
    \item \textbf{Probabilistic forecasting.}\ The deterministic ESFM is extended into an ensemble with negligible costs through latent embedding modulation based on adaptive layer norms. 
    \item \textbf{An open community model.}\ ESFM is fully open, including the training scripts, weights and pre-processing code. 
    This allows the community to finetune it for their specific tasks, add additional features, and advance FMs in the weather and climate domain.
\end{enumerate}

In the following, we go over the design and training considerations of ESFM in more detail. 
Then, we present the forecasting performance of ESFM under different evaluation scenarios. 
We compare forecast predictions with other state-of-the-art (SotA) methods and examine trade-offs for partially missing observations.

\section{ESFM framework design}
\label{sec:motivation}

\begin{figure}[h!]
    \centering
    \includegraphics[width=0.9\linewidth]{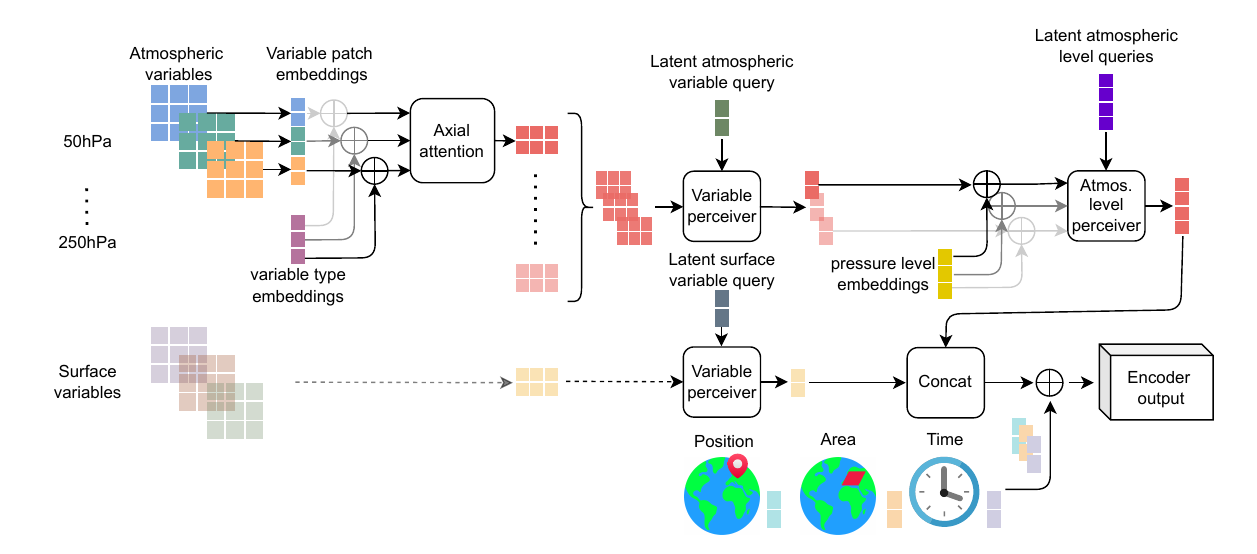}
    \caption{\textbf{ESFM encoder:} ESFM tokenizes each input variable individually, then performs self-attention across variable tokens (i.e., axial attention). 
    The size of the variable dimension is then reduced using a perceiver module for both atmospheric and surface variables.
    Observations across different pressure levels are tokenized separately and corresponding tokens across different pressure levels are aggregated to latent atmospheric pressure levels after the atmospheric level perceiver module.
    We give further details on the components in the Appendix~\ref{sec:ESFM_hyperparameters}.
    }
    \label{fig:schematic-encoder1}
\end{figure}

Most SotA FM models in weather and climate science expect a standardization of the input data stream such that they have the same number of channels (i.e., variables) per dataset during training and testing. 
Although this is a fair assumption for heavily processed datasets such as ERA5~\citep{rasp2024weatherbench} -- and even then, not always valid~\citep{era5bias} -- most climate data are not as clean, causing significant developments in the weather domain to be out of reach for other domains in environmental sciences. 
A typical workaround is to train a new encoder and decoder for every set of data, even if there is a significant overlap of the variables across them. 
However, this approach only works for sufficiently large sets of unique data (i.e., still consistent within itself), and only during training time, making models useless for previously non-encountered set of variables at test time. 
Treating different variables of climate datasets similar to RGB channels in the natural image domain is an unnecessary limiting factor. 
Instead, ESFM encodes observations of different variables separately, already proposed previously in other works such as ClimaX \cite{nguyen2023climax} and Stormer.
However, through individual tokenization of global input at high (pixel) resolution, the memory footprint of patch embeddings quickly saturates available resources, even on HPCs. 
In ClimaX, the authors apply a cross-attention layer to token embeddings with a learnable vector.
In ESFM, we use axial-attention~\citep{ho2019axial} across variable tokens, a concept that has been proven successful, for example, in the diffusion-based model SEEDS~\citep{li2023seeds}.
Axial-attention across variable tokens is a self-attention mechanism over the variable dimension, making the context length extremely short. 
Following the axial-attention layer, we reduce the number of tokens along the variable dimension using a perceiver module~\citep{jaegle2021perceiver}. 
This is done separately for both atmospheric variables and surface variables. 
Perceiver is a sequence of cross-attention layers where we use learnable variable embeddings as query and token embeddings of observed variables as key and values. 
An additional reduction of tokens along atmospheric pressure levels is achieved using another perceiver module, as was originally proposed in Aurora.
The schematic of the ESFM encoder is shown in Fig.~\ref{fig:schematic-encoder1}.

\begin{figure}[t!]
    \centering
    \includegraphics[width=0.7\linewidth]{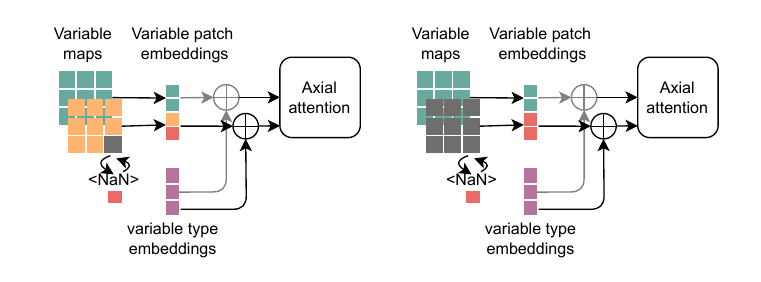}
    \caption{\textbf{ESFM training with missing data:} Partial (left) or completely missing (right) input variable embeddings (shown in gray) are exchanged with a learnable \textit{NaN} token (shown in red) in ESFM.
    }
    \label{fig:schematic-nantoken}
\end{figure}

\subsection{Training with missing data}
\label{sec:missing_data}

In various climate datasets, parts of the observations across different variables could be missing.
An obvious example of this is with satellite imagery, where only a regional observation is available at any given time step.
Furthermore, combining observational data across multiple satellites would imply different regional observations across different variables.

A related situation can happen when a certain observation is simply non-existent in certain regions at specific times. 
For example, some climate models are based on terrain-following vertical coordinates but the output is interpolated to a default stack of pressure levels (e.g, 850, 500, and 300\,hPa).
This means that in certain mountainous regions missing values can be present in the data for a limited time period if a pressure level intersects the topography.
Another example of this is studied in Aurora, where certain variables in the ECMWF Wave Model can be absent at different times and at different locations. 
\cite{bodnar2025foundation} overcome this challenge by introducing an additional \textit{density} channel for every such variable that may partially have missing data.
Although this approach works well for the wave modeling example, it (1) is unsuitable for inferring missing input variables when their absence stems from observational limitations or assimilation procedures, and (2) does not scale efficiently, as each additional variable doubles the number of input and output channels.

ESFM introduces learnable \textit{NaN} tokens, which replace input patches where part or all of the region within an input patch consists of missing observations. 
Internally, \textit{NaN} tokens get positional encodings of the corresponding variable-type embeddings. 
This makes ESFM flexible in handling arbitrarily missing input variables, whether they are partially or completely absent (cf.\ Fig.~\ref{fig:schematic-nantoken}).
This feature naturally extends to missing pressure levels and would extend to the temporal dimension as well.

\subsection{Multi-resolution patch embedding}
\label{sec:multi_res_patch_embedding}

ESFM can be trained seamlessly across very different data resolutions, having very different ranges of grid resolutions and even station data. 
We achieve this through partitioning the input resolutions into discrete bins, where variables of each range are assigned to their dedicated tokenizer.
For example, gridded data for all streams at a \textit{very fine} resolution, say between $(0^\mathrm{o},0.05^\mathrm{o}]$, will be tokenized using the set of patch tokenizers for each variable defined at this resolution interval.
In doing so, a patch tokenizer for any given variable is expected to be flexible to account for (and hence tokenize) only slightly different resolutions of input data across various data sources.
Our decision to do so can be further supported with the fact that the weather variables resolve different phenomena at different resolutions, yielding observations with distinct local patterns, for example local hail hot spots~\citep{schemm2016link,nisi201815}.
In this work, we use six discrete bins, ranging from (i) \textit{very coarse} data streams that are 1.5$^\mathrm{o}$ or coarser, to (v) \textit{very fine} data streams mentioned above, and a separate one for (vi) \textit{station data}, explained in the next paragraph. 
We list the ranges of resolutions and tokenizer details further in the Appendix~\ref{sec:ESFM_hyperparameters}.

\begin{figure}[h!]
    \centering
    \includegraphics[width=0.99\linewidth]{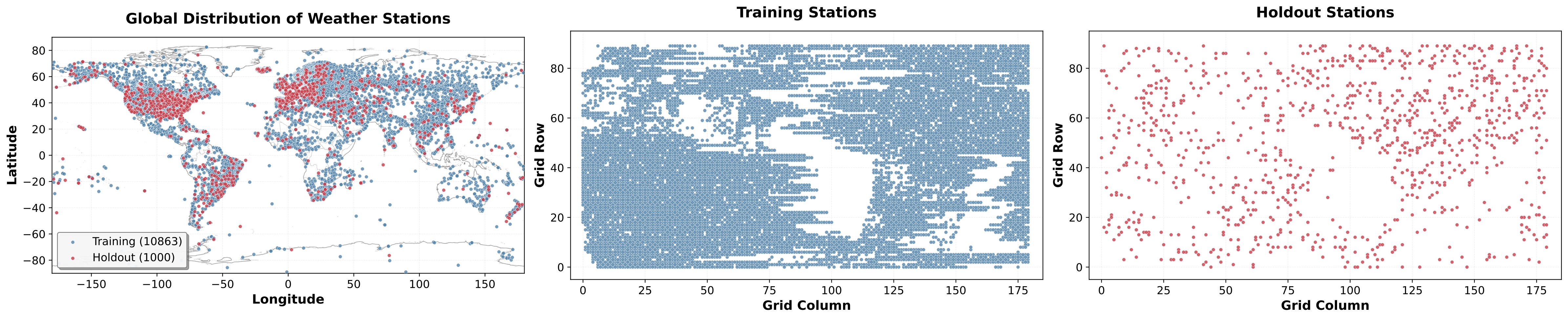} %
    \caption{\textbf{ESFM training with station data:}
    ECMWF 11k station data (left) is greedily mapped onto a compact irregular grid (middle \& right).
    Stations are marked to indicate those that will be used for training (blue) and kept as holdout sets (red).
    }
    \label{fig:station-ecmwf11k}
\end{figure}

Multi-resolution patch embeddings allow for an intuitive way to tokenize non-gridded station data using patch resolution of 1$\times$1 pixels.
This means that ESFM can be trained on station data without modifying the 3D Swin UNet backbone used in Aurora. 
To achieve this, we map the list of in-situ stations greedily to an irregular longitude-latitude grid, as shown in Fig.~\ref{fig:station-ecmwf11k} for one of the station datasets.
Our only soft constraint is to ensure monotonically increasing latitudes within each column and longitudes within each row.
This allows neighboring pixels to still be the most relevant stations for windowed self-attention later in the backbone.
We give more details on the greedy mapping algorithm in the Appendix~\ref{sec:station_greedy_mapping}.
In order to respect the locality of the station data, we reduce the minimum patch area positional embedding from previously defined $0.01^\mathrm{o}$ to $10^{-5}$ degrees, which roughly corresponds to $1.2$\,m$^2$ as opposed to $1.2$\,km$^2$ at the equator.
The actual position of each station is still accurately embedded with positional encoding without any further changes, unaffected by the greedy irregular grid transformation.

\subsection{Deterministic to ensemble simulations}
\label{sec:deterministic-to-ensemble}

\begin{figure}[t!]
    \centering
    \includegraphics[width=0.9\linewidth]{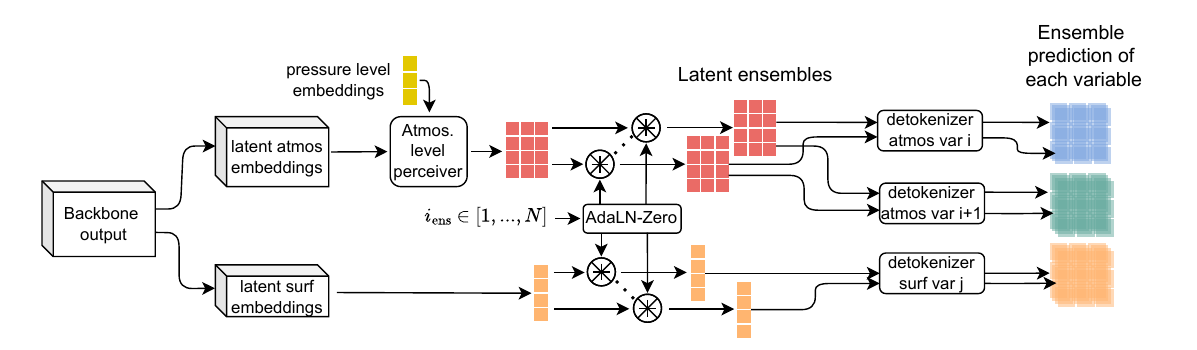}
    \caption{\textbf{ESFM decoder:} The decoder structure of ESFM utilizes perceiver module to map latent atmospheric embeddings to queried set of target pressure levels, which does not have to be the same levels at observation.
    Using a set of queried ensemble members, latent ensembles are formed from latent embeddings using AdaLN-Zero layer, which are then passed through detokenizer of each variable to reconstruct variable patches.
    }
    \label{fig:esfm-adaLN-decoder}
\end{figure}

Probabilistic prediction capability is crucial for most applications in the environmental sciences. 
We propose a simple means to extend the deterministic ESFM model to potentially very large numbers of ensembles through adaptive layer normalization (AdaLN-Zero; ~\citet{peebles2023scalable}) on the latents of the network backbone output.
We first construct a conditioning vector $c_{i_\mathrm{ens}}$ based on the identifier of an ensemble member.
This is then used to apply a scale and shift modulation on the token embeddings $z$ as:
\begin{equation}
    z' = z + \gamma(c_{i_\mathrm{ens}})\mathrm{LN}(z)+ \beta (c_{i_\mathrm{ens}}) \mathrm{\, ,}
\end{equation}
where we do not add a scale bias and $\mathrm{LN}$ is a layer normalization layer.
The modulated token embeddings of different ensembles then pass through the shared detokenizers that map them to patches in the pixel space. 
Thanks to the initialization of the AdaLN parameters $\gamma$ and $\beta$ with zeroes, the extension from deterministic ESFM to ensemble ESFM is very stable and non-destructive, as all ensembles will initially simply generate the same output as the deterministic ESFM. 
A great benefit of this approach is that during training, one can sample any subset of the total number of ensemble members $N$; allowing one to trivially extend ESFM to thousands of ensembles, if not more, with only the small cost of linearly increasing AdaLN parameters. 
Furthermore, the backbone does not need to be queried multiple times, making the additional memory overhead significantly more manageable. 
In Fig.~\ref{fig:esfm-adaLN-decoder}, we visualize the ensemble ESFM decoder.
We train the set of ensemble members using almost fair CRPS loss~\citep{lang2024aifs-crps} in combination with MAE on the ensemble mean to retain the deterministic skill with equal weights on both objectives. 
We apply latitude-weighting on all datasets that are saved on a regular latitude-longitude grid.

\section{ESFM training protocols} %
\label{sec:training_protocols}

To maximize the flexibility and generalizability of ESFM, various randomized masking and pre-training strategies are incorporated into the training process.

\subsection{Pretraining strategies: CMIP6, ERA5 and knowledge distillation}

The first pre-training is done using eight different CMIP6 datasets at resolutions between $0.7^\mathrm{o}$ to $1.9^\mathrm{o}$ (see Appendix~\ref{sec:CMIP6_preprocesing}). 
The second pretraining is the alignment process of the ESFM encoder with the pretrained Aurora encoder, using knowledge distillation (KD;~\citet{hinton2015distilling}) to benefit from the pretrained Aurora backbone (Appendix~\ref{sec:knowledge-distillation}). 
A randomly initialized ESFM serves as the baseline to compare with the other pretraining strategies. 

\subsection{Random masking of variables, levels and regions}
\label{sec:masking}

As motivated in Sec.~\ref{sec:missing_data}, partially or completely missing variables at times is not a negligible problem setting.
In order to account for this, it is crucial to force the network into learning inter-token relations (e.g., inter-region and inter-variable). 
Therefore, we extend the training protocol of ESFM to include random masking at the input.
Although this is a common technique used in the natural image domain for self-supervised learning~\citep{he2021masked}, posed as a reconstruction task of the masked inputs, it has also been shown to work successfully in the climate domain in AtmoRep~\citep{lessig2023atmorep}.

Variable specific tokenization in ESFM allows for a very versatile range of masking possibilities to promote learning different ranges of inter-dependencies.
Accordingly, we propose three masking protocols;
\begin{enumerate}
    \item \textbf{Variable masking}: Randomly mask a variable with probability $p_v$, 
    \item \textbf{Pressure level masking}: Randomly mask a pressure level with probability $p_l$,
    \item \textbf{Spatial masking}: Randomly mask a contiguous region with probability $p_s$ \citep{baevski2023efficient}.
\end{enumerate}
Additionally, we also explore a training scheme with randomly selecting $n_l$ pressure levels out of all available pressure levels, instead of having them fixed to a specific set of pressure levels, separately for both observations and forecast predictions.
In Appendix~\ref{sec:masking_details}, we provide exact details on the different masking we employ during training and also give further details of the additional training scheme we explored with random selection of pressure levels.

\section{ESFM performance with dense gridded data} %
\label{sec:results}
This section explores the forecasting performance of ESFM models based on different pretraining schemes and compares it with SotA models under both deterministic and ensemble setups.
ESFM is also benchmarked for an extreme tropical cyclone event and a sudden stratospheric warming (SSW) event, both appearing outside of the training periods, showcasing that the model captures the surface influence in the weeks following the SSW event. 

The default training setup with the ERA5 dataset covers the period from 1979 to 2020 with a six-hour lead time using five atmospheric variables across 13 pressure levels, four surface variables, and three static 2D variables. 
This choice ensures comparison with other published models. 
Additional details on the ERA5 variables, as well as variables and time periods from other datasets are described in the Appendix~\ref{sec:datasets_and_variables}. 
We also provide autoregressive rollout forecast results of ESFM up to seven days in the Appendix~\ref{sec:rollout_forecasts}.

Unless stated otherwise, all results refer to forecasts throughout four full weeks, initialized on January 2nd, April 2nd, July 2nd, and October 2nd at midnight, for both 2023 and 2024.
We refer to ESFM parameter configuration with $\sim$110\,M parameters as ESFM\_s, as it corresponds to small size Aurora. 
We limit all our comparative experiments to ESFM\_s in order to keep compute costs manageable. 
For brevity, we report the most commonly used variables in this section; atmospheric variables at 850, 500, and 100\,hPa such as temperature (T), eastward component of the wind (U), northward component of the wind (V), specific humidity (Q), geopotential (Z), and surface variables such as 2-meter temperature (T2m), 10-meter eastward and northward components of the wind (U10m, V10m).

\subsection{Impact of pretraining on model performance}

\begin{figure}[t!]
\centering
\begin{minipage}{0.55\textwidth}%
\centering%
\resizebox{\textwidth}{!}{%
\begin{tabular}{@{}lllllll@{}}%
\toprule%
& T850 & Z500 & Q850 & T2m & U10m & V10m \\%
\midrule%
ESFM\_s,ri & 0.300 & 26.676 & 2.88e-4 & 0.306 & 0.322 & 0.332 \\%
ESFM\_s,ci & 0.281 & 26.218 & 2.67e-4 & 0.290 & 0.301 & 0.313 \\%
ESFM\_s,kd & 0.248 & 15.910 & 2.48e-4 & 0.269 & 0.280 & 0.290 \\%
\midrule
ESFM\_s,kd* & 0.224 & 12.535 & 2.24e-4 & 0.248 & 0.253 & 0.262 \\
\bottomrule%
\end{tabular}%
}%
\end{minipage}%
\hfill%
\begin{minipage}{0.449\textwidth}%
\centering%
\includegraphics[width=\textwidth]{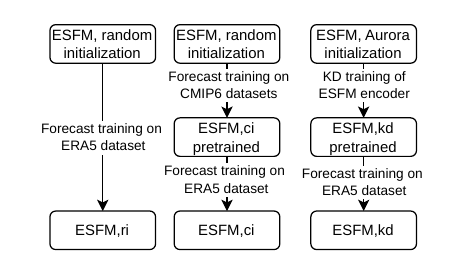}%
\end{minipage}%
\caption{\textbf{Impact of pretraining on forecast performance:} Six hour lead time forecast performance of ESFM\_s based on the different pretraining, shown in mean absolute error, and detailed in the schematic. 
Models are random initialization (ri), pretraining on 8 CMIP6 models (ci), and KD on pretrained Aurora (kd).
First three rows show models trained with masking protocol.
ESFM\_s,kd* in last row shows the performance gain when final ERA5 finetuning is done without the masking protocol.
}
\label{fig:pretraining-schematic}
\end{figure}

In Fig.~\ref{fig:pretraining-schematic}, we show six hour lead time forecast performance of different ESFM\_s on the ERA5 test set.
Each ESFM\_s is initialized from a different set of pretrained weights, and trained with a masking configuration involving all three masking protocols on the ERA5 training set for 100\,k steps.
\begin{enumerate}
    \item \textbf{ESFM\_s,ri:} Random initialization;
    \item \textbf{ESFM\_s,ci:} Weights after a pretraining of 92\,k steps on eight CMIP6 datasets: CMCC, MIROC6, TaiESM1, NESM3, AWI, MPI-M, EC-Earth3, and MRI-ESM2;
    \item \textbf{ESFM\_s,kd:}  40\,k steps of knowledge distillation (KD) training of the ESFM\_s encoder using pretrained small Aurora encoder on the ERA5 training set, and using weights of the pretrained small Aurora backbone and decoder afterwards.
\end{enumerate}
For reference, we also include ESFM\_s,kd*, which is ESFM\_s,kd, but the final training of 100\,k steps is done without the masking protocol.
The schematic in Fig.~\ref{fig:pretraining-schematic} shows an overview of the different pretraining strategies. 
ESFM\_s initialized with pretrained Aurora small weights through KD achieves the best performance. 
The superior performance of ESFM\_s,kd remains mostly stable over autoregressive rollouts, which we tested up to seven days (see also Appendix, Fig.~\ref{fig:pretraining_rollout_comparison}). 
Therefore, we continue to use ESFM\_s,kd as the initialization point for the following experiments, unless stated otherwise. 
Furthermore, we refer to ESFM\_s,kd simply as ESFM\_s herein and mention explicitly whether the ERA5 finetuning following the KD step was done with or without additional masking of variables, pressure levels or regions in the following experiments.

\subsection{Nowcasting performance}
\label{sec:6h_forecasting}

Here, we evaluate the performance of ESFM\_s against other SotA models in the nowcasting range with a lead time of six-hours (Table~\ref{tab:6h_forecast}). 
Furthermore, we also include IFS control NWP.
We only consider ESFM\_s that has not been trained with masking of variables (i.e., ESFM\_s,kd* from Fig.~\ref{fig:pretraining-schematic}), levels and regions during training, because other SotA models have also not been trained with a similar masking strategy. 
From Table~\ref{tab:6h_forecast} it becomes clear that ESFM\_s is on a par with AIFS and small Aurora for surface variables, and outperforms GraphCast, SFNO, and IFS control. 
As expected, large Aurora (Aurora\_l) leads the ranking. 
A fairly similar picture emerges for the upper air variables. 
ESFM small and Aurora small are comparable in their performance, with AIFS performing slightly better than both. 
Once again, large Aurora outperforms all other models. 
Because Aurora and ESFM share the same backbone, we are confident that ESFM can be scaled easily to the performance of the largest Aurora model, while retaining all of the higher flexibility of the ESFM framework.

For reference, we note that we observe a performance drop of ESFM\_s when trained with masked forecasting objective. 
The performance drop ranges between 8.5\% (T2m, 0.269) to 26.9\% (Z500, 15.91), with a median value of 10.7\% (V850, 0.454). 

\begin{table}[b!]
\caption{Six hour lead time forecast performance of ESFM\_s shown in mean absolute error in comparison with various SotA models.}
\label{tab:6h_forecast}
\resizebox{\textwidth}{!}{%
\begin{tabular}{@{}lccccccccccccc@{}}
\toprule
& T850 & T500 & U850 & U500 & V850 & V500 & Q850 & Q500 & Z850 & Z500 & T2m & U10m & V10m \\ \midrule
ESFM\_s & 0.224 & 0.156 & 0.406 & 0.473 & 0.410 & 0.486 & 2.24e-4 & 7.32e-5 & 11.831 & 12.535 & 0.248 & 0.253 & 0.262 \\
Aurora\_s & 0.218 & 0.146 & 0.396 & 0.451 & 0.402 & 0.458 & 2.18e-4 & 7.21e-5 & 12.314 & 13.825 & 0.246 & 0.256 & 0.262 \\
Aurora\_l & 0.167 & 0.119 & 0.309 & 0.368 & 0.312 & 0.375 & 1.76e-4 & 6.02e-5 & 10.090 & 11.687 & 0.194 & 0.195 & 0.201\\
AIFS & 0.182 & 0.124 & 0.351 & 0.410 & 0.351 & 0.416 & 1.90e-4 & 6.74e-5 & 14.406 & 14.679 & 0.219 & 0.232 & 0.238 \\
GraphCast & 0.208 & 0.136 & 0.381 & 0.443 & 0.382 & 0.453 & 2.08e-4 & 7.20e-5 & 13.068 & 13.608 & 0.304 & 0.269 & 0.272 \\
SFNO & 0.217 & 0.177 & 0.423 & 0.561 & 0.431 & 0.580 & 2.17e-4 & 8.64e-5 & 13.074 & 13.852 & 0.255 & 0.282 & 0.294 \\
IFS Control & 0.346 & 0.200 & 0.621 & 0.681 & 0.616 & 0.676 & 3.90e-4 & 1.11e-4 & 14.959 & 15.804 & 0.421 & 0.433 & 0.444 \\
\bottomrule
\end{tabular}
}
\end{table}

\subsection{Ensemble forecasting}

Next, consideration is given to the ensemble extension of ESFM\_s. 
To this end, we finetune the deterministic ESFM\_s model with the AdaLN based method proposed in Sec.~\ref{sec:deterministic-to-ensemble} to extend it to $N$=$8$ ensemble members. 
The finetuning is done for 10'000 steps using the loss function \mbox{$\mathcal{L}=\mathrm{MAE}+\mathrm{afCRPS}$} that includes the almost fair CRPS (introduced in~\cite{lang2024aifs-crps}).
Table~\ref{tab:ensembles_6h} lists the CRPS and MAE (calculated on the ensemble mean) scores of ESFM\_s in comparison with AIFS-CRPS~\citep{lang2024aifs-crps} and FourCastNet 3~\citep{bonev2025fourcastnet3}. 
For the compared models, we also produced eight ensemble predictions.
The ESFM\_s ensembles achieve a better CRPS score than both the AIFS-CRPS and the FourCastNet 3 ensembles for surface variables.
The CRPS scores for atmospheric variables are more varied. 
FourCastNet 3 tends to rank first. 
Nevertheless, the differences between all models for both surface and upper air variables are very comparable.

\begin{table}[b!]
\caption{
Six hour lead time forecast performance of ESFM\_s ensemble model (indicated with ESFM\_s$^+$) shown in comparison with AIFS-CRPS and FourCastNet 3.
Mean absolute error (MAE) is computed on the ensemble means.
}
\label{tab:ensembles_6h}
\resizebox{\textwidth}{!}{%
\begin{tabular}{@{}llllllllllllllll@{}}
\toprule
& \multicolumn{2}{c}{U10m} & \multicolumn{2}{c}{V10m} & \multicolumn{2}{c}{T2m} & \multicolumn{2}{c}{T850}  & \multicolumn{2}{c}{Z500} &  \multicolumn{2}{c}{Q850} \\
\cmidrule(lr){2-3} \cmidrule(lr){4-5} \cmidrule(lr){6-7} \cmidrule(lr){8-9} \cmidrule(lr){10-11} \cmidrule(lr){12-13}
 & CRPS $\downarrow$ & MAE $\downarrow$  & CRPS $\downarrow$ & MAE $\downarrow$  & CRPS $\downarrow$ & MAE $\downarrow$  & CRPS $\downarrow$ & MAE $\downarrow$  & CRPS $\downarrow$ & MAE $\downarrow$  & CRPS $\downarrow$ & MAE $\downarrow$ \\
\midrule
ESFM\_s$^+$ & 0.249 & 0.260 & 0.258 & 0.269 & 0.238 & 0.248 & 0.218 & 0.225 & 14.484 & 15.311 & 2.19e-4 & 2.28e-4 \\
AIFS-CRPS   & 0.259 & 0.246 & 0.264 & 0.251 & 0.265 & 0.238 & 0.202 & 0.188 & 15.946 & 14.098 & 2.08e-4 & 1.96e-4 \\
FourCastNet 3& 0.250 & 0.236 & 0.260 & 0.246 & 0.250 & 0.235 & 0.199 & 0.186 & 12.641 & 11.761 & 2.03e-4 & 1.90e-4 \\
\bottomrule
\end{tabular}%
}
\end{table}

\begin{figure}[t!]
    \centering
    \includegraphics[width=0.6\linewidth]{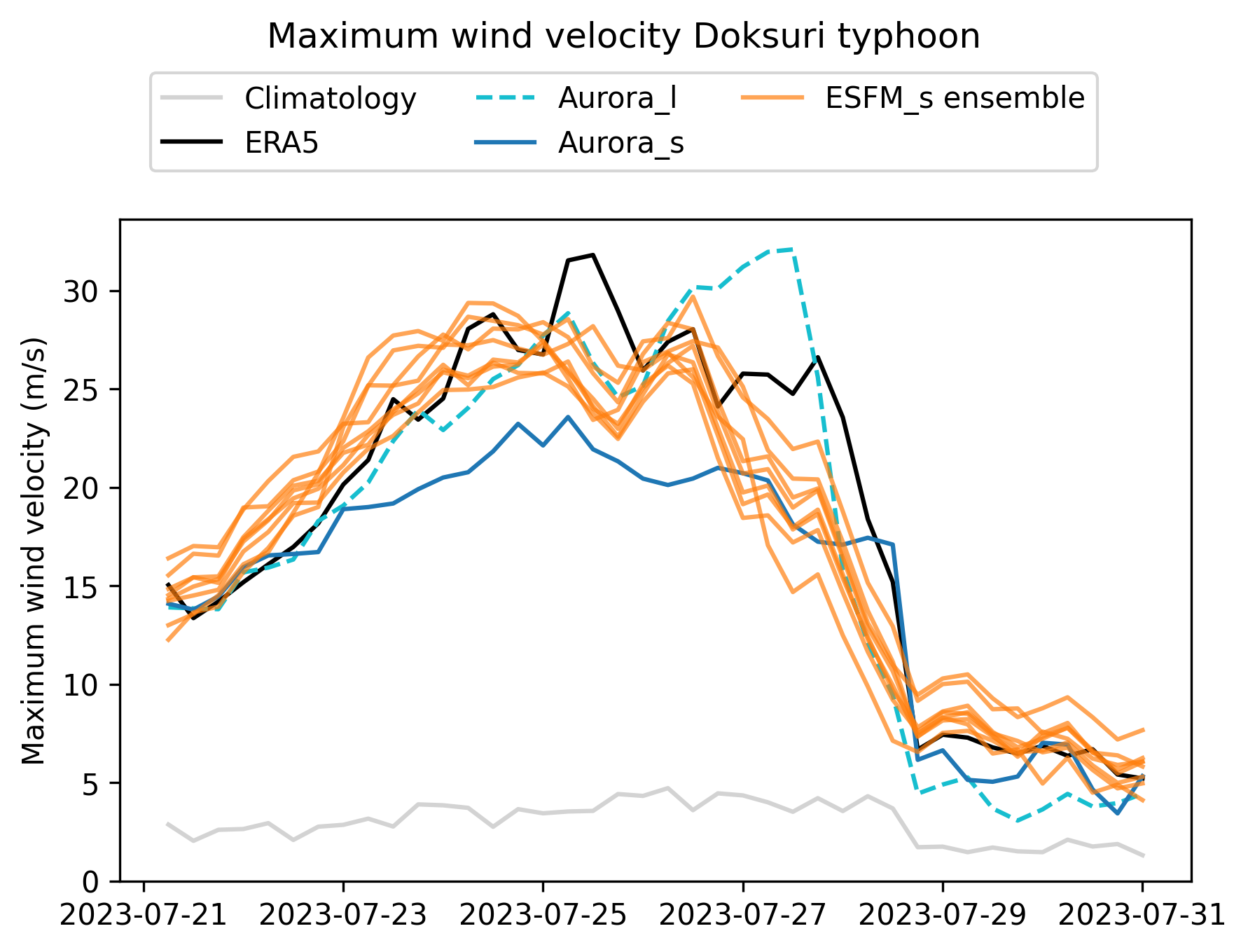}
    \includegraphics[width=0.24\linewidth]{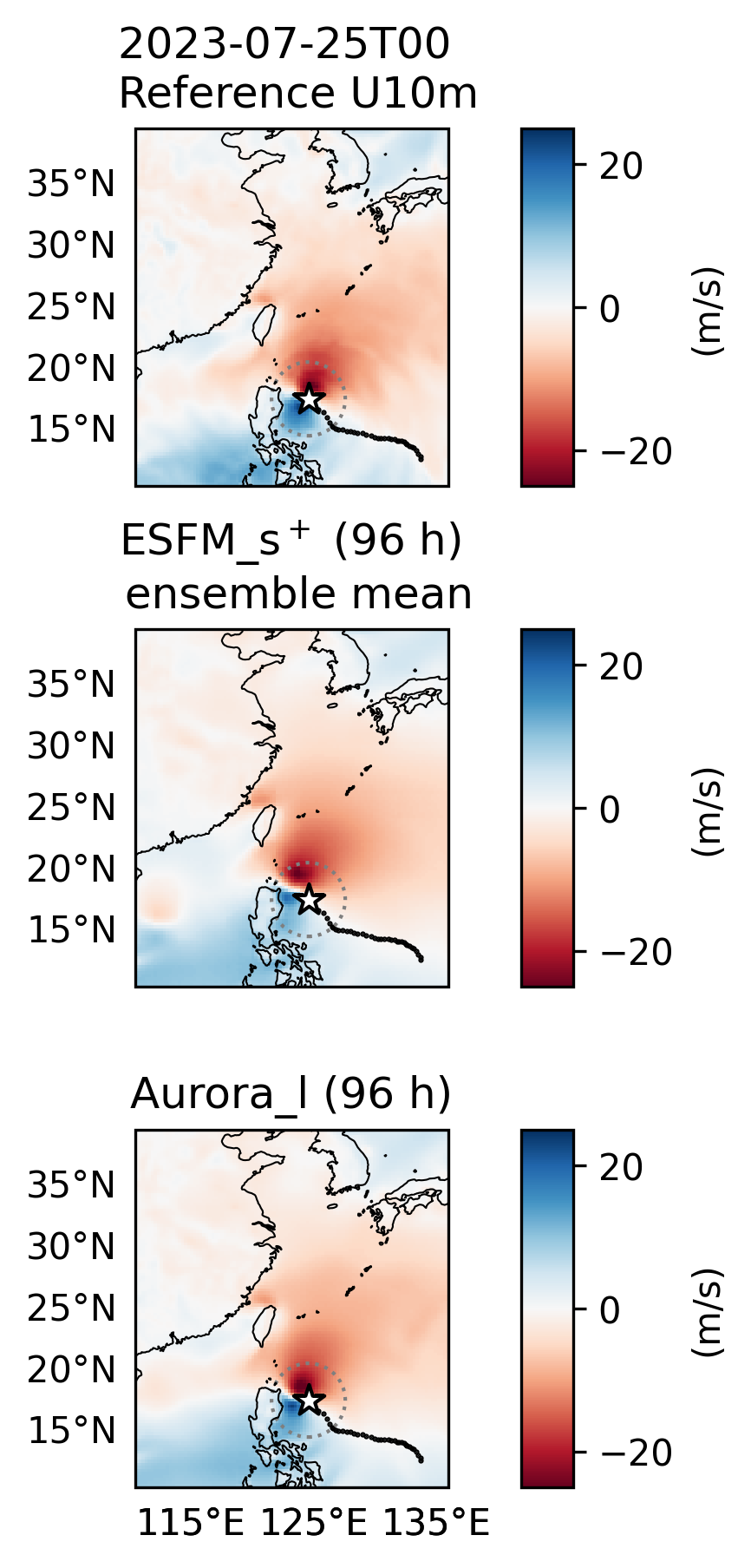}
    \caption{\textbf{Maximum wind velocity (left) and location (right) of Super Typhoon Doksuri 2023}: 
    (left) ERA5 (black),  ESFM\_s (orange) and Aurora models initialized on 21.07.2023 at midnight: Aurora large (cyan dashed line), Aurora small (blue solid line), ESFM small with 8 probabilistic decoders. 
    (right) Map of eastward wind velocity on 25.07.2023 (96-hour lead time). The IBTrACS best track is shown in all figures.}
    \label{fig:finetuning_doksuri_timeseries}
\end{figure}

\subsection{Case study I: Tropical cyclone Doksuri  (July 2023)}
\label{sec:doksuri}
Super Typhoon Doksuri (2023) is selected as an out-of-training and extreme weather test case. 
Tropical cyclones are events with high socio-economic impact and are therefore often used as test cases for ML weather prediction models. 
MLWP models tend to accurately forecast the position of cyclones, but underestimate their intensity \citep{dacre2026northern}. 
This bias relates partly to the resolution of the underlying training data. 
Doksuri is an exceptionally powerful typhoon that hit South-East Asia in summer 2023. 
We test ESFM\_s accuracy in reproducing the position and intensity of the typhoon and study the benefits of the ensemble ESFM\_s in this respect using ERA5 and Aurora as benchmarks. 
Intensity is defined here in terms of maximum wind speed.

Figure~\ref{fig:finetuning_doksuri_timeseries} shows the maximum wind speed for 10-day rollout initialized on 21.07.2023 at midnight, which is around the time Doksuri formed east of the Philippines, and the track forecast with a lead time of 96 hours. 
The maximum wind velocity is computed as the maximum amplitude in a radius of three degrees around the cyclone eye, as provided by IBTrACS~\citep{knapp_international_2010,kenneth_international_2019}. 
The typhoon location of ESFM\_s and Aurora after 96 hours agrees well with the reference (Fig.~\ref{fig:finetuning_doksuri_timeseries}). 
The timing of the landfall is also fairly well predicted (not shown). 
This is in agreement with the current literature. 
For the intensity, however, a different picture emerges. 
During the period between July 21st and July 27th, the ESFM\_s ensemble predicts maximum wind speed close to ERA5, which is also the case for the large Aurora model, while the small Aurora model, which has a comparable size to ESFM\_s, underestimates the tropical cyclone intensity (Fig.~\ref{fig:finetuning_doksuri_timeseries}). 
This highlights the efficiency of the ESFM framework. 
Maximum wind speeds tend to be underestimated by the ESFM\_s ensemble during landfall, which also seems to affect Aurora. 
After landfall on July 29th, the velocities again fall within the uncertainty range. 
It is worth noting, however, that ERA5 systematically underestimates tropical cyclone intensity, with observed wind speeds reaching nearly 60 m\,s$^{-1}$~\citep{tang2025monsoonal}, approximately twice those recorded in the reanalysis.

\begin{figure}[t!]
    \centering
    \includegraphics[width=\linewidth]{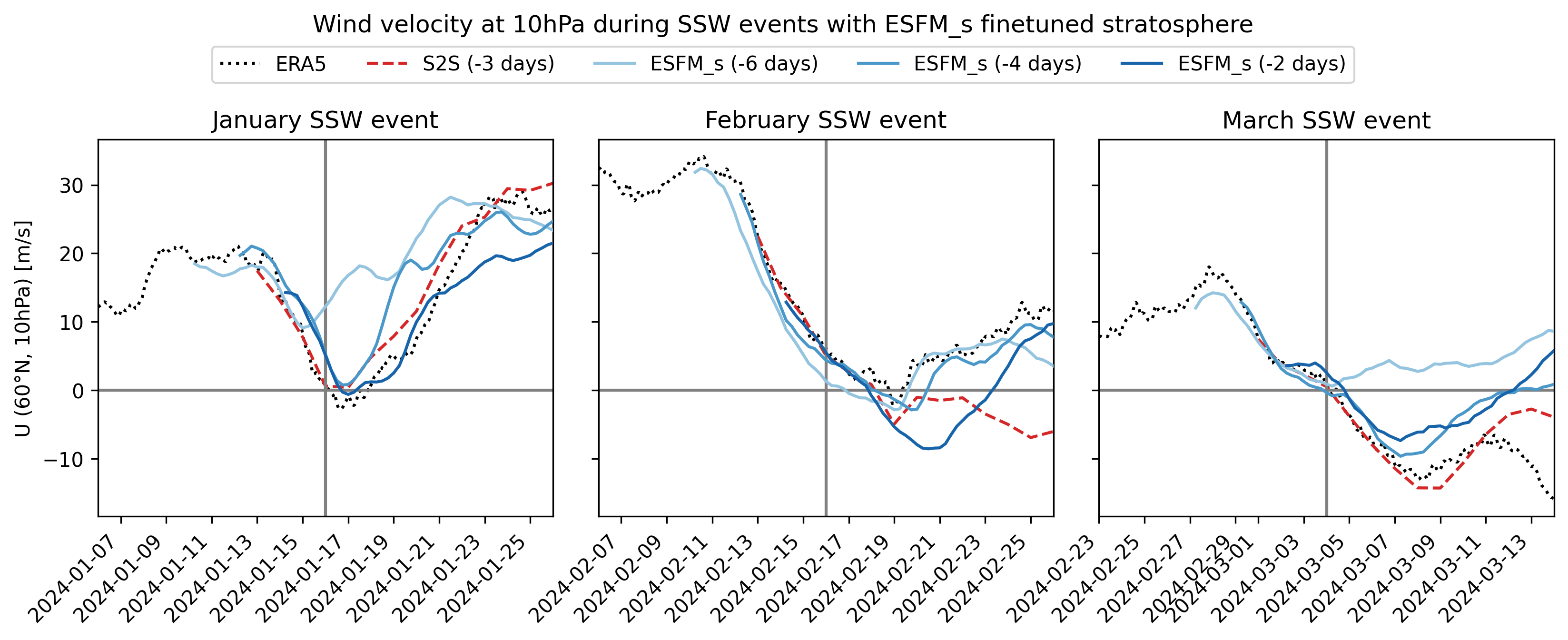}
    \caption{\textbf{Sudden stratospheric warming:} Eastward wind velocity at 10\,hPa and latitude 60$^{\circ}$N during three SSW events. 
    Each panel shows one event, with the vertical line denoting the start of the SSW event. 
    ESFM\_s is initialized on three dates up to 6 days before the SSW event (blue lines). 
    The numerical S2S baseline model is initialized three days before the event (red dashed line). 
    The wind velocity is averaged along the longitude dimension.}
    \label{fig:SSW_timeseries}
\end{figure}

\subsection{Case study II: Sudden stratospheric warmings 2023/24}
\label{sec:ssw}

The state of the stratosphere is a strong predictor for tropospheric weather on the subseasonal-to-seasonal (S2S) time scales \citep{baldwin2001_Stratospheric}. 
In particular, major stratospheric changes such as Sudden Stratospheric Warmings (SSW) may have strong impacts on mid-latitude surface weather during weeks and cause high-impact weather such as cold spells \citep{baldwin2021_Sudden}. 
However, numerical models struggle to represent the dynamics in the lower stratosphere and its coupling with the troposphere \citep{taguchi2018_Comparison,kuchar2024_Largeensemble}. 
Most data-driven models to date lack a representation of pressure above 50\,hPa since the expected gain is small for short lead times~(e.g., \cite{bodnar2025foundation,bi2022panguweather}).
In this section, we take advantage of the vertical masking training strategy to finetune ESFM\_s with 16 pressure levels instead of the standard 13 ones (adding 10, 20, and 70\,hPa). 
The benefits of including a representation of the stratosphere are explored studying the winter season 2023/24, which reported an unusually high number of SSW events~\citep{qian_enhanced_2024}.

First, Fig.~\ref{fig:SSW_timeseries} presents the prediction of SSW onset, length, and strength, in terms of the reversal in the zonal wind direction at 10\,hPa away from the otherwise predominant westerlies ($u>0$).
The predictive quality regarding the SSW onset is comparable to the numerical S2S model by ECMWF~\citep{vitartSubseasonalSeasonalPrediction2018}, with the usual lead time dependency. 
Performance during the days following the SSW is more inconsistent. 
While both the ECMWF S2S model and ESFM\_s capture the weak January SSW event fairly well, the former misses the return of the westerlies after the February event, and the latter underestimates the strength and duration of the March SSW.

\begin{figure}[t!]
    \centering
    \includegraphics[width=0.9\linewidth]{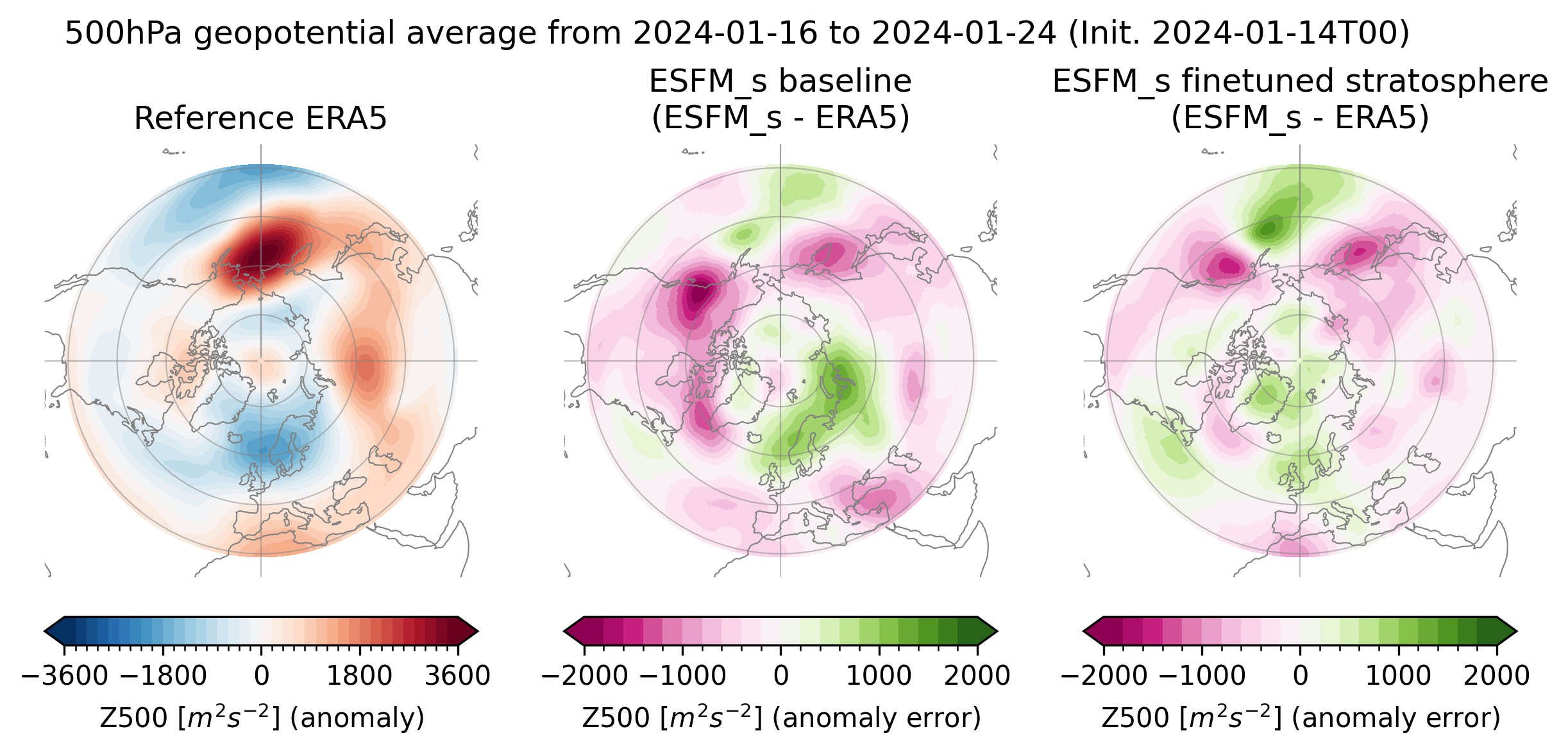}
    \caption{\textbf{Stratosphere-troposphere coupling:} 500\,hPa geopotential poleward of 30$^\mathrm{o}$N latitude averaged over the week following the January SSW, displayed as anomaly compared to the ERA5 climatology. 
    Models are initialized two days before the event. 
    (Left) ERA5 reference geopotential height anomaly, (middle) error in the anomaly predicted by the baseline ESFM, (right) error in the anomaly predicted by ESFM finetuned with stratospheric levels.}
    \label{fig:SSW_maps}
\end{figure}

Regarding the impact on tropospheric weather regimes, an accurate prediction of troughs and ridges (negative and positive anomalies in geopotential height (GH), respectively) at the steering level of 500\,hPa is essential.
As shown in Fig.~\ref{fig:SSW_maps}, the January SSW event is followed by a ridge anomaly over the North Pacific, which favors cold air outbreaks over North America, and a trough regime over Northern Europe, which is associated with cold and stormy conditions~\citep{baldwin2021_Sudden}.
The exact position and amplitude of both anomalies is predicted more accurately using the ESFM\_s that has been finetuned including stratospheric levels; however, both models tend to simulate a ridge anomaly that extends too much into the zonal direction.

These experiments indicate that equipping ESFM\_s with a stratospheric representation leads to a generally skillful prediction of SSW events at low stratospheric levels, as well as improved forecasts of sub-seasonal weather regimes in the troposphere.

\subsection{Case study III: Multidecadal runs and long-term stability}
\label{sec:climate_forcings}

In addition to medium-range weather forecasting and sub-seasonal predictions, the ability of foundation models to remain stable over several simulated decades is crucial for studying climate on long timescales. 
To foster the adoption of foundation models also for tasks related to decadal variability, we study the long-term stability of ESFM\_s on decadal time scales. %

\begin{figure}[t!]
    \centering
    \includegraphics[width=0.7\linewidth]{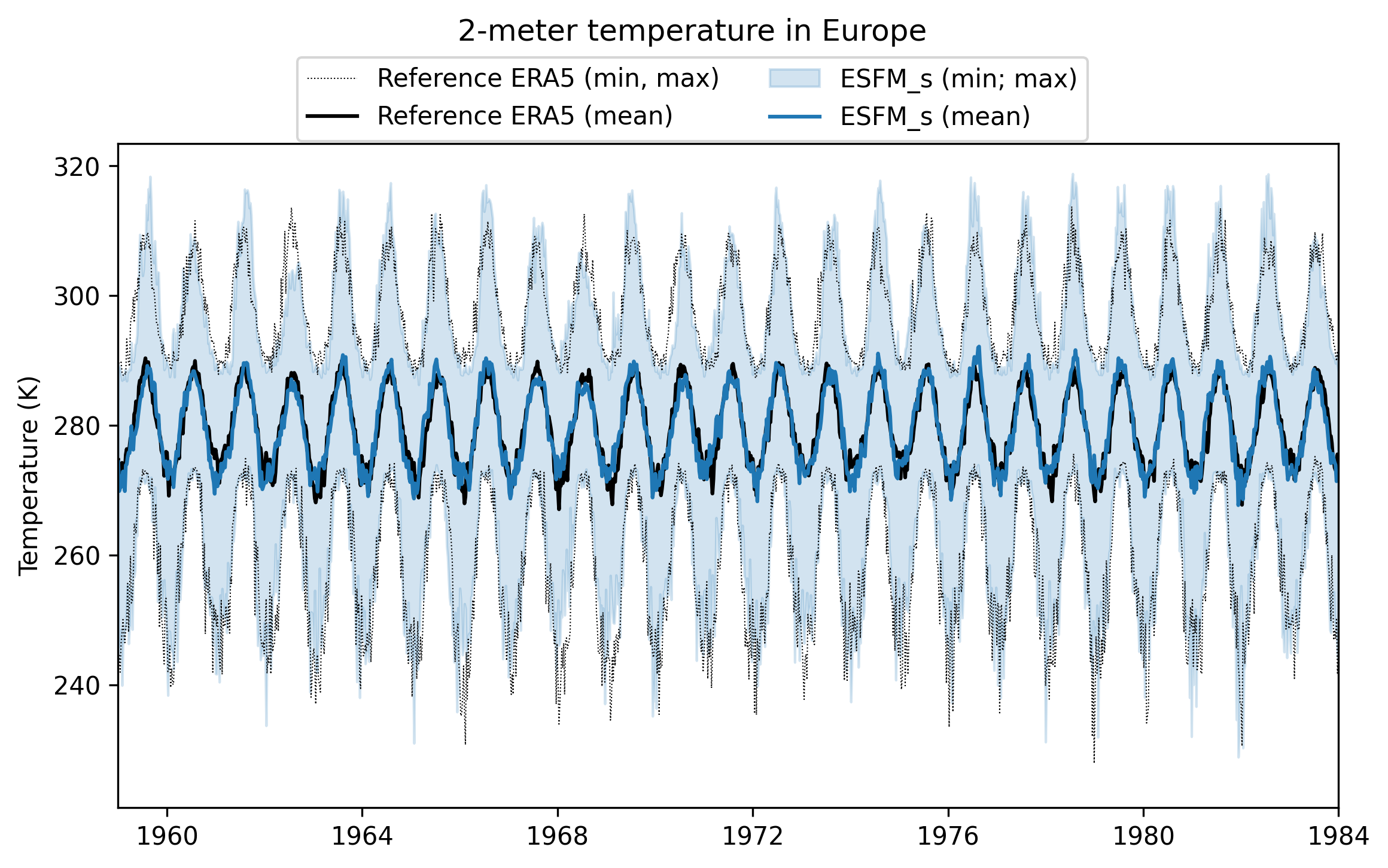}
    \caption{\textbf{Multidecadal rollout stability} The shown 25-year long rollout was initialized on 02.01.1959. 
    The solid line shows the spatial average over Europe, and the shaded area shows the range from the spatial minimum to the spatial maximum. 
    ERA5 is shown as a reference. 
    Points are sampled every seven days.}
    \label{fig:climate_forcings_rollout}
\end{figure}

Figure~\ref{fig:climate_forcings_rollout} shows the 25-year rollout forecasts for surface temperature over Europe, initialized on January 2nd, 1959. 
ESFM\_s remains stable over a multidecadal autoregressive rollout, eliminating the need for manual constraints on mass or the global energy balance.  
ESFM\_s preserves the seasonal cycle and maintains physically realistic temperature distributions, as indicated by the shaded areas, closely following the ERA5 reference. 
The extent to which it simulates atmospheric teleconnections driven by internal climate modes, such as ENSO~\citep{liu2007atmospheric}, or the exact energy and mass balance is left for future research. 

\section{ESFM performance with partially missing and sparse data}

\subsection{Forecasting with data gaps or missing variables}
\label{sec:masked_experiments}

This section examines the forecasting performance of the ESFM\_s model when specific regions, variables or levels are missing from the observations using the same ERA5 test set. 
The first task of the model is therefore to make regional short-term forecasts using only observations from data points outside the target region. 
The second task is to forecast a variable that is not present at the initial time point, for example due to a data gap. 
Instead, it forecasts the variable using all the other variables present in the data. 
A novel evaluation plot, adapted from~\citet{Brochet2023}, is used to assess the consistency of the relationship between the target variables and other variables.
The final task is to forecast atmospheric variables for a target pressure level when no data points are available across any variable at that pressure level in the observations.

\subsubsection{Regional data missing from observations}
\label{sec:regional-masking}

\begin{figure}
    \centering
    \includegraphics[width=0.5\linewidth]{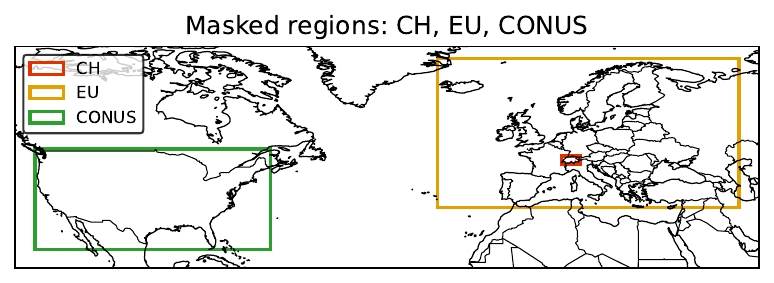}
    \caption{Bounding boxes of the regions masked from the test set in experiments under Sec.~\ref{sec:regional-masking}.}
    \label{fig:masked-regions}
\end{figure}

We evaluate six hour lead time forecast performance of three distinct regions (Fig.~\ref{fig:masked-regions}) when these regions are completely masked in the input observations across all surface and atmospheric variables: Switzerland (CH), Europe (EU), and the contiguous United States (CONUS) in Table~\ref{tab:consolidated_regional_masking}.

Global MAE performance drop in forecasts when a region is removed from observations ranges between 0.0-0.0\% in Switzerland, 5.2\% (V10m) - 19.7\% (Z500) in EU, and 4.5\% (V10m) - 11.7\% (Z500) in the CONUS, when compared to ESFM\_s that had all regions available in the observations.
In addition to showing global performance under regionally missing observations, we also show performance purely within the removed region for Switzerland, Europe, and the contiguous United States in Table~\ref{tab:regional_masking}.
Regional forecast MAE performance drop when all observations within the region are removed ranges between 9.7\% (Z500) - 33.2\% (Q850) in Switzerland, 107\% (Q850) - 419 (Z500)\% in EU, and 117\% (T2m) - 321\% (Z500) in the CONUS, when compared to all observations being available.
While relative errors are high, especially when the considered region that is removed from observations gets as large as a continent, the MAE across all variables is only strongly impacted for geopotential at 500\,hPa. 

\begin{figure}[b!]
    \centering
    \includegraphics[width=0.7\textwidth]{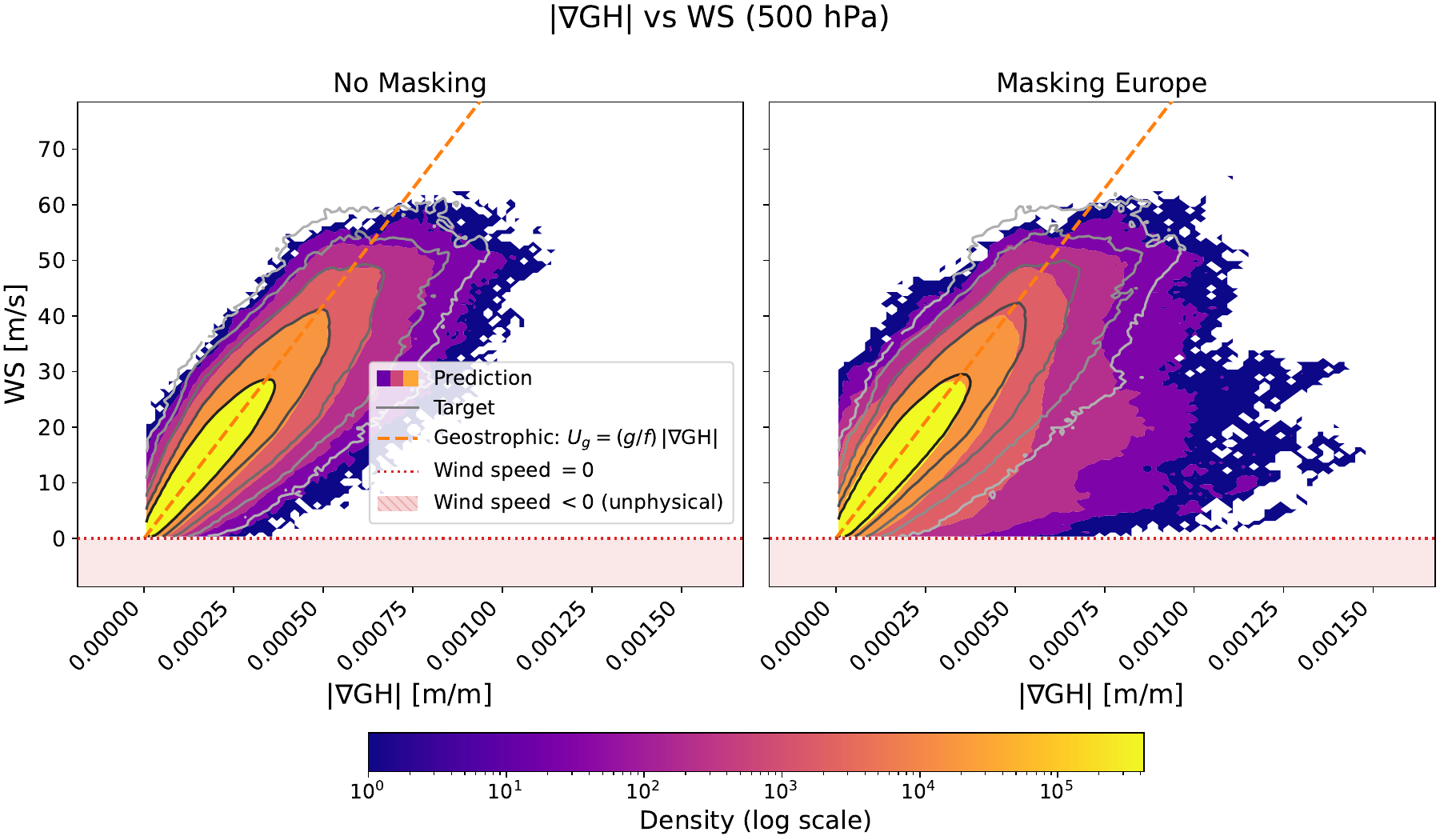}
    \caption{\textbf{Physical consistency:} Joint density of geopotential height (GH) gradients and wind speed 
    (WS) at 500\,hPa over Europe
    (35°N--72°N, 25°W--50°E) for ESFM\_s without masking (\textit{left}) and
    with the European region withheld from the initial conditions
    (\textit{right}).
    ERA5 reference (gray isolines); ESFM prediction (colors; note the log density scale). Additionally shown are geostrophic balance (orange line) and zero wind speed (red line).}
    \label{fig:bivariate_europe}
\end{figure}

To investigate whether masking of the initial conditions induces physically inconsistent behavior, we examine the joint distribution of geopotential height gradient ($|\nabla GH|$) and wind speed (WS) at 500\,hPa, computed \emph{within} the European domain where that is witheld at forecast initial time (Fig.~\ref{fig:bivariate_europe}).
Under the geostrophic approximation $U_g = (g/f)\,|\nabla GH|$, WS scales linearly with the local horizontal pressure gradient, providing a physically grounded reference against which model fidelity can be assessed.

The predicted distribution of ESFM\_s under both masked and fully observed initial conditions remains above zero and the bulk of the probability mass tracks the geostrophic slope (orange line in Fig.~\ref{fig:bivariate_europe}), confirming that withholding an entire continent's observations from the initial conditions does not force the model into physically implausible wind states and that the model has learned geostrophic wind balance to a certain extent.
Nevertheless, when European observations are withheld from the initial conditions, the predicted distribution shifts toward lower WS relative to the ERA5 reference, revealing a tendency to underestimate wind strength within the masked region for high gradients.
This effect appears more pronounced in the European domain than in the equivalent analysis over the contiguous United States (not shown).
Rather than a simple amplitude shift, the predicted distribution broadens, with a pronounced accumulation of probability mass in the region where WS falls below what $|\nabla GH|$ alone would predict.
This points to a flow regime in which European observations in the initial conditions are necessary to accurately reconstruct the wind field, as no such broadening is observed when the initial conditions are fully observed.
This may reflect the complexity of European flow along steep coastlines or topographic regions when $|\nabla GH|$ can be large.
Although the predicted WS is erroneously too low, further investigation is needed to shed light on this effect.

\begin{table}[t!]
\centering
\caption{Global six-hour lead time forecast performance of ESFM\_s with masked training when different regions are removed from observations (Switzerland, Europe, CONUS). 
For reference, we also include results when test observations are dense; ESFM\_s as ESFM\_s with masked training protocol and ESFM\_s* as ESFM\_s trained without the masking protocol.
Metrics shown are Mean Absolute Error (MAE) and Pearson Correlation Coefficient (PCC). 
}
\label{tab:consolidated_regional_masking}
\resizebox{\textwidth}{!}{%
\begin{tabular}{@{}llcccccccccc@{}}
\toprule
& & \multicolumn{2}{c}{T850} & \multicolumn{2}{c}{Z500} & \multicolumn{2}{c}{T2m} & \multicolumn{2}{c}{U10m} & \multicolumn{2}{c}{V10m} \\
\cmidrule(lr){3-4} \cmidrule(lr){5-6} \cmidrule(lr){7-8} \cmidrule(lr){9-10} \cmidrule(l){11-12}
Masked region & Model & MAE $\downarrow$ & PCC $\uparrow$ & MAE $\downarrow$ & PCC $\uparrow$ & MAE $\downarrow$ & PCC $\uparrow$ & MAE $\downarrow$ & PCC $\uparrow$ & MAE $\downarrow$ & PCC $\uparrow$ \\ \midrule
CH & ESFM\_s  & 0.248 & 0.999 & 15.908 & 1.000 & 0.269 & 0.999 & 0.280 & 0.993 & 0.290 & 0.991 \\
EU & ESFM\_s  & 0.267 & 0.999 & 19.046 & 1.000 & 0.291 & 0.999 & 0.296 & 0.992 & 0.305 & 0.989 \\
CONUS & ESFM\_s & 0.264 & 0.999 & 17.771 & 1.000 & 0.285 & 0.999 & 0.293 & 0.992 & 0.303 & 0.989 \\
\midrule
\multirow{2}{1.5cm}{Dense \\ Observations} 
& ESFM\_s & 0.248 & 0.999 & 15.910 & 1.000 & 0.269 & 0.999 & 0.280 & 0.993 & 0.290 & 0.991 \\
& ESFM\_s*& 0.224 & 0.999 & 12.535 & 1.000 & 0.248 & 1.000 & 0.253 & 0.994 & 0.262 & 0.992 \\
\bottomrule
\end{tabular}%
}
\end{table}

\begin{table}[b]\centering
\caption{Regional six hour lead time forecast performance of ESFM\_s trained with masking when corresponding region is removed across the surface and atmospheric variables of the observations, shown in mean absolute error (MAE) and Pearson Correlation Coefficient (PCC).
For reference, we also list results when test observations are fully available; ESFM\_s as ESFM\_s trained with masking and ESFM\_s* trained without the masking protocol.
}
\label{tab:regional_masking}
\resizebox{\textwidth}{!}{%
\begin{tabular}{@{}lllcccccccccccc@{}}
\toprule
 & & & \multicolumn{2}{c}{T850} & \multicolumn{2}{c}{Z500} & \multicolumn{2}{c}{Q850} & \multicolumn{2}{c}{T2m} & \multicolumn{2}{c}{U10m} & \multicolumn{2}{c}{V10m} \\
\cmidrule(lr){4-5}\cmidrule(lr){6-7}\cmidrule(lr){8-9}\cmidrule(lr){10-11}\cmidrule(lr){12-13}\cmidrule(lr){14-15}
\multirow{-2}{1.2cm}{Samples \\ masked?} & Evaluated region & Model & MAE $\downarrow$ & PCC $\uparrow$ & MAE $\downarrow$ & PCC $\uparrow$ & MAE $\downarrow$ & PCC $\uparrow$ & MAE $\downarrow$ & PCC $\uparrow$ & MAE $\downarrow$ & PCC $\uparrow$ & MAE $\downarrow$ & PCC $\uparrow$ \\
\midrule
\checkmark & CH & ESFM\_s & 0.423 & 0.993 & 19.464 & 1.000 & 3.69e-4 & 0.949 & 0.791 & 0.982 & 0.275 & 0.933 & 0.294 & 0.941 \\
\checkmark & EU & ESFM\_s & 0.660 & 0.982 & 80.762 & 0.993 & 3.93e-4 & 0.915 & 0.800 & 0.973 & 0.624 & 0.939 & 0.620 & 0.940 \\
\checkmark & CONUS & ESFM\_s & 1.059 & 0.980 & 88.363 & 0.994 & 8.90e-4 & 0.899 & 1.231 & 0.979 & 0.885 & 0.880 & 0.937 & 0.889 \\
\midrule
$\times$ & \multirow{2}{*}{CH}
 & ESFM\_s & 0.367 & 0.994 & 17.749 & 1.000 & 2.77e-4 & 0.965 & 0.637 & 0.987 & 0.249 & 0.945 & 0.267 & 0.951 \\
 $\times$& & ESFM\_s* & 0.346 & 0.995 & 13.474 & 1.000 & 2.61e-4 & 0.968 & 0.586 & 0.989 & 0.233 & 0.951 & 0.250 & 0.957 \\
 $\times$& \multirow{2}{*}{EU}
 & ESFM\_s & 0.242 & 0.998 & 15.562 & 1.000 & 1.90e-4 & 0.981 & 0.314 & 0.997 & 0.292 & 0.989 & 0.288 & 0.989 \\
 $\times$& & ESFM\_s* & 0.220 & 0.998 & 12.166 & 1.000 & 1.72e-4 & 0.984 & 0.289 & 0.998 & 0.265 & 0.991 & 0.262 & 0.991 \\
 $\times$ & \multirow{2}{*}{CONUS}
 & ESFM\_s & 0.391 & 0.997 & 20.978 & 1.000 & 3.56e-4 & 0.983 & 0.567 & 0.996 & 0.405 & 0.977 & 0.414 & 0.980 \\
 $\times$ & & ESFM\_s* & 0.356 & 0.998 & 15.959 & 1.000 & 3.25e-4 & 0.985 & 0.520 & 0.996 & 0.370 & 0.980 & 0.380 & 0.983 \\
\bottomrule
\end{tabular}%
}
\end{table}

The thermodynamic consistency of predictions under masked initial conditions is further confirmed in Appendix~\ref{sec:bivariate_appendix} (Fig.~\ref{fig:bivariate_europe_tq}), where the joint T--Q distribution at 500\,hPa remains below the saturation curve in both the masked and unmasked configurations.
This shows that ESFM has learned the nonlinear relationship between Q, T, and P, which allows it to skillfully reproduce regions absent from the initial conditions.

\subsubsection{Predicting variables absent from observations}

In Table~\ref{tab:variable_masking}, we present the six-hour lead time forecast performance of individual variables when that particular variable was removed from the input across all spatial positions. 
It is important to note that, for atmospheric variables, all pressure levels of a given variable are removed from the observations. 

\begin{figure}[htbp]
    \centering
    \includegraphics[width=0.7\textwidth]{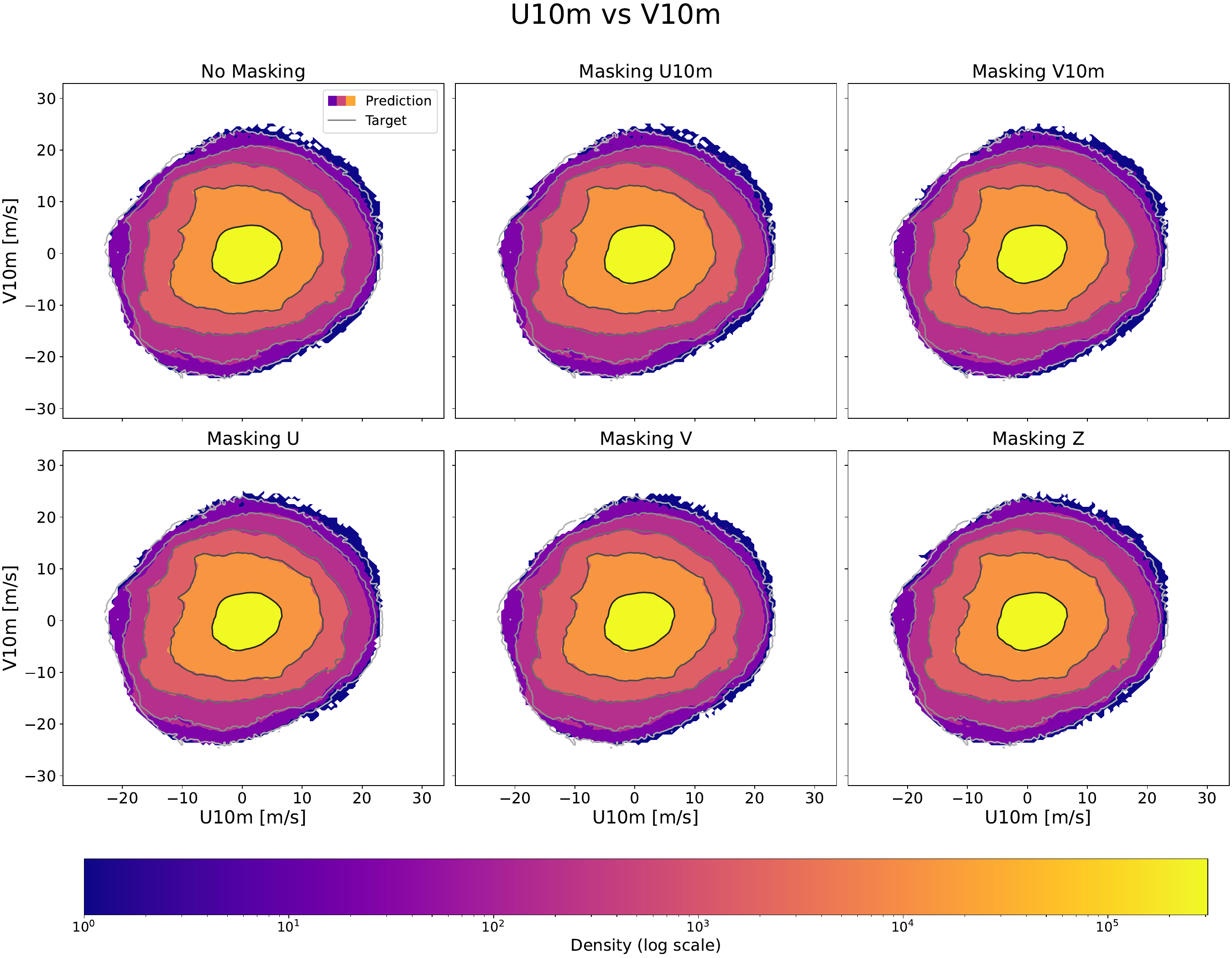}
    \caption{\textbf{Physical consistency:} Joint density of U10m and V10m evaluated over Europe (35°N--72°N, 25°W--50°E) for ESFM\_s under five distinct variable-masking configurations.
    In each panel a different variable is withhold from the initial conditions across all pressure and surface levels: U10m, V10m, U, V and Z.
    ESFM prediction  (color); ERA5 reference (gray isolines). Note the logarithmic density scale.}
    \label{fig:bivariate_10m_wind_var_masking}
\end{figure}

We further examine the joint distribution of U10m and V10m as one possible diagnostic for cross-variable consistency under variable masking, evaluated over the European domain (Fig.~\ref{fig:bivariate_10m_wind_var_masking}). 
Surface winds are coupled to U and V through, e.g., vertical momentum transport in the boundary layer, and Z via, e.g., the pressure-gradient force driving the large-scale circulation. 
Therefore, masking any of these fields should, in principle, affect near-surface wind predictions. 
Yet the predicted distributions are nearly indistinguishable across all panels, including when U10m or V10m are directly withheld from the initial conditions. 
The same visual indistinguishability holds for the joint distribution of U and V at 500\,hPa (not shown).
This indicates that ESFM\_s recovers the wind joint distribution from the remaining variables with sufficient fidelity that variable masking leaves no discernible imprint on the wind statistics over Europe. 
Identical bivariate distributions do not, however, imply identical point-wise accuracy: the histogram captures the joint distribution of two variables pooled over all locations and time steps, whereas MAE in Table~\ref{tab:variable_masking} measures sample-by-sample correspondence between prediction and ground truth at each location and time step. 
The two are therefore complementary, with the bivariate analysis serving as a diagnostic of physical consistency and the table quantifying forecast accuracy.

\begin{table}[]
\caption{Six hour lead time forecast performance of ESFM\_s trained with masking when different variables are removed from observations, across all pressure levels, shown in mean absolute error.
Therefore, each atmospheric and surface variable in the row of ESFM\_s is computed in a separate evaluation, where the corresponding variable is removed from observations.
For reference, we also list ESFM\_s trained with masking and ESFM\_s* trained without the masking protocol, where none of the variables in the observations are removed.
}
\label{tab:variable_masking}
\resizebox{\textwidth}{!}{%
\begin{tabular}{@{}llllllllllllllllllll@{}}
\toprule
&\\
\multirow{-2}{1.cm}{Samples \\ masked?}& & T850 & T500 & T100 & U850 & U500 & U100 & V850 & V500 & V100 & Q850 & Q500 & Q100 & Z850 & Z500 & Z100 & T2m & U10m & V10m \\
\midrule
\checkmark & ESFM\_s & 0.284 & 0.210 & 0.398 & 0.517 & 0.662 & 0.864 & 0.510 & 0.648 & 0.816 & 3.70e-4 & 1.63e-4 & 1.36e-7 & 16.731 & 21.260 & 39.298 & 0.321 & 0.286 & 0.294 \\
\midrule
$\times$ & ESFM\_s & 0.248 & 0.179 & 0.297 & 0.450 & 0.527 & 0.639 & 0.454 & 0.541 & 0.656 & 2.48e-4 & 8.02e-5 & 4.09e-8 & 14.368 & 15.910 & 29.319 & 0.269 & 0.280 & 0.290 \\
$\times$ & ESFM\_s* & 0.224 & 0.156 & 0.259 & 0.406 & 0.473 & 0.565 & 0.410 & 0.486 & 0.580 & 2.24e-4 & 7.32e-5 & 3.54e-8 & 11.831 & 12.535 & 21.873 & 0.248 & 0.253 & 0.262 \\
\bottomrule
\end{tabular}%
}
\end{table}

\subsubsection{Predicting atmospheric variables on pressure levels absent from observations}

Table~\ref{tab:plev_masking} lists six hour lead time forecast performance at specific pressure levels when all atmospheric variables at those levels are masked entirely in input observations.
In particular, we selected levels near the top of the atmosphere (100\,hPa), at half the surface pressure (500\,hPa), and closer to the surface (850\,hPa) to evaluate forecasting performance under different environmental conditions.

\begin{figure}[h!]
    \centering
    \includegraphics[width=0.7\textwidth]{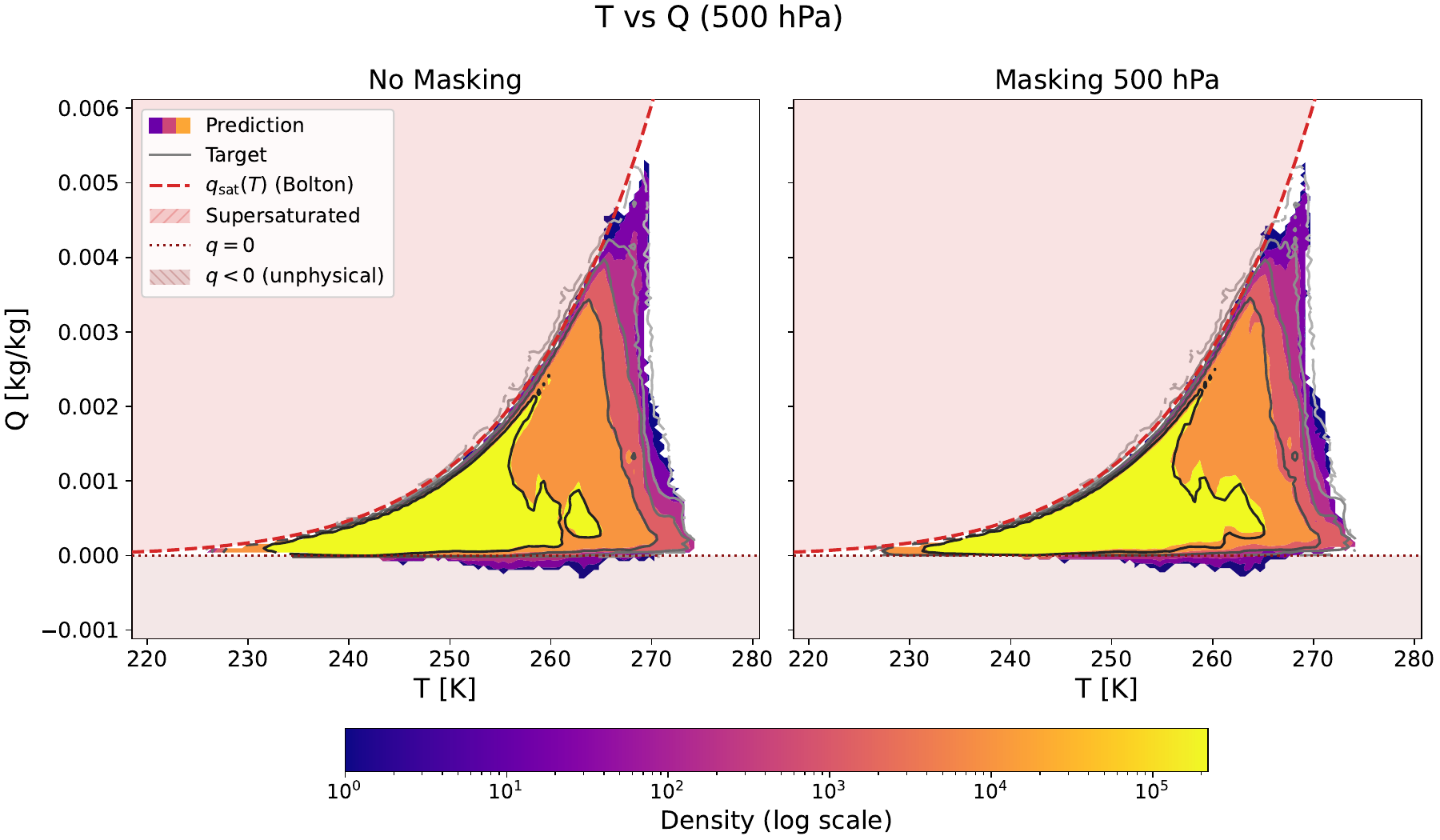}
    \caption{\textbf{Physical consistency:} Joint density of T and Q at
    500\,hPa evaluated over Europe (35°N--72°N, 25°W--50°E) for
    ESFM\_s without masking (\textit{left}) and with the 500\,hPa level withheld from the initial conditions (\textit{right}).
   ERA5 reference (Gray isolines); ESFM prediction (color). Note the logarithmic
    density scale.
    The red dashed curve is $q_\mathrm{sat}(\mathrm{T})$ following~\citet{Bolton1980}.}
    \label{fig:bivariate_plev500_tq}
\end{figure}

Beyond scalar error metrics, we assess whether the model preserves key thermodynamic constraints at the fully withheld 500\,hPa level.
Figure~\ref{fig:bivariate_plev500_tq} shows the joint distribution of T and Q at 500\,hPa, with the saturation specific humidity $q_\mathrm{sat}(T)$~\citep{Bolton1980} overlaid as a physical upper bound.
Remarkably, even when the entire 500\,hPa level is withheld from the initial conditions, the predicted distribution remains (mostly) below the saturation curve and above zero, meaning the model does not generate supersaturated or negative-humidity states at this unobserved level.
The configuration with masked initial conditions closely mirrors the fully observed baseline, with the colored model prediction contours tracking the ERA5 reference along the Clausius-Clapeyron envelope in both cases.

A similarly consistent picture emerges from the geostrophic wind balance at the same level (Fig.~\ref{fig:bivariate_plev500_geo} in Appendix~\ref{sec:bivariate_appendix}): the proportionality between $|\nabla GH|$ and WS is retained under pressure level masking of the initial conditions, with no systematic collapse toward physically implausible states.
Interestingly, the reconstructed relationship is even tighter here than in the case of regional masking of the initial conditions discussed in the previous section.
Together, these results suggest that ESFM\_s is able to reconstruct physically consistent atmospheric states at pressure levels entirely absent from its initial conditions.
Note that neither the geostrophic wind balance nor the Clausius-Clapeyron envelope are directly enforced during training.

\begin{table}[!t]
\caption{Six-hour lead time forecast performance of ESFM\_s for when different pressure levels (indicated in second row, [hPa]) are removed from observations, individually, shown in mean absolute error.
For reference, we also show the performance of two ESFM\_s models where observations are dense and no pressure levels are removed from observations: trained with masking, ESFM\_s, and trained without masking protocol, ESFM\_s*.}
\label{tab:plev_masking}
\resizebox{\textwidth}{!}{%
\begin{tabular}{@{}llccccccccccccccc@{}}
\toprule
 &  & \multicolumn{3}{c}{T} & \multicolumn{3}{c}{U}& \multicolumn{3}{c}{V} & \multicolumn{3}{c}{Q} & \multicolumn{3}{c}{Z}\\
\cmidrule(lr){3-5}\cmidrule(lr){6-8}\cmidrule(lr){9-11}\cmidrule(lr){12-14}\cmidrule(lr){15-17}
\multirow{-2}{1.cm}{Samples \\ masked?}  & & 850 & 500 & 100 & 850 & 500 & 100 & 850 & 500 & 100 & 850 & 500 & 100 & 850 & 500 & 100 \\ \midrule
\checkmark & ESFM\_s & 0.327 & 0.234 & 0.525 & 0.579 & 0.802 & 1.294 & 0.566 & 0.778 & 1.173 & 3.79e-4 & 1.57e-4 & 1.43e-7 & 16.103 & 17.449 & 44.114  \\
\midrule
$\times$ & ESFM\_s & 0.248 & 0.179 & 0.297 & 0.450 & 0.527 & 0.639 & 0.454 & 0.541 & 0.656 & 2.48e-4 & 8.02e-5 & 4.09e-8 & 14.368 & 15.910 & 29.319 \\
$\times$ & ESFM\_s* & 0.224 & 0.156 & 0.259 & 0.406 & 0.473 & 0.565 & 0.410 & 0.486 & 0.580 & 2.24e-4 & 7.32e-5 & 3.54e-8 & 11.831 & 12.535 & 21.873 \\
\bottomrule
\end{tabular}
}
\end{table}

\subsection{MODIS sparse satellite data}
\label{sec:modis_experiments}

Many key atmospheric observations are obtained from instruments onboard low-Earth-orbit satellites. 
Although these sensors provide valuable measurements, their spatial and temporal coverage is inherently limited: at any given time, they only observe a relatively narrow swath of the Earth, resulting in sparse instantaneous sampling on the global scale. 
A prominent example is the Moderate Resolution Imaging Spectroradiometer (MODIS), carried by the Terra and Aqua satellites. 
MODIS observes the Earth from an orbital altitude of about 705 km, covering a swath of approximately 2330 km and revisiting the globe in successive orbital segments. 
Based on these observations, total precipitable water (PWV) can be retrieved, but only over the observed swaths rather than continuously across the globe~\citep{gao2015modis}. 
As a result, users who need global PWV fields typically face two unsatisfactory options: 
either aggregate MODIS retrievals over one to two days to approximate global coverage, which still remains spatially incomplete and temporally inconsistent, or rely on computationally expensive data assimilation systems, such as the Integrated Forecasting System (IFS), to propagate the satellite information into a globally continuous product. 

\begin{figure}[h!]
    \centering
    \includegraphics[width=0.99\linewidth]{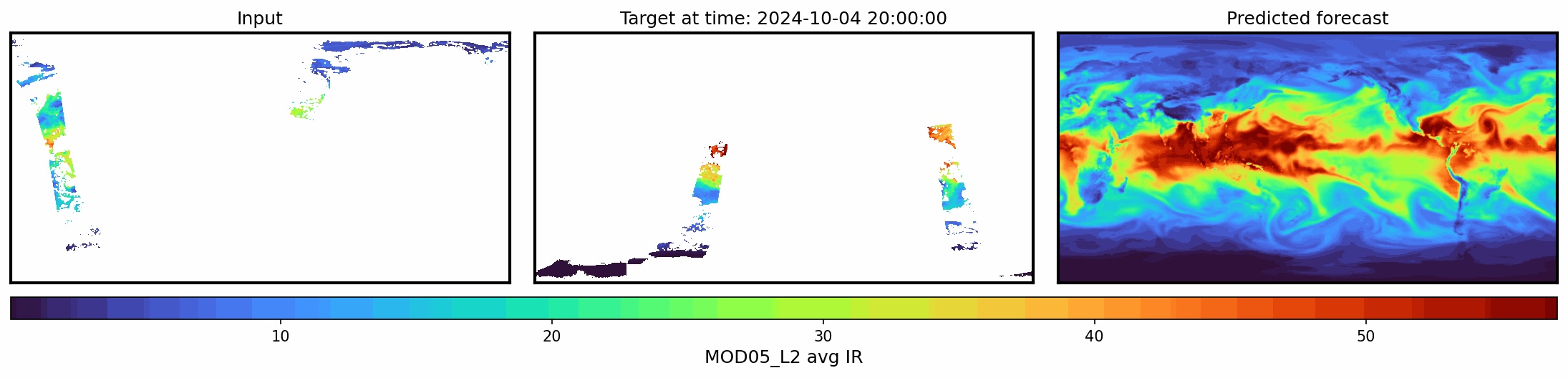} 
    \caption{\textbf{ESFM can handle very sparse observations for its prediction:} Sample PWV observation (MOD05 IR) from MODIS at times $t$ (input), $t$+$1$ (target), and forecasted PWV at time $t$+$1$ (six-hour lead time) by ESFM trained with masking protocol.}
    \label{fig:modis_sample}
\end{figure}

\subsubsection{Fine tuning with sparse data}
To construct an hourly MODIS-based dataset, we aggregate swath observations collected within each ~1-hour window (e.g., from HH:00 to HH:55) and assign them to the subsequent hourly timestamp (HH+1). 
The resulting fields are then interpolated onto the 0.25$^{\mathrm{o}}$ grid used by ERA5. 
Despite this temporal aggregation, the hourly MODIS fields remain highly sparse. 
For reference, in 2020 the valid pixel occupancy of the gridded MODIS PWV fields ranges from 0.0\% to 5.0\%, with a median of 3.3\%, depending on the observation time and retrieval type, namely infrared (IR) and near-infrared (NIR).

Using this MODIS dataset, we finetune ESFM\_s trained with masking to forecast four MODIS PWV\,[mm] variables simultaneously: MOD05 IR and NIR retrieved from the Terra platform, and MYD05 IR and NIR retrieved from the Aqua platform. 
To help the model better capture the dynamical evolution of these sparse satellite-derived PWV fields, we additionally include ERA5 atmospheric temperature and specific humidity as auxiliary variables during finetuning. 
These auxiliary fields are subjected to the same masking strategy described in Sec.~\ref{sec:training_protocols}. %
We highlight the dense forecasting capabilities of ESFM despite the very sparse observations and forecast targets through a visualization for a single qualitative sample from the test set in Fig.~\ref{fig:modis_sample}.

\begin{figure}[t!]
    \centering
    \begin{minipage}[t]{0.424\linewidth}
        \vspace{0pt}
        \includegraphics[width=\linewidth]{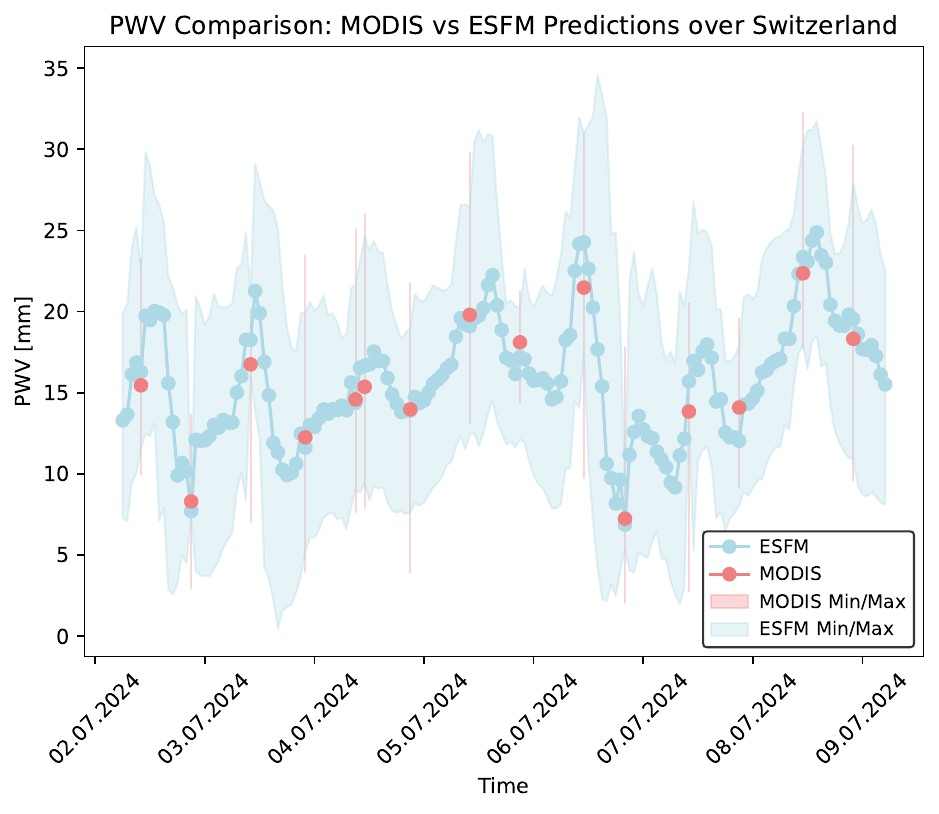}
    \end{minipage}%
    \hfill%
    \begin{minipage}[t]{0.424\linewidth}
        \vspace{0pt}
        \includegraphics[width=\linewidth]{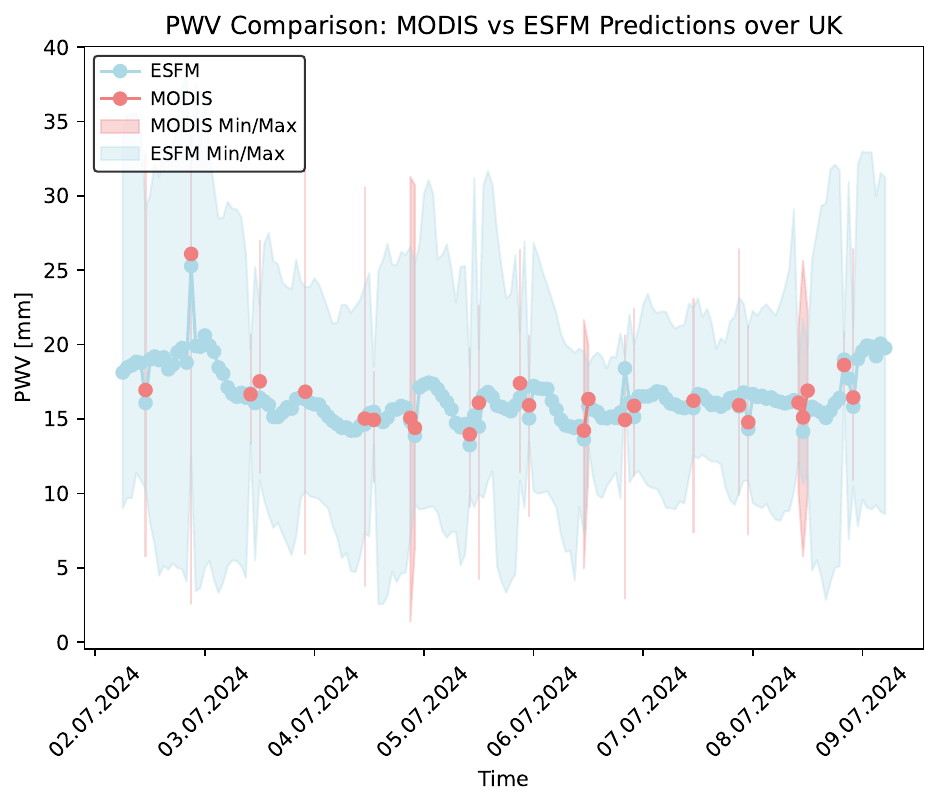}
    \end{minipage}%
    \hfill%
    \begin{minipage}[t]{0.15\linewidth}
        \vspace{0pt}
        \includegraphics[width=\linewidth]{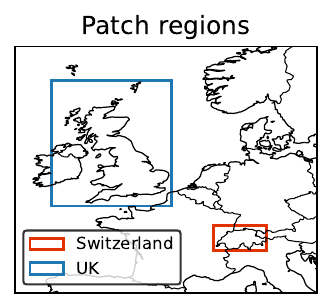}
    \end{minipage}
    \caption{\textbf{ESFM forecast finetuned on MODIS satellite data:} Comparison of six hour forecast of ESFM vs satellite observations (MOD05 IR based PWV) aggregated over Switzerland (left) and United Kingdom (right) for a sample week in 2024.}
    \label{fig:MODIS_patches}
\end{figure}

In Fig.~\ref{fig:MODIS_patches}, we compare six hour lead time forecast predictions within a patch over Switzerland (latitudes 45.75$^{\mathrm{o}}$ to 48$^{\mathrm{o}}$, longitudes 5.75$^{\mathrm{o}}$ to 10.5$^{\mathrm{o}}$) and the United Kingdom (latitudes 49.75$^{\mathrm{o}}$ to 61$^{\mathrm{o}}$, longitudes -8.75$^{\mathrm{o}}$ to 2$^{\mathrm{o}}$) with satellite observations for a sample week from the test set for the variable MOD05 IR. 
Since these observations are typically not available throughout the full patch, we also crop the predictions to the pixels where satellite data is available at a given time step, which sometimes yields the minor jumps on the plotted prediction curves.
The gaps along the time axis for MODIS samples showcase the sparsity of data, where not a single pixel within the considered patch had observations.

In Table~\ref{tab:MODIS_6h}, we show the global forecasting performance of ESFM\_s finetuned for the MODIS variables at a six-hour lead time. 
We compare the forecasts of ESFM\_s against PWV numerically derived from ERA5 atmospheric temperature, specific humidity, and geopotential at the target time. 
Specifically, ERA5 PWV is computed by reconstructing the vertical atmospheric structure from these variables and integrating the water vapor content throughout the atmospheric column~\citep{haase2003accuracy}.

\begin{table}[b!]
\caption{Six-hour lead time forecast performance of ESFMs\_s on MODIS variables shown in mean absolute error (MAE), Pearson correlation coefficient (PCC), and relative MAE (rMAE).
ERA5-derived PWV is computed from ERA5 atmospheric temperature, specific humidity, and geopotential through vertical integration of the atmospheric water vapor content. 
}
\label{tab:MODIS_6h}
\resizebox{\textwidth}{!}{%
\begin{tabular}{@{}lcccccccccccc@{}}
\toprule
             & \multicolumn{3}{c}{MOD05 L2 IR}
             & \multicolumn{3}{c}{MOD05 L2 NIR}
             & \multicolumn{3}{c}{MYD05 L2 IR}
             & \multicolumn{3}{c}{MYD05 L2 NIR} \\
\cmidrule(lr){2-4}
\cmidrule(lr){5-7}
\cmidrule(lr){8-10}
\cmidrule(lr){11-13}
             & MAE $\downarrow$ & PCC $\uparrow$ & rMAE $\downarrow$
             & MAE $\downarrow$ & PCC$\uparrow$ & rMAE $\downarrow$
             & MAE $\downarrow$ & PCC$\uparrow$ & rMAE $\downarrow$
             & MAE $\downarrow$ & PCC$\uparrow$ & rMAE $\downarrow$ \\
\midrule
ESFM\_s & 0.718 & 0.987 & 0.039 & 1.574 & 0.977 & 0.071 & 0.849 & 0.981 & 0.045 & 1.546 & 0.974 & 0.070 \\ %
ERA5-derived PWV & 1.470 & 0.964 & 0.080 & 2.800 & 0.938 & 0.129 & 1.556 & 0.956 & 0.082 & 2.251 & 0.944 & 0.106 \\
\bottomrule
\end{tabular}
}
\end{table}

\begin{figure}[bt!]
    \centering
    \includegraphics[width=0.499\linewidth]{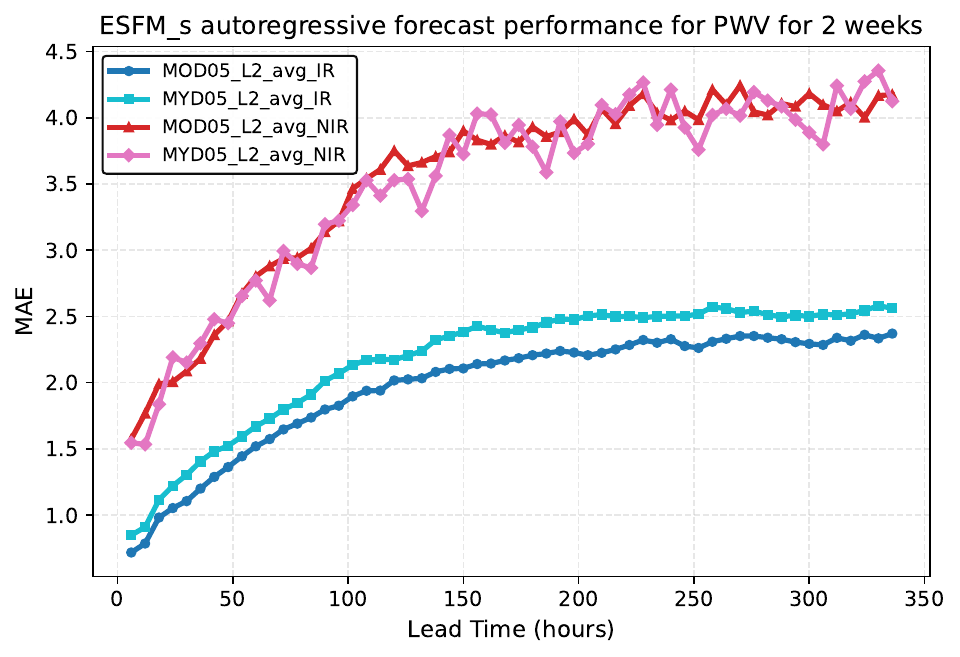}%
    \hfill%
    \includegraphics[width=0.499\linewidth]{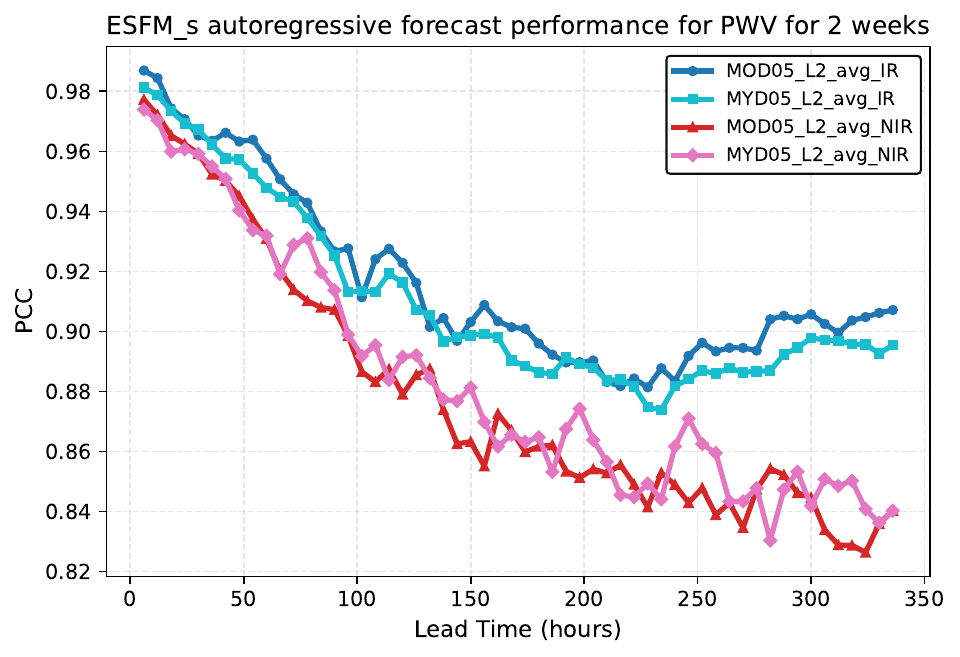}%
    \caption{\textbf{Stable rollout forecast initialized from sparse MODIS data:} Two-week lead time forecast performance of MODIS IR and NIR based PWV variable throughout the test set of 2023--2024 measured in terms of mean absolute error (MAE, left) and Pearson correlation coefficient (PCC, right).
    Autoregressive rollouts are generated through 6 hour forecast steps.
    }
    \label{fig:modis_rollout}
\end{figure}

Despite the sparsity of the data, ESFM\_s is able to achieve stable forecasting performance on MODIS data in longer lead time experiments as well.
For example, in Fig.~\ref{fig:modis_rollout}, we show the forecasting performance of ESFM\_s for all four MODIS variables up to two weeks of lead time without rollout finetuning across the test set.
It is important to note that after the initial sparse observation at time step $t$, all subsequent ESFM\_s inferences have dense input, produced by ESFM\_s, starting from $t$+$1$.

\subsection{Weather station data}

\subsubsection{ECMWF 11k station data}
In-situ (station) data sits at the heart of weather and broader environmental sciences applications as the highest spatiotemporal resolution observations.
We downloaded station data from ECMWF~\citep{ecmwf11kstation} corresponding to 11'863 stations between 2000 and 2024. 
This dataset is not imputed and therefore is also of sparse nature. 
To name a few, within the downloaded period, the availability of the temperature variable ranges between $\sim$\,0\% to 99.3\% across different stations, with a median occupancy of 66.1\%.
Some variables, such as sea-level pressure (MSL), are even sparser at 29.1\% median occupancy.
Following the greedy mapping logic introduced in Sec.~\ref{sec:multi_res_patch_embedding}, the station data is mapped onto an irregular grid of size (90$\times$180). 
We randomly select 1\,k stations to be the holdout set and finetune ESFM\_s on the remaining $\sim$11\,k stations until the end of 2020. 
This implies that the holdout stations are completely masked from both observations and targets during the entirety of the training.
We explain more details about the station dataset as well as the nature of the data in Appendix~\ref{sec:station_greedy_mapping}. 

In Table~\ref{tab:ecmwf_11k}, we showcase three different experimental results on ECMWF 11k station data. 
First, we show forecast performance of ESFM\_s on the same set of station positions for the test time period in 2023 and 2024.
This is the most typical form of evaluations done in the literature for station level forecasting, with the only exception to the best of our knowledge being MetNet-3~\citep{andrychowicz2023deep}, where the authors also incorporate dense regional input in addition to station observations.
Second, we show forecast performance of ESFM\_s on the holdout stations for the test time period.
This experiment shows generalization capability of ESFM\_s to previously unseen stations, when they are introduced for the first time at test period (i.e., in the wild).
Finally, we show the forecast performance of ESFM\_s for the holdout stations when providing only trained station observations during the test time period. 
This experiments shows extrapolation capabilities of ESFM\_s to novel locations without even providing any observations at these new coordinates at test time.

\begin{table}[t!]
\caption{Forecast performance of ESFM\_s with 6, 12, and 24 hour lead times on ECMWF station variables shown in mean absolute error (MAE).
Variables are temperature (T), dew point temperature (dT), Eastward wind (U), Northward wind (V), mean sea level pressure (MSL), air pressure (P), wind speed (WS), and wind direction (WD).
Each block of three rows corresponds to a different evaluation setting and lead time.
Namely, \textbf{regular evaluation} means the trained station locations are evaluated on the test time period.
\textbf{Holdout station I/O} is when a novel set of stations are introduced at test time (i.e., in the wild), which have not been seen during training period.
\textbf{Extrapolating holdout stations} measures performance on coordinates of the novel set of stations when only previously trained station locations are provided at test time period.}
\label{tab:ecmwf_11k}

\resizebox{\textwidth}{!}{%
\begin{tabular}{@{}llcccccccccccccccc@{}}
\toprule
& & \multicolumn{2}{c}{T} & \multicolumn{2}{c}{dT} & \multicolumn{2}{c}{U} & \multicolumn{2}{c}{V} & \multicolumn{2}{c}{MSL} & \multicolumn{2}{c}{P} & \multicolumn{2}{c}{WS} & \multicolumn{2}{c}{WD} \\
\cmidrule(lr){3-4} \cmidrule(lr){5-6} \cmidrule(lr){7-8} \cmidrule(lr){9-10}
\cmidrule(lr){11-12} \cmidrule(lr){13-14} \cmidrule(lr){15-16} \cmidrule(lr){17-18}

Evaluation & Lead time &
MAE & PCC & MAE & PCC & MAE & PCC & MAE & PCC &
MAE & PCC & MAE & PCC & MAE & PCC & MAE & PCC \\

\midrule

\multirow{3}{*}{Regular evaluation}
& 6h  & 1.07 & 0.98 & 1.18 & 0.98 & 1.18 & 0.86 & 1.18 & 0.85 & 76.89 & 0.98 & 89.65 & 1.00 & 1.04 & 0.83 & 34.26 & 0.68 \\ 
& 12h & 1.24 & 0.98 & 1.38 & 0.97 & 1.32 & 0.82 & 1.32 & 0.82 & 126.63 & 0.97 & 134.88 & 1.00 & 1.13 & 0.80 & 36.68 & 0.65\\
& 24h & 1.56 & 0.97 & 1.75 & 0.96 & 1.63 & 0.72 & 1.63 & 0.71 & 248.58 & 0.90 & 250.94 & 1.00 & 1.34 & 0.72 & 44.58 & 0.58 \\ %

\midrule

\multirow{3}{*}{Holdout station I/O}
& 6h  & 1.87 & 0.97 & 1.63 & 0.97 & 1.77 & 0.75 & 1.76 & 0.72 & 148.75 & 0.96 & 1247.59 & 0.86 & 1.57 & 0.68 & 47.85 & 0.54\\
& 12h & 2.05 & 0.97 & 1.93 & 0.96 & 1.98 & 0.68 & 1.95 & 0.64 & 196.65 & 0.94 & 1252.43 & 0.87 & 1.74 & 0.60 & 51.92 & 0.51\\
& 24h & 2.20 & 0.96 & 2.21 & 0.95 & 2.18 & 0.59 & 2.16 & 0.53 & 308.78 & 0.85 & 1297.88 & 0.87 & 1.83 & 0.53 & 57.48 & 0.46\\

\midrule

\multirow{3}{*}{Extrapolating holdout stations}
& 6h  & 2.32 & 0.94 & 2.37 & 0.92 & 2.02 & 0.64 & 1.97 & 0.61 & 177.09 & 0.92 & 1642.21 & 0.80 & 1.81 & 0.55 & 53.69 & 0.43 \\
& 12h & 2.36 & 0.94 & 2.44 & 0.93 & 2.09 & 0.62 & 2.03 & 0.60 & 211.29 & 0.90 & 1652.05 & 0.80 & 1.87 & 0.54 & 55.06 & 0.41 \\
& 24h & 2.50 & 0.94 & 2.64 & 0.92 & 2.28 & 0.54 & 2.20 & 0.52 & 302.12 & 0.85 & 1672.58 & 0.80 & 1.96 & 0.48 & 60.12 & 0.36 \\

\bottomrule
\end{tabular}%
}
\end{table}

\subsubsection{Weather-5K station data}

In addition to the sparse ECMWF dataset we procured, there are also imputed station datasets such as Weather-5K~\citep{han2024weather5k}, which provide dense observations in a smaller set of stations distributed globally, spanning nine years for training between 2013 to 2021, and one year for the test set of 2023. 
The Weather-5K also has a benchmark with nine different models for different lead time forecast tasks ranging from 24 to 168 hours, from which we show the five best models in Table~\ref{tab:weather5k}. 
These models are provided observations with a temporal history of 48 hours and they predict directly the target lead times without autoregressive rollouts. 
Despite having trained with the default ESFM\_s setup of (t-$\Delta$t, t) observations, ESFM\_s performs competitively in short lead times as shown in Table~\ref{tab:weather5k}.

\begin{table}[t!]
\caption{Forecast performance of ESFM\_s on the Weather-5K station dataset, shown in mean absolute error (MAE).
Wind speed and direction for ESFM\_s forecasts are derived from Eastward and Northward wind variables.
Overall score is calculated as the average of the five variables.
}
\label{tab:weather5k}
\resizebox{\textwidth}{!}%
{%
\begin{tabular}{@{}lccccccc@{}}
\toprule
 & Lead time & Temperature & Dewpoint & Wind speed & Wind direction & Sea level pressure & Overall \\
\midrule
ESFM\_s & 6  & 1.10 & 1.17 & 1.06 & 36.44 & 0.62 & 8.1 \\
ESFM\_s & 12 & 1.32 & 1.45 & 1.16 & 39.32 & 1.23 & 8.9 \\\midrule
ESFM\_s & 24 & 1.76 & 1.86 & 1.40 & 46.29 & 2.55 & 10.8  \\
Pyraformer~\citep{liu2022pyraformer} & 24 & 1.75 & 1.83 & 1.30 & 61.8 & 1.90 & 13.7 \\
iTransformer~\citep{liu2023itransformer} & 24 & 1.82 & 1.93 & 1.32 & 63.2 & 1.99 & 14.1 \\
Informer~\citep{zhou2021informer} & 24  & 1.88 & 1.94 & 1.30 & 60.7 & 2.01 & 13.6 \\
Autoformer~\citep{wu2021autoformer} & 24 & 1.93 & 2.06 & 1.42 &66.5 &2.26 & 14.8 \\
FEDformer~\citep{zhou2022fedformer} & 24 & 1.98 & 2.02 & 1.36 & 66.0& 2.13 & 14.5 \\
\bottomrule
\end{tabular}%
}
\end{table}

\begin{table}[h]
\caption{Comparison of ESFM\_s finetuned on CNRM-CM6-1-HR dataset after pretraining on CMIP6 (8 datasets), ERA5, combination of 8 CMIP6 datasets and ERA5, no pretraining. 
Surface variables are sea ice cover (CI), sea surface temperature (SST), terrestrial water storage (TWS), mean sea level pressure (MSLP), while atmospheric variables u component of wind (U), v component of wind (V), geopotential (Z), temperature (T) are indicated with the corresponding pressure levels in [hPa] next to them. Best results are in bold.
}
    \label{tab:holdout_cmip6_val}
    \resizebox{\textwidth}{!}{%
    \begin{tabular}{l cccc cccc}
    \toprule
    & \multicolumn{4}{c}{\textbf{MAE}} & \multicolumn{4}{c}{\textbf{PCC}} \\
    \cmidrule(lr){2-5} \cmidrule(lr){6-9}
    Variable
    & CMIP6 & CMIP6 + ERA5 & ERA5 & No pretrain.
    & CMIP6 & CMIP6 + ERA5 & ERA5 & No pretrain. \\
    \midrule
    
    CI
    & \textbf{0.007} & 0.008 & 0.011 & \textbf{0.007}
    & \textbf{0.995} & 0.994 & 0.992 & \textbf{0.995} \\
    
    SST
    & \textbf{0.203} & 0.236 & 0.391 & 0.220
    & \textbf{0.999} & \textbf{0.999} & \textbf{0.999} & \textbf{0.999} \\
    
    TWS
    & \textbf{14.035} & 20.616 & 31.339 & 15.918
    & \textbf{0.940} & 0.826 & 0.671 & 0.909 \\
    
    MSLP
    & 49.406 & \textbf{49.305} & 120.921 & 61.112
    & \textbf{0.998} & \textbf{0.998} & 0.990 & 0.997 \\
    
    U500
    & \textbf{1.069} & 1.070 & 1.425 & 1.253
    & \textbf{0.993} & \textbf{0.993} & 0.988 & 0.990 \\
    
    U850
    & \textbf{0.882} & 0.890 & 1.187 & 1.024
    & \textbf{0.990} & \textbf{0.990} & 0.983 & 0.986 \\
    
    V500
    & 1.098 & \textbf{1.096} & 1.515 & 1.303
    & \textbf{0.988} & \textbf{0.988} & 0.979 & 0.983 \\
    
    V850
    & \textbf{0.885} & 0.890 & 1.175 & 1.035
    & \textbf{0.983} & \textbf{0.983} & 0.973 & 0.977 \\
    
    Z500
    & 4.029 & \textbf{3.032} & 9.307 & 4.852
    & \textbf{1.000} & \textbf{1.000} & 0.999 & \textbf{1.000} \\
    
    Z850
    & 3.090 & \textbf{2.874} & 7.521 & 3.814
    & \textbf{0.998} & 0.997 & 0.995 & 0.997 \\
    
    T500
    & 0.387 & \textbf{0.375} & 0.512 & 0.450
    & \textbf{0.999} & \textbf{0.999} & 0.998 & \textbf{0.999} \\
    
    T850
    & \textbf{0.503} & 0.522 & 0.718 & 0.580
    & \textbf{0.998} & \textbf{0.998} & 0.997 & \textbf{0.998} \\
    
    \bottomrule
    \end{tabular}%
    }%
\end{table}

\subsection{Unseen climate models and variables}

Beyond its flexibility for exploiting sparse and non-gridded data, ESFM\_s is also a skillful FM when it comes to other challenges, such as coupling forecasts with climate forcings and generalizing to unseen numerical models upon finetuning.
This section explores how the choice of pretraining datasets and pretraining variables impacts finetuning performance, in terms of accuracy after the finetuning phase. 

\subsubsection{Modeling unseen climate models}
\label{sec:novel_CMIP6}

To study the influence of the pretraining datasets, we finetune ESFM\_s on the CNRM-CM-6-1-HR CMIP6 dataset that was held out from our set of pretraining CMIP6 datasets \citep{voldoireEvaluationCMIP6DECK2019}. 
CNRM-CM6-1-HR has a resolution of 0.5$^\mathrm{o}$, which is higher than any CMIP6 dataset used during pretraining (see Section~\ref{sec:CMIP6_preprocesing} for the description of CMIP6 datasets).  
In addition to having some common variables with the pretraining such as atmospheric wind velocity, air temperature, geopotential, and sea level pressure, there are also previously unseen variables: sea surface temperature, sea ice cover, and terrestrial water storage. 
The temporal resolution is six hours and variables were linearly interpolated to six-hour time steps when necessary. 
Training was conducted for 23 epochs with a batch size of 8, with data from 1950 to 2013 (2014 and 2015 are kept as validation and test data).

\begin{figure}[b!]
    \centering
    \includegraphics[width=0.85\linewidth]{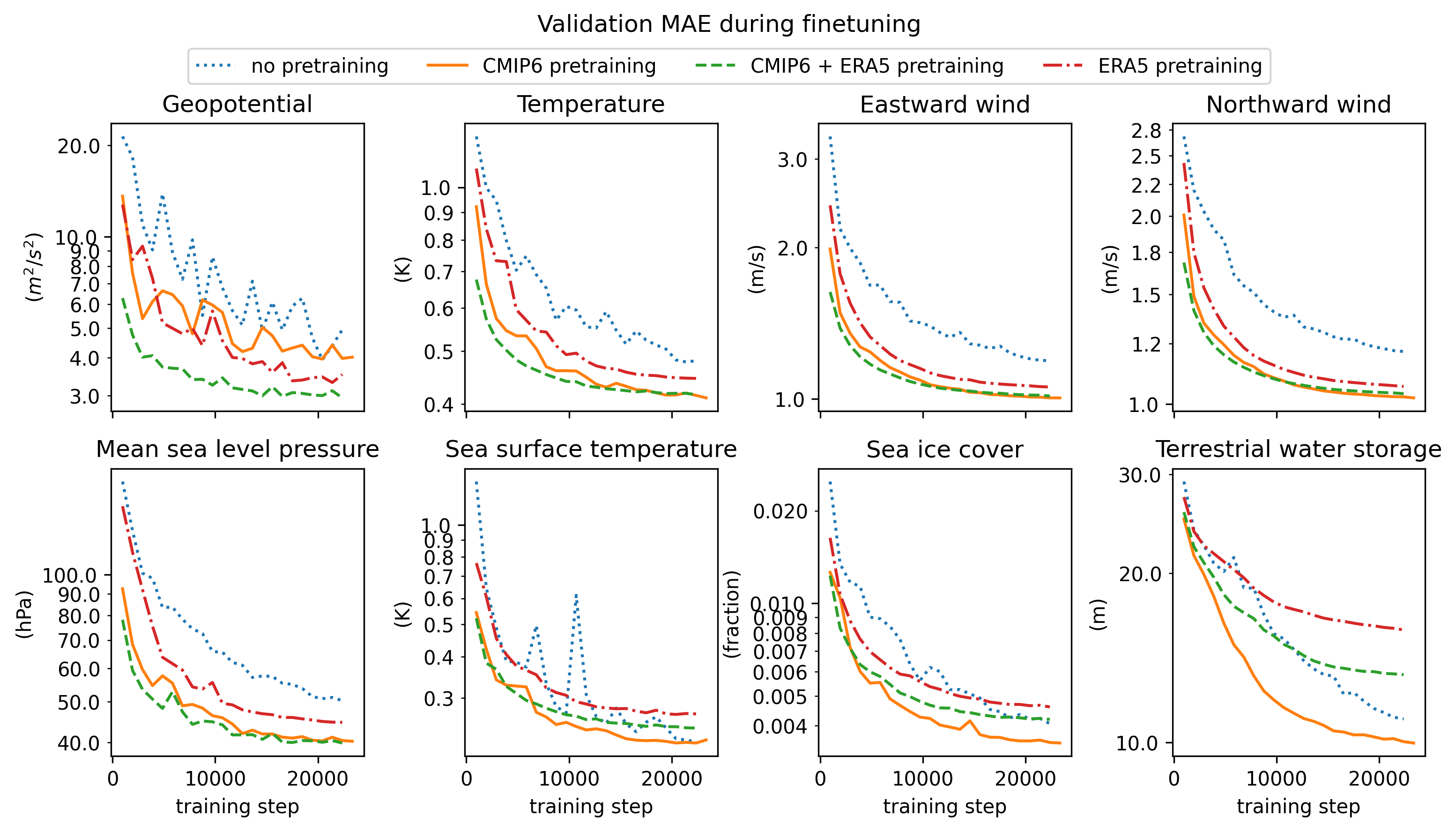} 
    \caption{\textbf{Validation MAE loss (year 2014) for CNRM-CM6-1-HR dataset} when finetuned without pretraining (blue), CMIP6 pretraining (orange), CMIP6 + ERA5 pretraining (green), ERA5 pretraining (red).}
    \label{fig:holdout_cmip6_val_curve}
\end{figure}

Figure~\ref{fig:holdout_cmip6_val_curve} shows validation set loss curves of ESFM\_s with four different initializations: (i) no pretraining, (ii) pretraining on eight CMIP6 datasets, (iii) pretraining on 8 CMIP6 datasets and ERA5, and (iv) pretraining on ERA5. 
Pretraining on CMIP6 datasets, whether alone or in addition to ERA5, yields a faster decrease of the loss function and stabilization to better accuracy than pretraining on ERA5 alone or without pretraining. 
These results are confirmed by the detailed metrics in Table~\ref{tab:holdout_cmip6_val}.

The advantages of CMIP6 pretraining are not restricted to the original variable set. 
Even for previously unseen variables such as sea surface temperature, sea ice cover, and terrestrial water storage, CMIP6-pretrained initializations yield lower error metrics and more stable convergence than models trained from scratch or pretrained only on ERA5.

\subsubsection{Modeling unseen variables}
\label{sec:novel_variables}
\begin{figure}[]
    \centering
    \includegraphics[width=0.75\linewidth]{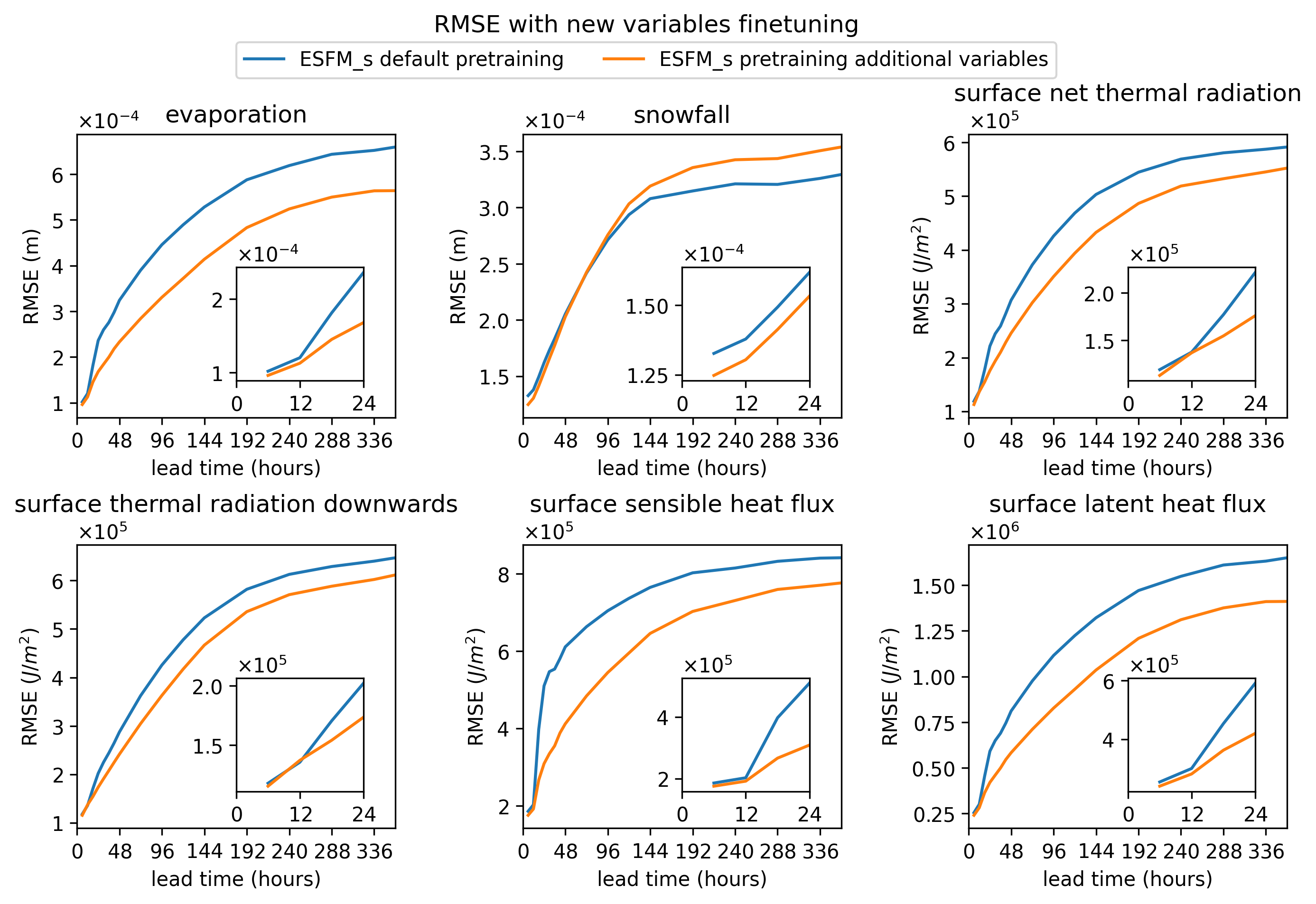}
    \caption{\textbf{RMSE on the test year 2023 when finetuning ESFM\_s with six new surface variables, starting from two pretrained models}: baseline ESFM\_s (blue) and ESFM\_s pretrained with a subset $S_{new,1}$ of additional variables (orange). 
    The plot shows variables from $S_{new,2}$ that are unseen for both pretrained models. 
    The inner boxes show a zoom on lead times shorter than 24 hours.}
    \label{fig:metrics_newvariables}
\end{figure}

In this section, we examine how the diversity of variables included in pretraining affects the finetuning accuracy. 
To do so, we constitute two sets of surface variables that are generally absent from AI weather models: $S_{new,1}$ with total precipitation, potential evaporation, surface net solar radiation, surface solar radiation downwards, top net thermal radiation, top net solar radiation, and $S_{new,2}$ with evaporation, snowfall, surface net thermal radiation, surface thermal radiation downward, surface sensible heat flux, surface latent heat flux. 
We continue pretraining baseline ESFM\_s with its original variables and $S_{new,1}$ to create a richer pretrained model. 
Then, we finetune the baseline ESFM\_s (blue curves in Figs.~\ref{fig:metrics_newvariables} and \ref{fig:loss_finetuning_newvars}) and this richer ESFM\_s (orange curves in Figs.~\ref{fig:metrics_newvariables} and \ref{fig:loss_finetuning_newvars}) on the large set of original variables, $S_{new,1}$, $S_{new,2}$. 

Figure~\ref{fig:metrics_newvariables} shows that ESFM\_s with the richer pretraining variables outperforms the baseline ESFM\_s, except for snowfall at long lead times (longer than 96\,hours). 
The benefits of pretraining with additional variables are more pregnant at long lead times, which suggest that richer pretraining datasets improve the physical consistency of the model. 

\section{Summary and discussion}
This study introduces ESFM, a foundation model capable of skillfully forecasting both traditional and unconventional variables, including those sparsely available as satellite data or irregularly observed, such as station data, while accepting gridded data of various resolutions and missing data. 
This flexibility has potentially profound implication for the adoption of foundation models in the broader climate science domain. By implementing individual variable tokenization alongside a masked training protocol, the model extends the capabilities of foundation models for the environmental sciences, surpassing previous experimental boundaries. 
The architectural adjustments and masked training pipeline incur only minimal tradeoffs, as demonstrated through global metrics and targeted case studies. Specifically, our results highlight the importance of different initialization weights for ESFM\_s training (Sec.~\ref{sec:results}). 
Depending on the variable, the performance gain compared to random initialization ranges between 12\% to 40\% of MAE relative to knowledge distillation based on weights of Aurora, while saving significant compute costs as a pretraining objective.

The small ESFM models (ESFM\_s) matches the performance of leading state-of-the-art models. %
However, most existing models rely on fixed sets of variables, pressure levels, and dense inputs during both training and inference, optimizing their performance within these constraints. 
ESFM incorporates an additional encoder overhead to aggregate variables and pressure levels, enabling robust operation even with partially or fully missing observations. 
While this flexibility results in a marginal reduction in some performance metrics such as MAE or RMSE, the model’s scalable backbone architecture positions it for future advancements beyond current benchmarks.

\paragraph{Forecasting with no initial data} Training ESFM with structured masking applies controlled dataset mismatch as a proxy for real-world data gaps.
This reveals the model’s ability to function under partial observability across three dimensions: spatial, entire variable, and pressure level. 
Section~\ref{sec:masked_experiments} highlights the competitive performance of small ESFM relative to state-of-the-art models in multi-dimensional masked test scenarios using ERA5 data. Regional masking of variables results in only a minimal reduction in global forecasting accuracy, while forecasting performance within the masked regions remains robust across all variables. As a result ESFM can forecast variables even in areas lacking data at initial time (Tab.~\ref{tab:consolidated_regional_masking} and \ref{tab:regional_masking}).  ESFM successfully forecasts not only variables absent at the initial time, independent of whether the data is absent at the surface or in the atmosphere. This outcome confirms the capture of meaningful statistics across atmospheric variables (Tab.~\ref{tab:variable_masking}). The forecasted variables retain key inter-variable relationships (Fig.~\ref{fig:bivariate_europe}). Performance decreases when all atmospheric variables at a pressure level are masked, requiring ESFM\_s to infer variables from other pressure levels (Tab.~\ref{tab:plev_masking}).

\paragraph{Weather extremes and Sudden Stratospheric Warming (SSW)}
Two case studies illustrate the capabilities of small ESFM in extreme event anticipation with the probabilistic ESFM (Super Typhoon Doksuri in Section~\ref{sec:doksuri}) and finetuning for more stratosphere levels (SSW events in Section~\ref{sec:ssw}). 
The small ESFM accurately predicts the maximum sustained wind speed during the entire duration of the Doksuri typhoon and even predicts higher amplitudes than the large Aurora despite the reduced model complexity of small ESFM. 
Thanks to its probabilistic extension, ESFM gives a reasonable spread of maximum wind speed, with the higher tail of the distribution overestimating wind velocity by around 3\,m/s compared to ERA5 for the first four days. 
The localization of the cyclone is also visually satisfying for around four days, although detailed cyclone tracking was out of the scope of this work. 

Regarding the predictability of the stratosphere during SSW events, we found that the small ESFM was effective in predicting wind velocity reversal at 10\,hPa during the three SSW events investigated. 
While the impact of representing the stratosphere closer to the surface is encouraging, as the geopotential anomaly is more realistic than the baseline small ESFM, this needs to be confirmed with more accurate predictions on timescales beyond ten days.

Our empirical studies suggest that ESFM is a promising model for training and inference with heterogeneous datasets where the set of available variables and pressure levels differ across sources, without requiring ad-hoc imputation or requiring strict variable completeness.

\paragraph{Satellite data}
ESFM accepts highly sparse satellite data as input for a dense forecast. 
The predicted precipitable water vapor quantities are very respectable, with autoregressive rollout forecast errors plateauing roughly at a Pearson Correlation Coefficient of 0.9 for IR and 0.83 for NIR at two-week lead time (Fig.~\ref{fig:modis_rollout}).
This is achieved without finetuning small ESFM with rollout forecasts. 

In Table~\ref{tab:MODIS_6h}, we additionally included scores for PWV fields derived from ERA5 variables. 
The predicted PWV fields from small ESFM show closer agreement with MODIS than with ERA5 at the global scale. 
A more detailed examination of the forecasted value ranges over regional domains further supports this behavior. 
For both Switzerland and the UK, the forecasted and observed MODIS PWV distributions overlap almost perfectly for a representative test case in 2024 (Fig.~\ref{fig:MODIS_patches}). 
This indicates that the small ESFM captures the spatiotemporal characteristics of the MODIS product, rather than merely reproducing a smoother reanalysis background. 
Nevertheless, this relatively superior agreement should be interpreted with care, because closer consistency with MODIS does not necessarily imply a closer representation of the true atmospheric state, as MODIS-specific retrieval characteristics and biases would also be learned.

Beyond statistical agreement, the generated dense forecast fields appear physically consistent, suggesting that ESFM can propagate PWV information from sparse MODIS observations to the global scale while preserving the effective characteristics of the original retrievals. 
The model goes beyond spatial reconstruction by also forecasting their temporal evolution. 
These results highlight the potential of ESFM not only for global MODIS-based PWV reconstruction and prediction, but also for other satellite-derived geophysical variables from observing systems such as the Metop, Sentinel, and Feng-Yun series.

\paragraph{Weather station data}
ESFM\_s demonstrates strong competitiveness on Weather-5K station data, matching state-of-the-art models specialized for station-based forecasting up to a 24-hour lead time (Table~\ref{tab:weather5k}). 
The model ranks first in average MAE, and achieves second, second, seventh, first, and ninth place for temperature, dewpoint, wind speed, wind direction, and sea level pressure, respectively, in 24-hour forecasts\footnote{Last verified against Weather-5K benchmark results on 17.04.2026.}.

Notably, ESFM attains this performance using only a single historical input step.
For example, 6-hour lead time forecast is produced using observations at ($t-6$h, $t$).
Similarly, ($t-12$h $t$), and ($t-24$h,$t$) for 12- and 24-hour lead time forecasts, and not using all of the 48 hours of history for different lead time forecasting tasks.

Performance further improves on the sparse ECMWF 11k station dataset compared to the denser Weather-5K set (Table~\ref{tab:ecmwf_11k}), confirming the ability of ESFM small to leverage additional stations for forecasting, regardless of data sparsity.

The model also excels in generalizing to previously unseen stations and extrapolating forecasts for new locations with high accuracy. While air pressure experiences the most significant performance drop in generalization and extrapolation tests, this limitation could likely be mitigated by incorporating static variables such as surface geopotential during finetuning for station data.

\paragraph{Data unseen during training}
The advantages of ESFM as a foundation model extend beyond managing data heterogeneity, as demonstrated through finetuning experiments on additional CMIP6 datasets and ERA5 variables that go beyond standard weather parameters (Section~\ref{sec:novel_CMIP6}). 
These experiments highlight the value of pretraining with well-chosen datasets: the original eight CMIP6 datasets provide an optimal starting point for finetuning with the additional CNRM-CM6-1-HR CMIP6 dataset.

When finetuning with additional ERA5 variables, two pretraining configurations were compared: one using default variables and the other with an expanded set. 
The latter approach yields superior results for unseen variables, indicating that foundation models benefit from greater diversity in physical variables during pretraining. 
While computational constraints may limit the volume of data processed in each training step, ESFM can address this by optimizing iteratively with random variable subsets.

\paragraph{Additional notes}
In addition to the changes we introduced in the encoder and decoder for ESFM, we have made several other subtle changes mentioned before.
First, we reduced the minimum values for the patch area and absolute time positional encodings from the order of km$^2$ to m$^2$ and from hours to minutes, respectively.
These changes made negligible differences in the default evaluation setup for ERA5 0.25$^{\circ}$ resolution, providing the same MAE overall for 6 hour lead time forecasts.
However, they are crucial to make ESFM ready for significantly finer resolution data such as station datasets.
Additionally, we introduced 2D sine-cosine positional embeddings at the decoder input, which yielded improved detokenized patch borders. 

Prior to AdaLN based ensembles, we experimented with tail ensembles, introduced already in~\cite{lessig2023atmorep}.
Tail ensembles, initialized with deterministic ESFM detokenizer weights as well as Gaussian perturbed versions of the deterministic ESFM work well.
However, they were prone to overfitting a distinct bias embedded into the detokenizer, which becomes more visible when forecasts are limited to a narrow dynamic range of intensity values. 

The initial release of ESFM features approximately 115 million parameters, aligning with the backbone size of Aurora small (Aurora\_s). 
Although the available computing infrastructure supports training models with up to 1.3 billion parameters, early instability in the HPC system led to prohibitive computational costs. 
Consequently, the focus shifted to the refinement of training strategies and design choices using the smaller ESFM variant. 
Plans to release larger models are underway, contingent on securing adequate computational resources.

\paragraph{Future work}

Our focus for future work will be on enhancing the flexibility of ESFM for additional downstream tasks for the environmental sciences community, exploiting information from additional streams simultaneously when finetuning on more niche variables.
In addition, we will further explore the setup that produces the optimal balance between performance and efficiency for autoregressive rollouts of ESFM with ensembles.
While we have tested training ESFM with $N=1000$ ensemble members while subsampling eight members to optimize in each step without any issues, the required finetuning steps are longer than the 10'000 steps that were sufficient for $N=8$ members; therefore, we leave experiments with larger ensembles for future work.

\acks{This work was supported under project IDs a01 and a122 as part of the Swiss AI Initiative, through a grant from the ETH Domain and computational resources provided by the Swiss National Supercomputing Centre (CSCS) under the Alps infrastructure. 
Contributions of Fanny Lehmann were primarily supported by the ETH AI Center through their ETH AI Center postdoctoral fellowship. 
Contributions of Leonardo Trentini were supported by the Swiss National Science Foundation (SNSF) under grant number 225851.
The authors thank sincerely Andrin Zoller for running ensemble rollout forecasts of the compared SotA models and Langwen Huang for insightful discussions throughout the development of ESFM.
}
\clearpage

\vskip 0.2in
\bibliography{bib}

\newpage

\appendix \section{Training details}

\subsection{Knowledge distillation for training ESFM encoder}
\label{sec:knowledge-distillation}
In this work, our proposed changes reflect to the encoder and decoder, while building directly on the architectural backbone choice proposed in Aurora.
We can therefore save significant computational overhead by aligning ESFM encoder embeddings with the pretrained encoder from Aurora, allowing us to finetune the pretrained backbone for different experiments.
For this purpose, we use knowledge-distillation (KD)~\citep{hinton2015distilling}, where Aurora encoder is the teacher and a randomly initialized ESFM encoder is the student. 
Specifically, we used KD to train the ESFM encoder while setting large (1.3\,B) and small (117\,M) Aurora parameter configurations as the teacher. 
We used L1 loss between the logits of the teacher and student networks, using the ERA5~\citep{rasp2024weatherbench} dataset for time steps between 1979 and 2020.
We trained for 40k steps using 32 GPUs, with a linear warmup of 1k steps up to learning rate 5e-4, followed by a cosine decay down to 4e-4.

\subsection{Additional details on the masking protocols used in training}
\label{sec:masking_details}

\begin{figure}
    \centering
    \includegraphics[width=\linewidth]{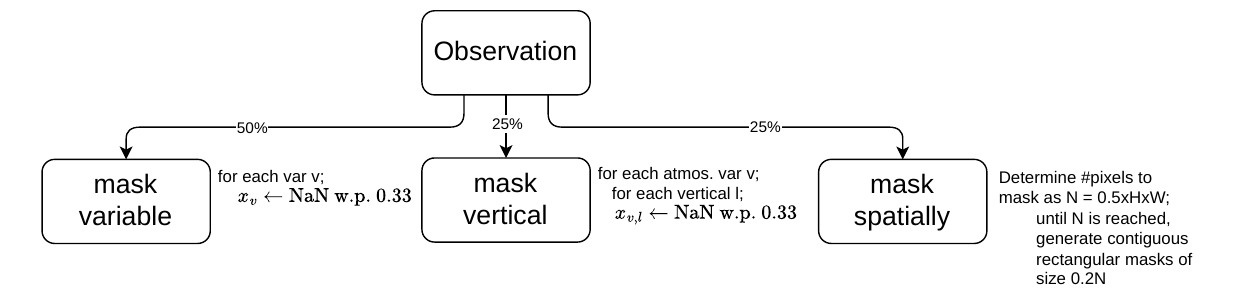}
    \caption{Schematic of the masking applied to the observations for masked ESFM\_s training.}
    \label{fig:masking-protocol}
\end{figure}

In Sec.~\ref{sec:masking}, we briefly describe the different kinds of masking protocols we used for masked ESFM\_s training. 
By default, the first three; (i) variable masking with probability $p_v=0.5$, (ii) pressure level masking with probability $p_l=0.25$, and (iii) contiguous spatial masking with a probability $p_s=0.25$ are applied in the masked ESFM\_s training over both training on 8 CMIP6 datasets as well as the subsequent ERA5 training. 
We show the masking process as a schematic in Fig.~\ref{fig:masking-protocol}.

Additionally, we have experimented with an additional randomized selection of pressure levels from the ERA5 dataset.
This means that, instead of always using the same 13 pressure levels typically used in weather forecasting literature (50, 100, 150, 200, 250, 300, 400, 500, 600, 700, 850, 925, 1000\,hPa), we randomly sample $n_l=13$ pressure levels from the 37 pressure levels served in WeatherBench2 ERA5. 
Furthermore, we sample $n_l$ pressure levels separately for the observations and the prescribed targets. 
ESFM\_s trained with this additional random pressure level set strategy gets a substantial performance drop when evaluated with the 13 pressure levels typically used in the literature. 
In particular, the MAE performance drop for six hour lead time forecasts ranges between 85.4\% (Q850 \& Q500) and 500\% (Z500) compared to ESFM\_s trained with the default masking strategy on the fixed 13 pressure levels, as shown in more detail in Table~\ref{tab:esfm_rplv_perf_drop}.
We suspect that this is due to a limitation of the perceiver module for pressure level aggregation. 
Accordingly, we have explored variations in the perceiver module, increasing the number of Perceiver blocks, trying newer Perceiver modules, but observed a similar limitation. 
We will investigate this shortcoming further in the future.

\begin{table}[]
    \centering
    \resizebox{\textwidth}{!}{%
    \begin{tabular}{lllllllllll}
    \toprule
    variable & T2m & U10m & V10m & T850 & Q850 & Z850 & U850 & V850 & Q500 & Z500 \\
    \midrule
    ESFM\_s & 0.268 & 0.281 & 0.290 & 0.249 & 2.487e-4 & 14.428 & 0.452 & 0.455 & 8.051e-5 & 15.957 \\
    ESFM\_s + rand plev & 0.653 & 0.686 & 0.692 & 0.654 & 4.385e-4 & 69.167 & 1.078 & 0.983 & 1.716e-4 & 99.297 \\ 
    \midrule
    MAE difference &0.385 & 0.405 & 0.402 & 0.405 & 1.898e-4 & 54.739 & 0.626 & 0.527 & 9.114e-5 & 83.340 \\
    \bottomrule
    \end{tabular}%
    }
    \caption{Six hour lead time forecast MAE performance comparison between ESFM\_s trained with masking (ESFM\_s) vs ESFM\_s trained with masking and random sets of 13 pressure levels out of the 37 pressure levels (ESFM\_s + rand plev).
    }
    \label{tab:esfm_rplv_perf_drop}
\end{table}

\subsubsection{Physical consistency after masking}
\label{sec:bivariate_appendix}

\begin{figure}[htbp]
    \centering
    \includegraphics[width=\textwidth]{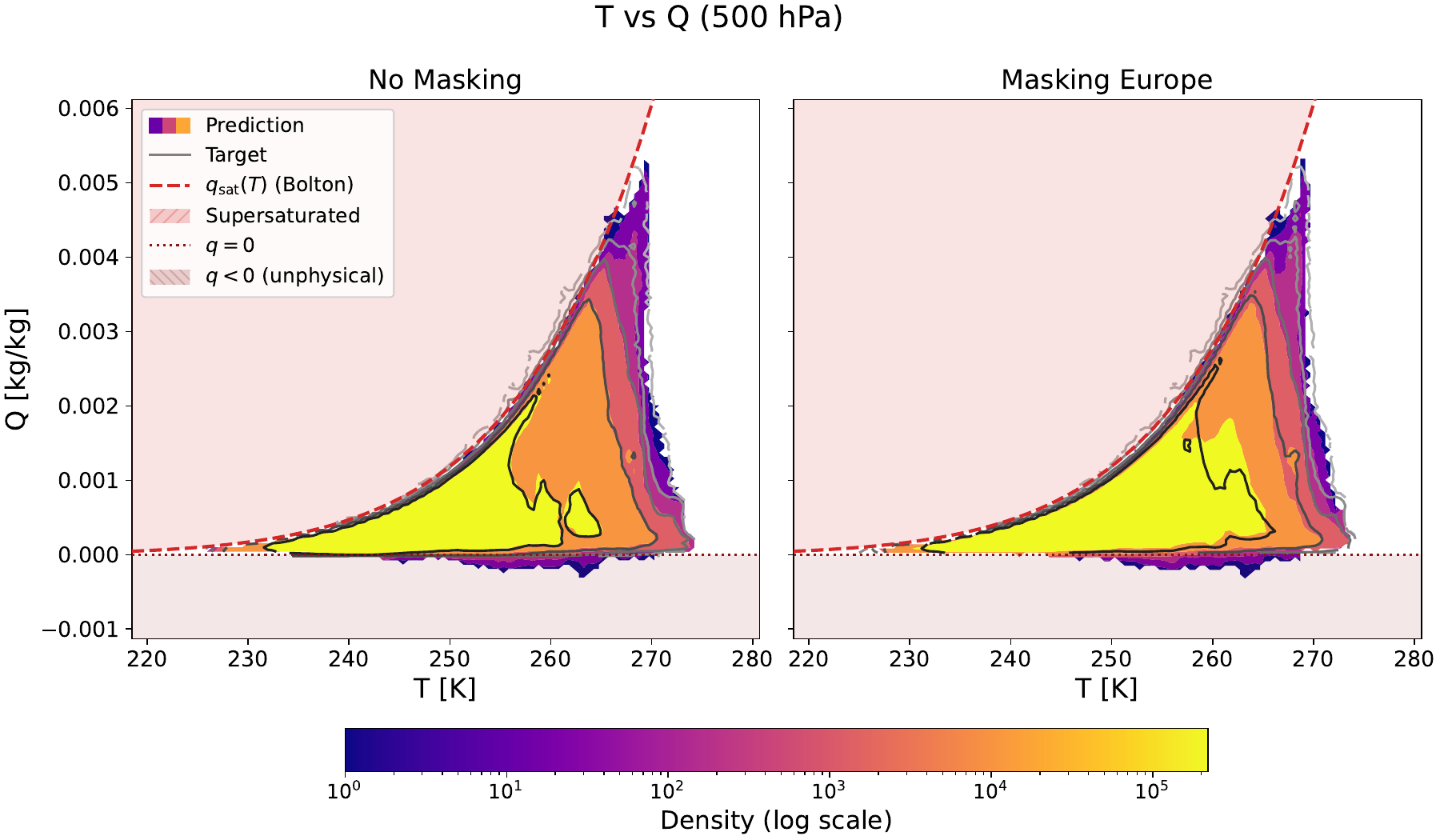}
    \caption{As Fig.~\ref{fig:bivariate_europe}, but for the joint distribution of T and Q. 
    The red dashed curve is $q_\mathrm{sat}(T)$ from \citet{Bolton1980} at 500\,hPa; the pink hatched region above it marks the highly unlikely supersaturated regime. 
    The red dotted line marks $q = 0$; the gray hatched region below indicates negative Q values, which are implausible.}
    \label{fig:bivariate_europe_tq}
\end{figure}

Fig.~\ref{fig:bivariate_europe_tq} extends the geostrophic analysis of Sec.~\ref{sec:masked_experiments} to the thermodynamic dimension, showing the joint distribution of T and Q at 500\,hPa evaluated within the withheld European domain.
As in the pressure level masking case, the predicted distribution remains below the Bolton saturation curve and above zero in both configurations, confirming that the model does not generate thermodynamically inconsistent states within a regionally masked domain.

\begin{figure}[htbp]
    \centering
    \includegraphics[width=\textwidth]{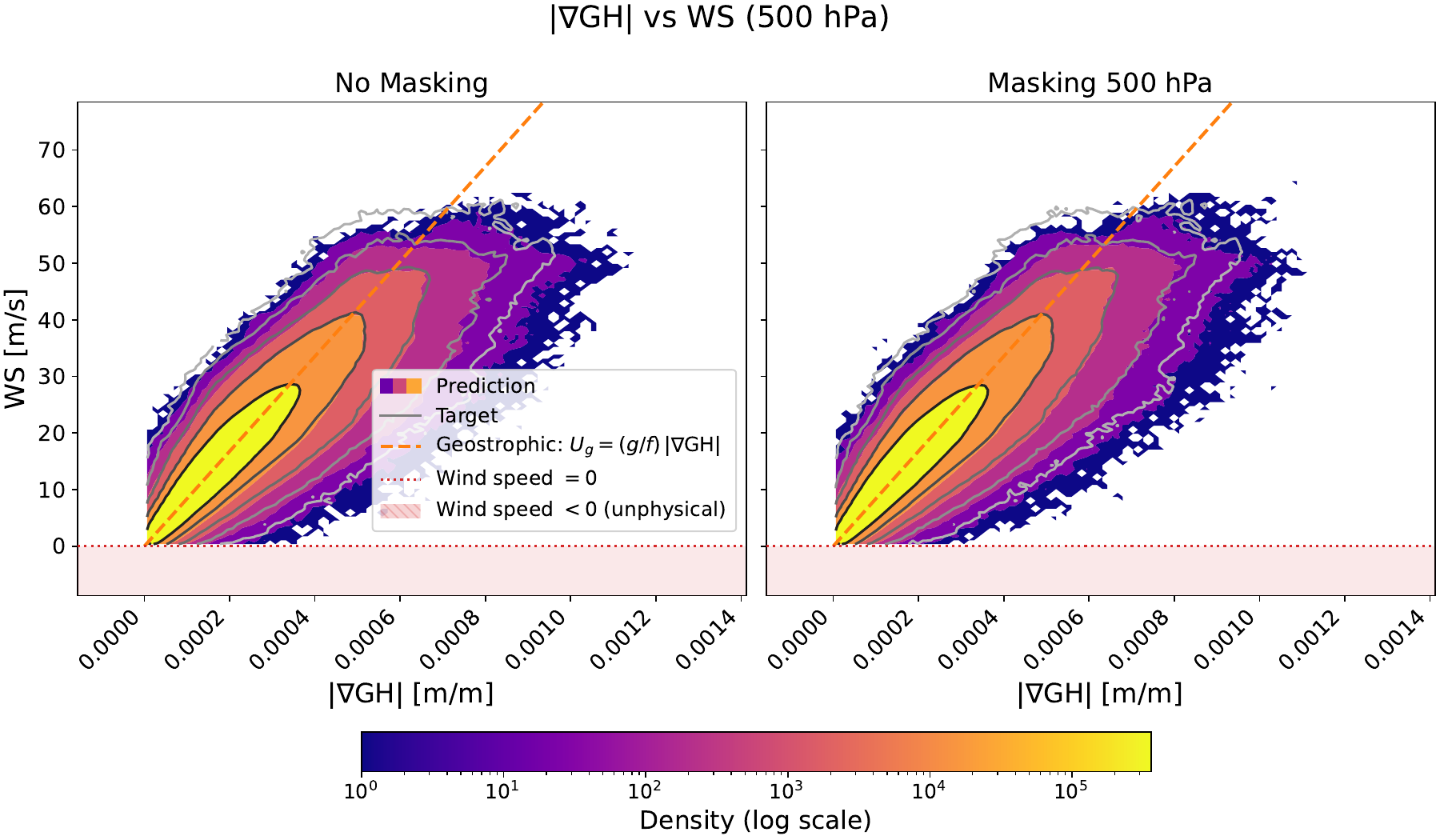}
    \caption{As Fig.~\ref{fig:bivariate_plev500_tq} but for $|\nabla GH|$ and WS at 500\,hPa.
    The orange dashed line is the geostrophic reference
    $U_g = (g/f)\,|\nabla GH|$ with $f = 1.17 \times 10^{-4}$\,s$^{-1}$
    (mean latitude $\approx$54°N).
    The red dotted line marks wind speed $= 0$; the pink hatched region
    below serves only as a visual reference for the axis extent, as negative
    WS is unphysical by construction.}
    \vspace{-1.75em}%
    \label{fig:bivariate_plev500_geo}
\end{figure}

Fig.~\ref{fig:bivariate_plev500_geo} complements the thermodynamic analysis in Sec.~\ref{sec:masked_experiments} by showing the geostrophic wind balance at 500\,hPa under pressure level masking.
The predicted joint distribution of $|\nabla GH|$ and WS tracks the geostrophic reference $U_g = (g/f)\,|\nabla GH|$ in both the baseline and the fully masked configuration, consistent with the finding that physical constraints are preserved even when the entire pressure level is absent from the observations.

\subsection{Hyperparameters for ESFM\_s}
\label{sec:ESFM_hyperparameters}

\begin{figure}[bt]
    \centering
    \includegraphics[width=0.7\linewidth]{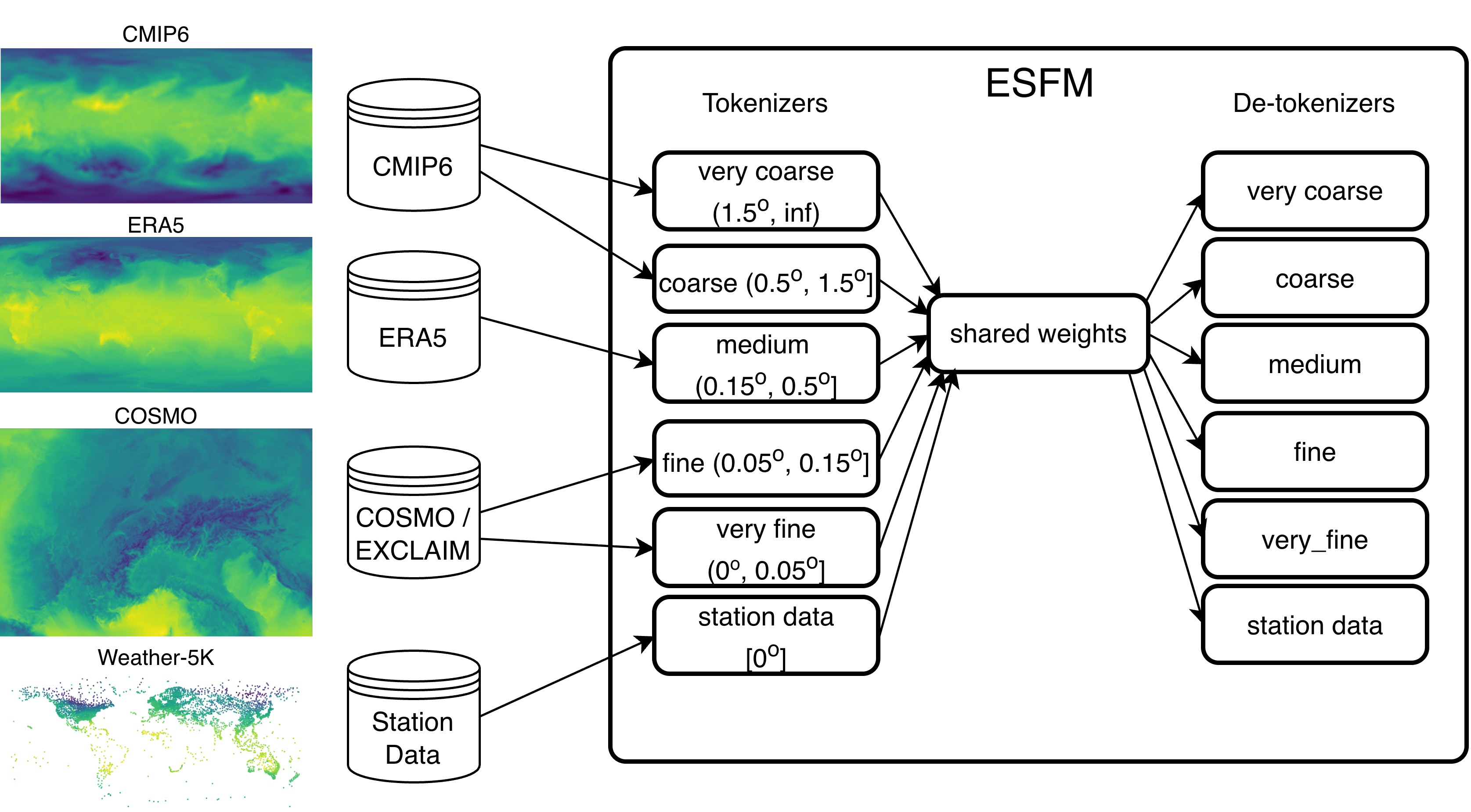}
    \caption{Multi-tokenizer structure of ESFM.}
    \label{fig:multi-res-patch-embed}
\end{figure}

In all ESFM\_s experiments, we used network configuration with 115\,M parameters, corresponding to Aurora small (Aurora\_s). 
Unless stated otherwise, we kept the default parameterization presented in Aurora and released by the authors:
\begin{itemize}
    \item Embedding dimension of 256, which becomes 512 at the output of backbone due to the final skip-connection,
    \item the backbone consists of 3 levels of token merges and token splits 
    \item stochastic drop path with $p=0.2$
\end{itemize}
We use the additional pre-LayerNorm to query and key tokens in the encoder perceiver modules for stabilization, proposed later in the repository of Aurora.
In all experiments, we use a single layer of axial attention in the encoder only along the dimension of variables.
Multi-resolution patch embeddings are used in all experiments involving training beyond purely single resolution data (i.e., ERA5).
When used, our choice of multi-resolution patch embeddings have been the following (also shown in Fig.~\ref{fig:multi-res-patch-embed}):
\begin{itemize}
    \item Very coarse gridded inputs, for resolution ranges between (1.5$^\mathrm{o}$, $\inf$),
    \item coarse gridded inputs, for resolution ranges between (0.5$^\mathrm{o}$, 1.5$^\mathrm{o}$],
    \item medium gridded inputs, for resolution ranges between (0.15$^\mathrm{o}$, 0.5$^\mathrm{o}$],
    \item fine gridded inputs, for resolution ranges between (0.05$^\mathrm{o}$, 0.15$^\mathrm{o}$],
    \item very fine gridded inputs, for resolution ranges between [10$^{-5\mathrm{o}}$, 0.05$^\mathrm{o}$],
    \item station data, for resolution at 0$^\mathrm{o}$.
\end{itemize}
We add a 2D sine-cosine positional embedding to every latent atmospheric pressure level along latitude-longitude token embeddings to enforce neighborhood information to detokenizers.
For ESFM\_s with ensemble prediction, we use $N$=8 AdaLN based ensemble modulation parameters. 
For larger ensembles such as $N$=$1000$, we implemented a routine that randomly selects a subset of $N_s$=$8$ ensemble parameters to optimize for each training step.
We use a combination of almost fair CRPS loss and MAE on the ensemble means with equal weight.

\subsubsection{Variable perceivers}
Token embeddings of both surface and atmospheric variables go through a separate perceiver module for aggregating tokens to latent number of variable tokens. 
For this purpose, we used the same hyperparameters as the perceiver module used for pressure level aggregation in Aurora. 
Namely, one perceiver layer consists of a cross attention layer between latent and embedded tokens, residually added to the latent embeddings, followed by an MLP with GELU activation, also residually added.
The query vectors for both surface and atmospheric variable perceivers are learnable parameters throughout the training of ESFM, each of shape $(1 \times d_\mathrm{emb})$.

\subsubsection{Positional embeddings}

In addition to the positional embeddings introduced in Aurora (i.e., pressure level, patch position, patch area, time of the year, lead time), we introduce unique positional embeddings for each surface and atmospheric variable, which we refer to as variable type embeddings in Fig.~\ref{fig:schematic-encoder1}. 
Specifically, we define two embedding layers of size 201, one for surface (and static) variables and one for atmospheric variables. 
Throughout different stages of (pre-)training across different datasets, the list of used and hence trained variable embeddings increases. 

\section{Long lead time forecasting performance}
\label{sec:rollout_forecasts}

ESFM shows stable behavior over autoregressive rollout forecasts. 
It should be noted that any partial observation for a given initial set of observations would be succeeded with dense forecasts from ESFM from the second step onward. 

\begin{figure}[htbp]
\centering
\begin{tabular}{ccc}
\includegraphics[width=0.32\textwidth]{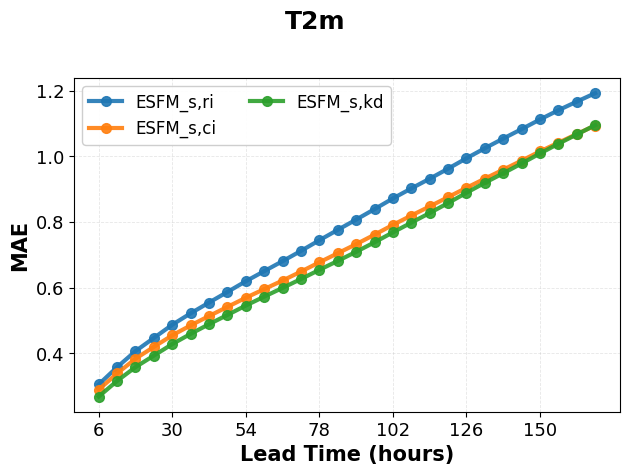} &
\includegraphics[width=0.32\textwidth]{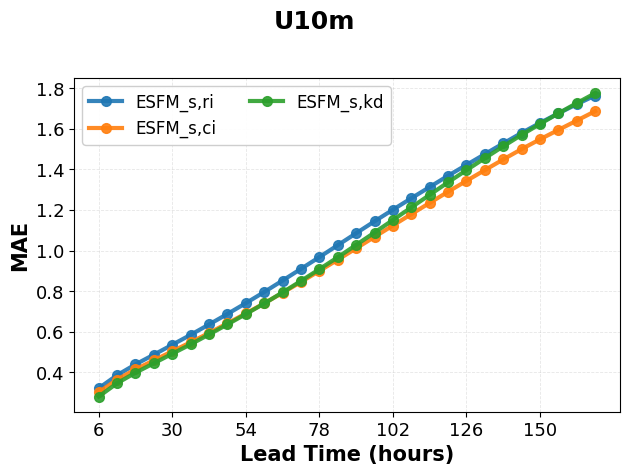} &
\includegraphics[width=0.32\textwidth]{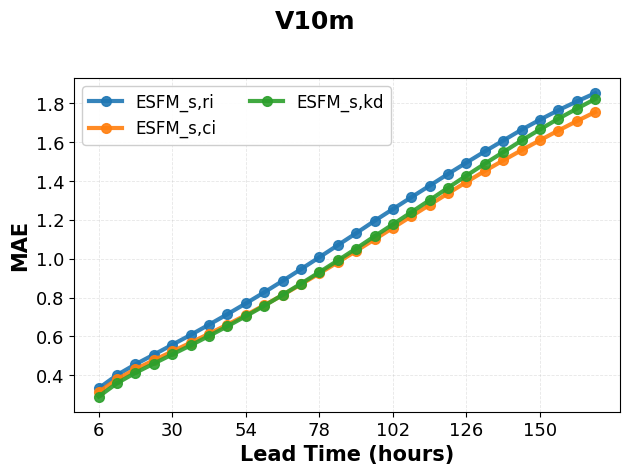} \\

\includegraphics[width=0.32\textwidth]{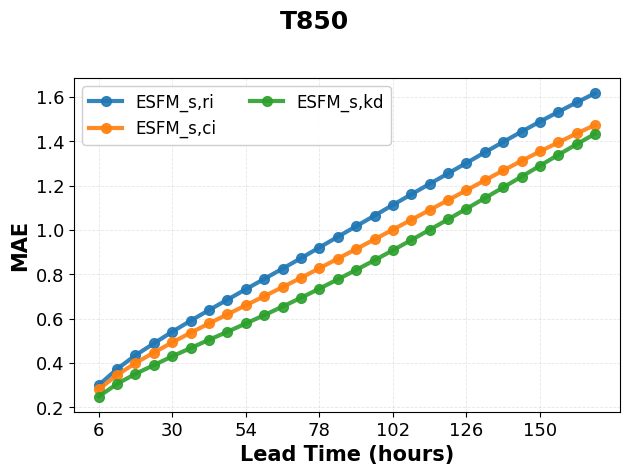} &
\includegraphics[width=0.32\textwidth]{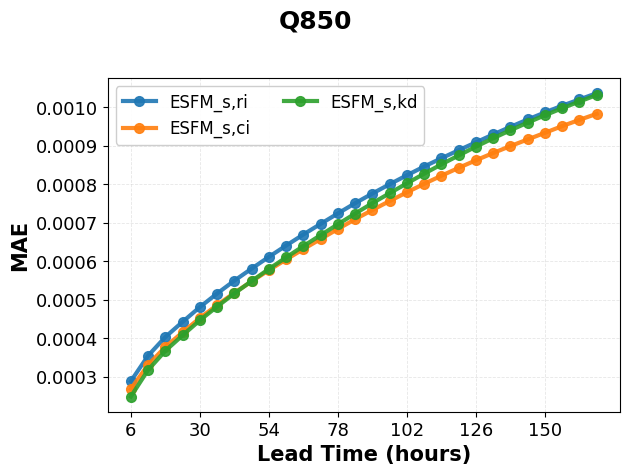} &
\includegraphics[width=0.32\textwidth]{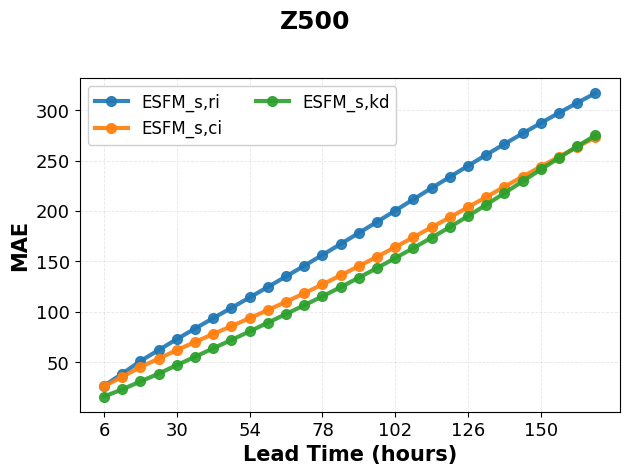}
\end{tabular}
\caption{Rollout forecasting MAE performance of ESFM\_s models depending on their pretraining objectives, up to seven days lead time, without any autoregressive finetuning.}
\label{fig:pretraining_rollout_comparison}
\end{figure}

In Fig.~\ref{fig:pretraining_rollout_comparison}, we show seven day rollout forecast performance of different ESFM\_s models, initialized from different weights; (i) random initialization (ESFM\_s,ri), (ii) initialization from ESFM\_s weights trained on eight CMIP6 datasets for 92\,k steps (ESFM\_s,ci), (iii) initialization from ESFM\_s encoder after aligning with Aurora encoder using knowledge distillation and Aurora pretrained backbone and decoder (ESFM\_s,kd).
ESFM\_s,kd is superior to the other initializations for all examined variables in general. 
However, upon closer inspection, it looks like the curve of increasing MAE over longer horizons is shallower for ESFM\_s,ci compared to ESFM\_s,kd for some variables.
This might suggest that the initialization of ESFM\_s,ci using CMIP6 datasets allowed it to converge to a more stable set of weights after finetuning on ERA5. 

\begin{figure}[htbp]
\centering
\begin{tabular}{ccc}
\includegraphics[width=0.32\textwidth]{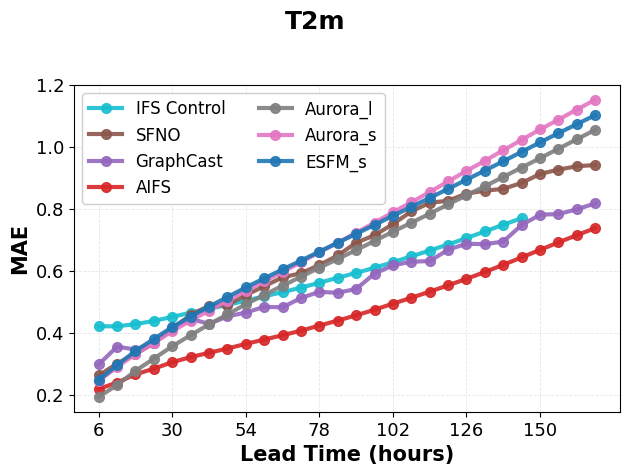} &
\includegraphics[width=0.32\textwidth]{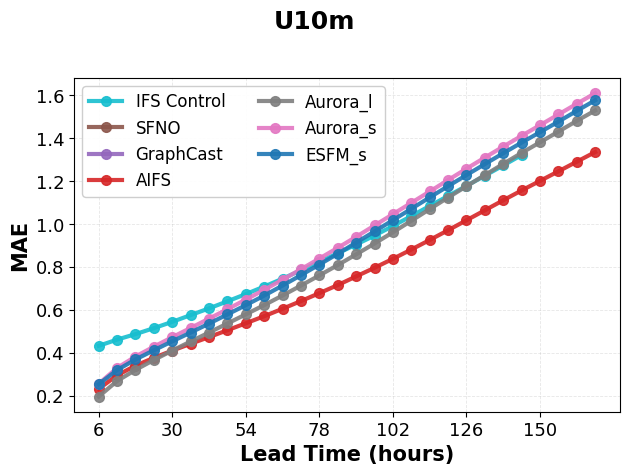} &
\includegraphics[width=0.32\textwidth]{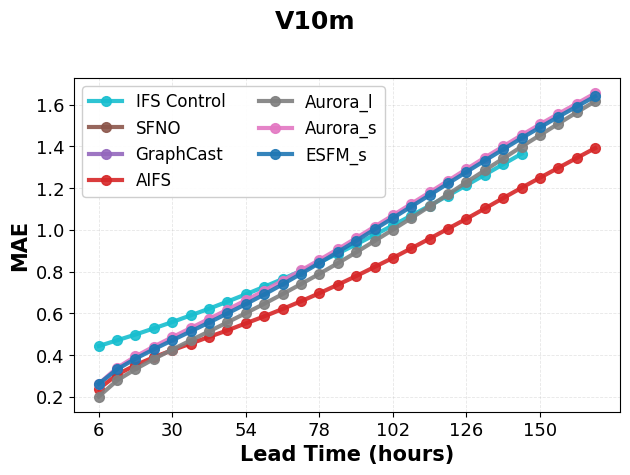} \\

\includegraphics[width=0.32\textwidth]{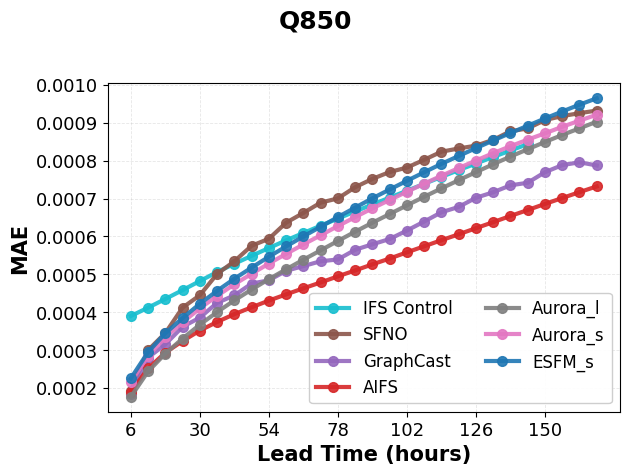} &
\includegraphics[width=0.32\textwidth]{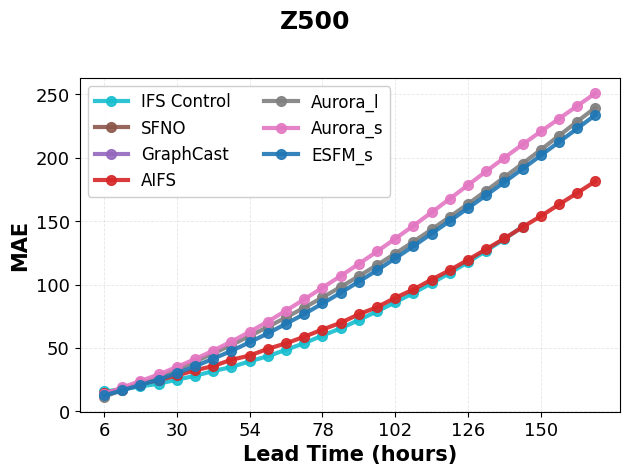} &
\includegraphics[width=0.32\textwidth]{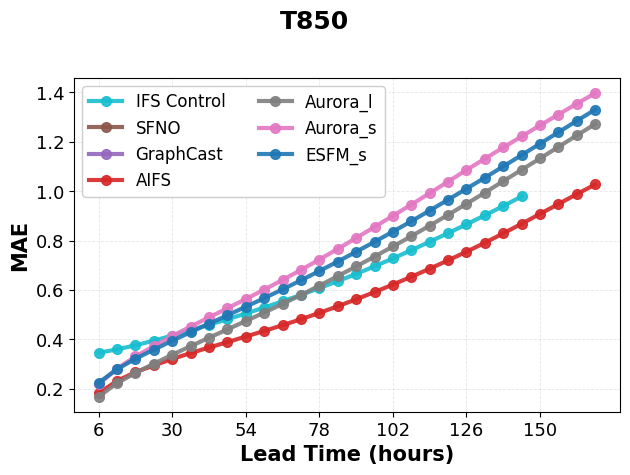}
\end{tabular}
\caption{Rollout forecasting MAE performance of ESFM\_s trained without masking in comparison with SotA models up to seven days lead time.
}
\label{fig:deterministic-rollouts-mae}
\end{figure}

We also show autoregressive rollout based forecasting performance of ESFM\_s with respect to SotA models in Fig.~\ref{fig:deterministic-rollouts-mae}.
It is important to note that both GraphCast and AIFS are operational models, finetuned on rollout forecasting. 
This becomes evident with the fact that while they are comparable (and sometimes worse) than ESFM\_s at six hour lead time. 
Errors of both GraphCast and AIFS increase significantly slower than the compared models, becoming especially more visible beyond two days lead time. 

\begin{figure}[htbp]
\centering
\begin{tabular}{ccc}
\includegraphics[width=0.32\textwidth]{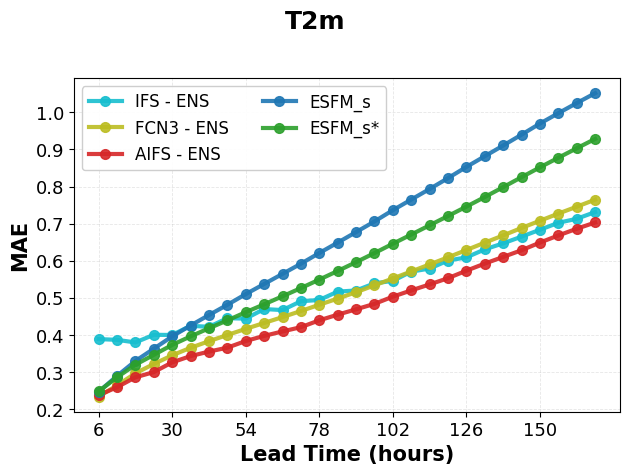} &
\includegraphics[width=0.32\textwidth]{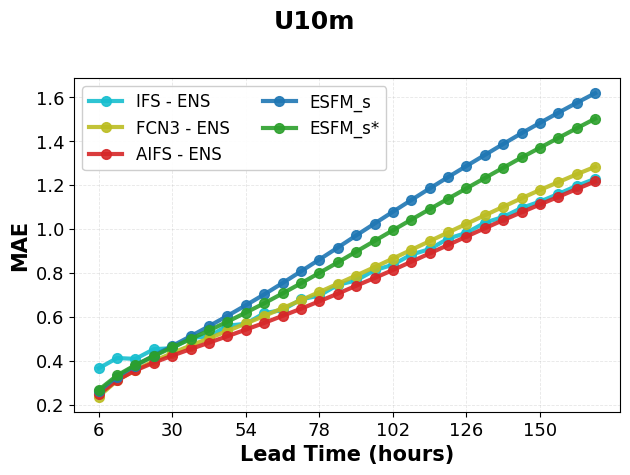} &
\includegraphics[width=0.32\textwidth]{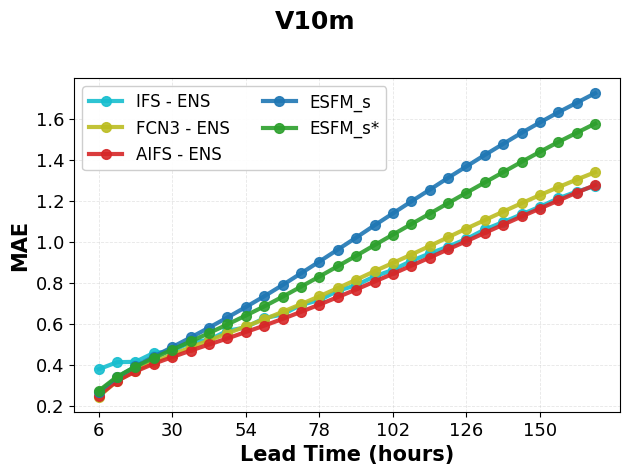} \\

\includegraphics[width=0.32\textwidth]{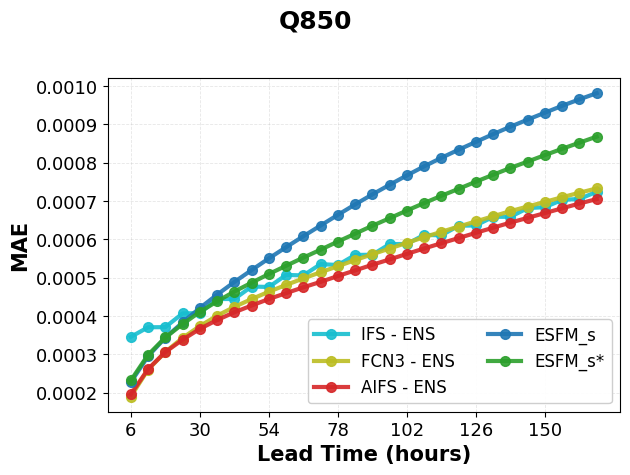} &
\includegraphics[width=0.32\textwidth]{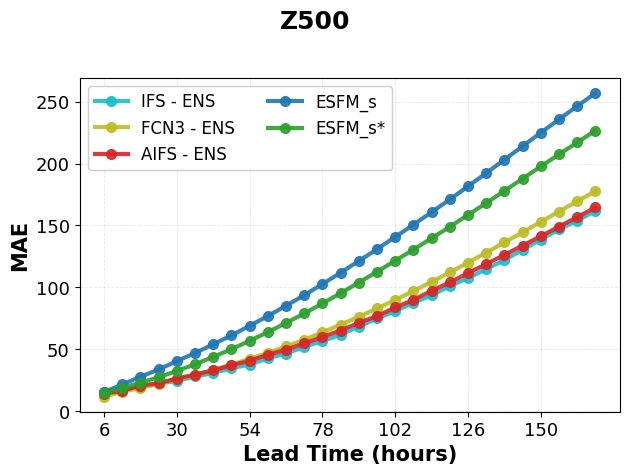} &
\includegraphics[width=0.32\textwidth]{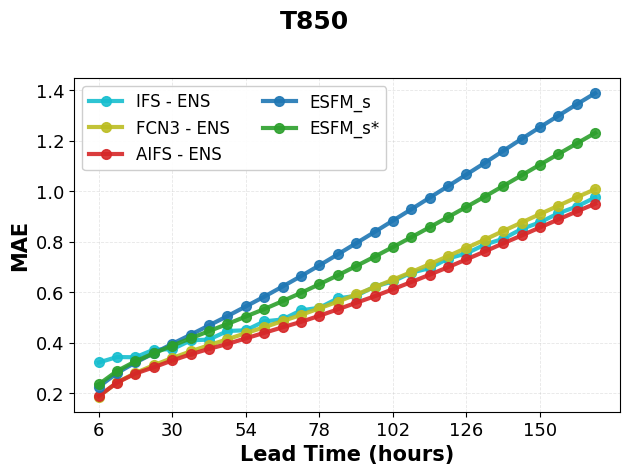}
\end{tabular}
\caption{Rollout forecasting performance of ensemble ESFM\_s trained without masking in comparison with SotA models up to seven days lead time.
For reference, we also include ESFM\_s*; autoregressively rollout finetuned ESFM\_s with a single LoRA layer for 3.4\,k steps, up to 10 steps (60 hours lead time).
Performance metric is MAE computed over the means of ensembles.
}
\label{fig:probabilistic-rollouts-MAE}
\end{figure}

\begin{figure}[htbp]
\centering
\begin{tabular}{ccc}
\includegraphics[width=0.32\textwidth]{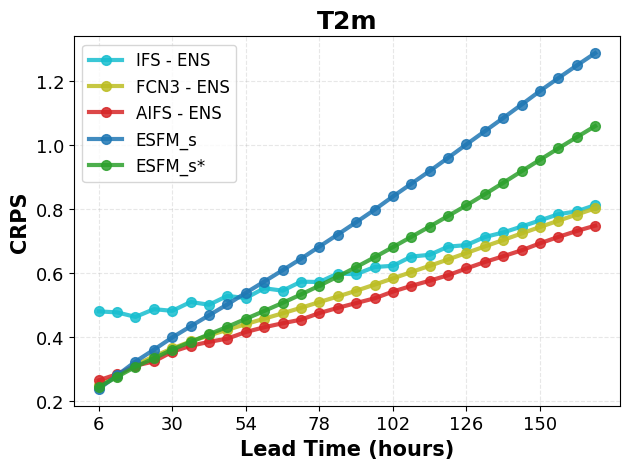} &
\includegraphics[width=0.32\textwidth]{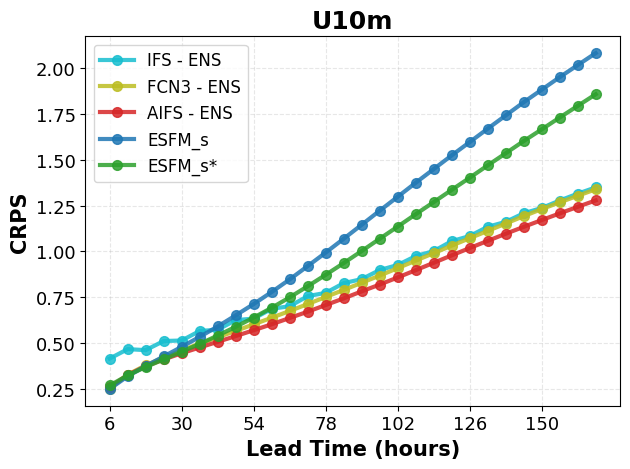} &
\includegraphics[width=0.32\textwidth]{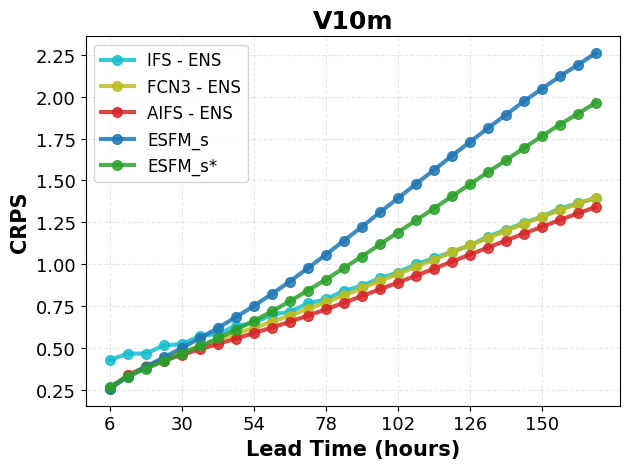} \\

\includegraphics[width=0.32\textwidth]{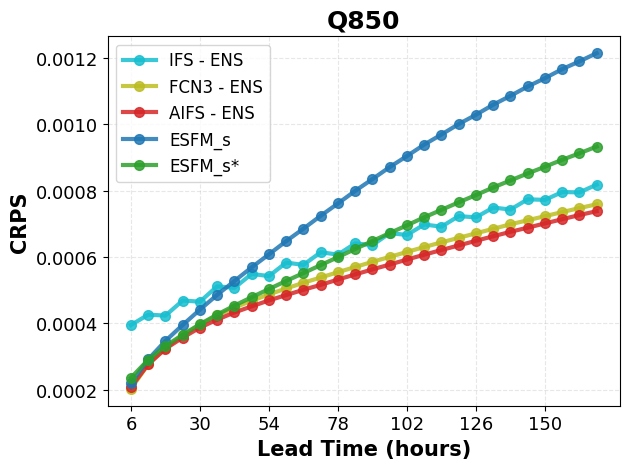} &
\includegraphics[width=0.32\textwidth]{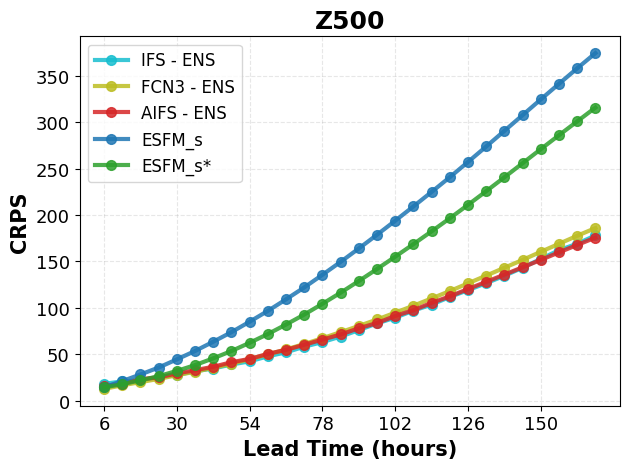} &
\includegraphics[width=0.32\textwidth]{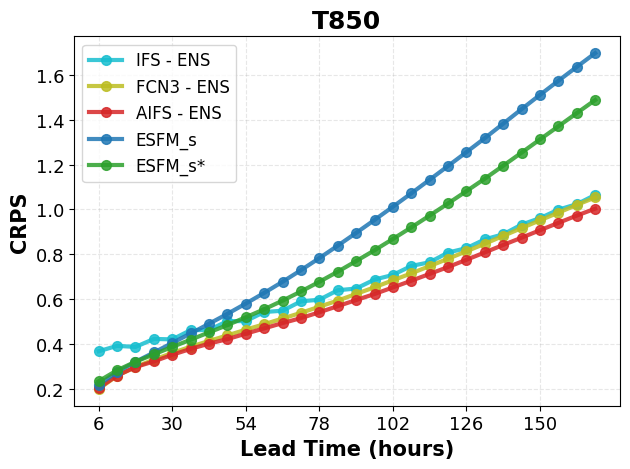}
\end{tabular}
\caption{Rollout forecasting CRPS performance of ensemble ESFM\_s trained without masking in comparison with SotA models up to seven days lead time.
For reference, we also include ESFM\_s*; autoregressively rollout finetuned ESFM\_s with a single LoRA layer for 3.4\,k steps, up to 10 steps (60 hours lead time).
}
\label{fig:probabilistic-rollouts-CRPS}
\end{figure}

Finally, we also show autoregressive rollout performance of ESFM\_s with ensembles in comparison with AIFS-CRPS and FourCastNet3 up to seven days lead time for MAE metric (calculated as ensemble means) in Fig.~\ref{fig:probabilistic-rollouts-MAE} and for CRPS metric in Fig.~\ref{fig:probabilistic-rollouts-CRPS}.
It can be seen that both ESFM\_s are very competitive with the SotA models at earlier lead times of up to 24 hours. 
However, beyond this time horizon, the performance metrics of the SotA models decay slower, presumably thanks to their autoregressive rollout finetuning. 
Furthermore, the SotA models generate ensembles independently, requiring a separate autoregressive rollout for each ensemble member. 
While we could do the same, we opted for a significantly more compute conservative approach in ESFM\_s, where we initialize each autoregressive inference on the ensemble means of the previous forecast, produced in a single inference step.
This means that our compute costs do not increase linearly with the number of ensemble members to generate. 
However, it seems to come at a performance cost, becoming increasingly evident with longer lead times.
We leave an investigation on finding a more optimal middle ground between compute cost and forecast performance for future work.
In Figs.~\ref{fig:probabilistic-rollouts-MAE} and~\ref{fig:probabilistic-rollouts-CRPS}, we also include a variant of ESFM\_s briefly rollout finetuned using low-rank adaptation (LoRA) for 3.4\,k steps for reference, denoted as ESFM\_s*. 
While this was not enough to close the gap between operationally finetuned SotA models over longer rollouts, it already made a significant impact.

\section{Datasets}
\label{sec:datasets_and_variables}

Here, we go over the contents of the datasets used in this work.

\subsection{ERA5 dataset}
\label{sec:ERA5_dataset}

Herein, we use a subset of variables as well as pressure levels that are available from the ERA5 dataset. 
In Table.~\ref{tab:era5_variables}, we list the names of the variables, corresponding pressure levels we used (for atmospheric variables), as well as their short names within the ERA5 dataset hosted in WeatherBench2.
Additional variables introduced in the finetuning study in Sec.~\ref{sec:novel_variables} are listed in Table.~\ref{tab:era5_surface_new}.

The training set comprises timeline between start of 1979 and end of 2020.
We sample training set randomly and do not actively finetune on the latest years of the training set for any of the experiments.
We select years 2023 and 2024 as the test set.
Due to prohibitive compute and storage costs, we limit our test set to a subset of these two years.
Namely, we pick four weeks that span the year, starting on 02.01, 02.04, 02.07, and 02.10 midnight UTC as described before. 
We further downsample these weeks to observation initialization time of every 6th hour, but shifting it by one hour every 3 steps; i.e., 00h, 06h, 12h, 18h, \textbf{01h}, 07h, 13h, 19h, \textbf{02h}, 08h, and so forth.
This is done to avoid any potential time-dependent bias of the compared models throughout this work.

\begin{table}[t]
\centering
\caption{ERA5 variables used for model training (1979--2020, sampled every 1 hour, $0.25^\circ$ spatial resolution).
The experiments listed in Sec.~\ref{sec:ssw} additionally use the pressure levels 10, 20, and 70\,hPa.
}
\label{tab:era5_variables}
\begin{tabular}{llll}
\toprule
\textbf{Category} & \textbf{Short Name} & \textbf{Full Name} & \textbf{Pressure Levels (hPa)} \\
\midrule
\multirow{4}{*}{Surface}
  & \texttt{2t}  & 2-meter temperature          & -- \\
  & \texttt{10u} & 10-meter U wind component    & -- \\
  & \texttt{10v} & 10-meter V wind component    & -- \\
  & \texttt{msl} & Mean sea-level pressure      & -- \\
\midrule
\multirow{5}{*}{Atmospheric}
  & \texttt{z}   & Geopotential                 & \multirow{5}{*}{\parbox{5cm}{50, 100, 150, 200, 250, 300, 400, 500, 600, 700, 850, 925, 1000}} \\
  & \texttt{u}   & U component of wind          & \\
  & \texttt{v}   & V component of wind          & \\
  & \texttt{t}   & Temperature                  & \\
  & \texttt{q}   & Specific humidity            & \\
\midrule
\multirow{3}{*}{Static}
  & \texttt{lsm} & Land-sea mask                & -- \\
  & \texttt{z}   & Geopotential (surface orography) & -- \\
  & \texttt{slt} & Soil type                    & -- \\
\bottomrule
\end{tabular}
\end{table}

\begin{table}[t]
\centering
\caption{New surface variables introduced for the finetuning experiment in Section \ref{sec:novel_variables}, aggregated over 6-hour intervals (1979--2020, sampled every 1 hour, $0.25^\circ$ spatial resolution).}
\label{tab:era5_surface_new}
\begin{tabular}{lll}
\toprule
\textbf{Set} & \textbf{Short Name} & \textbf{Full Name} \\
\midrule
\multirow{6}{*}{$\mathcal{L}_{\text{new},1}$}
  & \texttt{tp\_6hr}   & Total precipitation \\
  & \texttt{pe\_6hr}   & Potential evaporation \\
  & \texttt{ssr\_6hr}  & Surface net solar radiation \\
  & \texttt{ssrd\_6hr} & Surface solar radiation downwards \\
  & \texttt{ttr\_6hr}  & Top-of-atmosphere net thermal radiation \\
  & \texttt{tsr\_6hr}  & Top-of-atmosphere net solar radiation \\
\midrule
\multirow{6}{*}{$\mathcal{L}_{\text{new},2}$}
  & \texttt{e\_6hr}    & Evaporation \\
  & \texttt{sf\_6hr}   & Snowfall \\
  & \texttt{str\_6hr}  & Surface net thermal radiation \\
  & \texttt{strd\_6hr} & Surface thermal radiation downwards \\
  & \texttt{sshf\_6hr} & Surface sensible heat flux \\
  & \texttt{slhf\_6hr} & Surface latent heat flux \\
\bottomrule
\end{tabular}
\end{table}

\subsection{CMIP6 datasets}
\label{sec:CMIP6_preprocesing}

In order to assess the importance of initial weights when pretraining on ERA5, we included a \mbox{(pre-)pretraining} of ESFM\_s on eight CMIP6 datasets with masking protocol, also shown in the schematic Fig.~\ref{fig:pretraining-schematic}: 
CMCC, MIROC6, TaiESM1, NESM3, AWI-ESM, MPI-M, EC-Earth3, and MRI-ESM2.
We list the names and details of these datasets in Table~\ref{tab:cmip6_datasets}.
The naming convention of the variable abbreviations in CMIP6 differ from ERA5. 
In Table~\ref{tab:cmip6_variables}, we list the full set of variables we have used in this work.
During training, while the pressure levels of atmospheric variables are not uniform across different CMIP6 datasets, we used all pressure levels available at each individual dataset.

\begin{table*}[t]
\centering
\small
\setlength{\tabcolsep}{4pt}
\caption{Overview of the CMIP6 datasets used in ESFM\_s training.}
\resizebox{\textwidth}{!}{%
\begin{tabular}{llllllll}
\toprule
Dataset & Grid res. [$^\circ$] & Pixel res. & Pressure levels [hPa] & Surface vars & Atmos vars & Full name & Time range \\
\midrule
MPI-M    & 0.9375          & (192, 384) & 50, 250, 500, 600, 700, 850, 925 & tas, huss, uas, vas, psl & zg, ta, ua, va      & MPI-M/MPI-ESM1-2-HR           & 1850--2014 \\
AWI-ESM  & 1.875           & (96, 192)  & 50, 250, 500, 600, 700, 850, 925 & tas, huss, uas, vas, psl & zg, ta, hus, ua, va & AWI/AWI-ESM-1-1-LR            & 1850--2014 \\
CMCC     & 1.2153          & (192, 288) & 50, 250, 500, 600, 700, 850, 925 & psl                      & zg, ta, ua, va      & CMCC/CMCC-CM2-HR4             & 1850--2014 \\
TaiESM1  & 0.9375 / 1.2153 & (192, 288) & 250, 500, 850                    & tas, huss, psl           & ta, ua, va          & AS-RCEC/TaiESM1               & 1850--2004 \\
EC-Earth3& 0.7             & (256, 512) & 250, 500, 850                    & psl                      & zg, ta, ua, va      & EC-Earth-Consortium/EC-Earth3 & 1850--2014 \\
MIROC6   & 1.406           & (128, 256) & 50, 250, 500, 600, 700, 850, 925 & uas, vas, psl            & zg, ta, hus, ua, va & MIROC/MIROC6                  & 1850--2014 \\
MRI      & 1.125           & (160, 320) & 50, 250, 500, 600, 700, 850, 925 & psl                      & zg, ta, hus, ua, va & MRI/MRI-ESM2-0                & 1950--2014 \\
NESM3    & 1.875           & (96, 192)  & 250, 500, 850                    & psl                      & ta, ua, va          & NUIST/NESM3                   & 1950--2014 \\
\bottomrule
\end{tabular}%
}
\label{tab:cmip6_datasets}
\end{table*}

\begin{table*}[t]
\centering
\small
\setlength{\tabcolsep}{4pt}
\caption{CMPI6 dataset used in the finetuning experiment in Section \ref{sec:novel_CMIP6}. }
\resizebox{\textwidth}{!}{%
\begin{tabular}{llllllll}
\toprule
Dataset & Grid res. [$^\circ$] & Pixel res. & Pressure levels [hPa] & Surface vars & Atmos vars & Full name & Time range \\
\midrule
CNRM    & 0.5          & (360, 720) & 50, 250, 500, 600, 700, 850, 925 & psl, tos, tws, ci & zg, ta, ua, va      & CNRM-CM6-1-HR           & 1950--2014 \\
\bottomrule
\end{tabular}%
}
\label{tab:cmip6_dataset_cnrm}
\end{table*}

\begin{table}[t]
\centering
\caption{CMIP6 variable abbreviations and their full names used in this work.
}
\label{tab:cmip6_variables}
\begin{tabular}{lll}
\toprule
\textbf{Category} & \textbf{Short Name} & \textbf{Full Name} \\
\midrule
\multirow{4}{*}{Surface}
  & \texttt{tas}  & 2-meter temperature          \\
  & \texttt{uas} & 10-meter eastward wind component    \\
  & \texttt{vas} & 10-meter northward wind component    \\
  & \texttt{psl} & Air pressure at sea level      \\
  & \texttt{huss} & 2-meter specific humidity      \\
  & \texttt{tos} & Sea surface temperature    \\
  & \texttt{mrtws} & Terrestrial water storage      \\
  & \texttt{siconc} & Sea ice cover      \\
  
\midrule
\multirow{5}{*}{Atmospheric}
  & \texttt{zg}   & Geopotential height \\
  & \texttt{ua}   & Eastward wind          \\
  & \texttt{va}   & Northward wind          \\
  & \texttt{ta}   & Air temperature                  \\
  & \texttt{hus}   & Specific humidity            \\
\bottomrule
\end{tabular}
\end{table}

\subsection{Station datasets and greedy mapping}
\label{sec:station_greedy_mapping}

\paragraph{Weather-5K dataset.}
The Weather-5K dataset \cite{han2024weather5k} is a large-scale global collection of station weather data sourced from the NCEI Integrated Surface Database (ISD)~\citep{copernicus_insitu_land_2023}. 
It comprises hourly observations from 5,672 globally distributed stations spanning 2014–2023. 
To ensure data continuity, Weather-5K employs a 30-minute window replacement strategy, linear interpolation for gaps under 12 hours, and ERA5 reanalysis infilling for any remaining missing values. 
Consequently, the dataset only retains stations with $\geq$90\% valid hourly data.

\paragraph{ECMWF 11k dataset.}
Our ECMWF 11k dataset builds upon the source data used by Weather-5K but introduces significant changes; yielding more samples along time and space, without interpolating any missing values.
Unlike Weather-5K, which interpolates gaps to produce a dense time series using reanalysis data, our dataset retains only actual station observations. 

\begin{itemize}
\item \textbf{Temporal and spatial coverage.}
We extended the raw observation download via the Copernicus Climate Data Store (CDS) (\texttt{insitu-observations-surface-land}, v2.0.0) to cover the full period from January 2000 to December 2024. 
This results in a total of 11'863 stations and 219'168 hourly timesteps.

\item \textbf{Observation filtering.}
We do not apply spatial or temporal interpolation; all missing values are preserved as \texttt{NaN}.
This allows keeping station data free from the biases of reanalysis data, which would not be available at test time for station datasets.
Raw observations are snapped to the nearest UTC hour.
Any observation more than 15 minutes from the nominal hour is discarded.
When multiple reports fall in the same (hour, station, variable) bin,
the one with the smallest temporal offset is retained.

\item \textbf{Train/holdout split.}
To evaluate spatial generalization, 1000 holdout stations are withheld
from training.
Holdout stations are selected using uniform geographic sampling:
the globe is partitioned into regions and an equal number of stations is drawn
from each region, subject to a minimum temporal coverage requirement of
40\% over the evaluation year (2023).
The selected holdout stations span latitudes $-80.4$\textdegree\ to $78.2$\textdegree,
with 75.6\% in the Northern Hemisphere and 24.4\% in the Southern Hemisphere.
The remaining $\sim$10863 stations (occupying 15200 grid cells)
form the training set.
In the split files, withheld stations are masked to \texttt{NaN} in the training
array and vice versa, preserving the full $90\times180$ spatial grid for both subsets.

\end{itemize}

  \subparagraph{Variables.}
  The ECMWF 11k dataset contains nine surface variables listed in
  Table~\ref{tab:station_vars}.
  Compared to the five variables in Weather-5K (temperature, dew point, wind direction,
  wind speed, sea-level pressure), we additionally include surface air pressure (\texttt{pa})
  and decompose the scalar wind into explicit U and V components (\texttt{10u}, \texttt{10v})
  via $u = -ws\sin(\theta)$, $v = -ws\cos(\theta)$, where $\theta$ is the wind-from-direction
  in radians.
  An additional diagnostic variable \texttt{ts} (time shift) states, in minutes,
  how far the original observation timestamp deviates from the nominal hour,
  serving as a data-quality signal available to the model.
  All variables are z-score normalized using global location and scale statistics
  computed over all stations and all years (shown in Table~\ref{tab:station_vars}).

  \begin{table}[h]
  \centering
  \caption{Station dataset variables, units, global normalization statistics
  (location $\mu$ and scale $\sigma$ used for z-score normalization),
  and average number of valid grid cells per timestep.}
  \label{tab:station_vars}
  \begin{tabular}{llc}
  \toprule
  \textbf{Short name} & \textbf{Description} & \textbf{Unit}\\
  \midrule
  \texttt{pa}  & Surface air pressure          & Pa              \\
  \texttt{msl} & Mean sea-level pressure       & Pa              \\
  \texttt{2t}  & 2\,m air temperature          & K               \\
  \texttt{dt}  & 2\,m dew point temperature    & K               \\
  \texttt{10u} & 10\,m eastward wind (U)       & m\,s$^{-1}$     \\
  \texttt{10v} & 10\,m northward wind (V)      & m\,s$^{-1}$     \\
  \texttt{wd}  & Wind from direction           & deg             \\
  \texttt{ws}  & 10\,m wind speed              & m\,s$^{-1}$     \\
  \texttt{ts}  & Observation time shift        & min             \\
  \bottomrule
  \end{tabular}%
  \end{table}

\begin{table}[t]
\centering
\caption{Weather-5K and ECMWF 11k variable abbreviations and their full names used in this work.
Eastward and northward wind components are derived from wind speed and wind direction variables present in these datasets.
Time shift variable is calculated as the $\Delta$ in minutes from full hour to the actual station measurement. 
We calculate this variable in the curation of the ECMWF 11k dataset and use it only at inputs.
}
\label{tab:station_variables}
\begin{tabular}{lll}
\toprule
\textbf{Category} & \textbf{Short Name} & \textbf{Full Name} \\
\midrule
\multirow{4}{*}{Surface}
  & \texttt{T}  & Surface temperature          \\
  & \texttt{dT}  & Dew point temperature          \\
  & \texttt{WS} & Wind speed    \\
  & \texttt{WD} & Wind direction      \\
  & \texttt{MSL} & Air pressure at sea level      \\
  & \texttt{P} & Surface air pressure      \\
  & \texttt{U} & Surface eastward wind component    \\
  & \texttt{V} & Surface Northward wind component    \\
  & \texttt{TS} & Time shift from full hour    \\
\bottomrule
\end{tabular}
\end{table}

\subparagraph{Spatial structure.}
Both station observations are projected onto a more compact, sparse $90\times180$ grid following Algorithm~\ref{alg:build-grid}.
We map $N$ irregularly spaced weather stations onto a $H{\times}W$ grid by wrapping longitudes to $[0^\circ,360^\circ)$, partitioning stations into north-to-south latitude bands and assigning longitude-ordered columns via a monotonic greedy rule around the preferred position $c^*=\operatorname{round}(\lambda/360\cdot(W{-}1))$. 
We interpolate the cell coordinates without station data piecewise linearly, broadcasting row-wise reference latitudes while preserving exact station coordinates at cells with station data. 
We apply minimal adjustments to guaranty monotonicity favorable to ESFM.

Each occupied grid cell holds the measurement from its assigned station; while cells without observations are set to \texttt{NaN}.
Due to the sparse and irregular distribution of weather stations, valid observations cover approximately 3'000--7'000 of the 16'200 grid cells per timestep, depending on the variable (see Table~\ref{tab:station_vars}).
This approach results in exact latitude preservation and a maximum longitude displacement below $10^{-4}$ degrees.

\begin{algorithm}[t]                                      
  \small                       
  \caption{Station-to-Grid Mapping}           
  \label{alg:build-grid}                      
  \begin{algorithmic}[1]                      
  \Require Stations $\{(s,\varphi_s,\lambda_s)\}$, grid $H{\times}W$, half-width $\delta$                                                                          
  \Ensure  Grids $\mathbf{Lat}, \mathbf{Lon} \in \mathbb{R}^{H \times W}$, mapping $\mathcal{M}$                                                                   

  \State Wrap $\lambda_s \gets \lambda_s \bmod 360$; sort stations by $(-\varphi_s, \lambda_s)$
  \State Partition into $H$ equal bands $\mathcal{B}_0,\dots,\mathcal{B}_{H-1}$

  \For{each band $r$, station $s \in \mathcal{B}_r$ sorted by $\lambda_s$}
      \State $c^* \gets \operatorname{round}\!\left(\tfrac{\lambda_s}{360}(W{-}1)\right)$
          \Comment{preferred column}
      \State $c_s \gets \operatorname{clip}(c^*,\; c_{\mathrm{prev}}+1,\; W{-}1{-}n_{\mathrm{remaining}})$
          \Comment{greedy monotone assignment}
      \State $\mathcal{M}(s) \gets (r, c_s)$
  \EndFor

  \State $\mathbf{Lon}[r,:]$: interpolate linearly between station anchors $\{(c_s, \lambda_s)\}$ per row
  \State $\mathbf{Lat}[r,:]$: broadcast $\max\{\varphi_s \in \mathcal{B}_r\}$, overwrite station cells with $\varphi_s$
  \State Enforce strict decrease $\downarrow$ in $\mathbf{Lat}$ and strict increase $\rightarrow$ in $\mathbf{Lon}$ via minimal nudges
  \State $\mathbf{Lat}_{\min} \gets \mathbf{Lat} - \delta$;\quad $\mathbf{Lat}_{\max} \gets \mathbf{Lat} + \delta$;\quad
         $\mathbf{Lon}_{\min} \gets \mathbf{Lon} - \delta$;\quad $\mathbf{Lon}_{\max} \gets \mathbf{Lon} + \delta$

  \State \Return $\mathbf{Lat}, \mathbf{Lon}, \mathbf{Lat}_{\min}, \mathbf{Lat}_{\max}, \mathbf{Lon}_{\min}, \mathbf{Lon}_{\max}, \mathcal{M}$
  \end{algorithmic}
  \end{algorithm}

\section{Additional results on finetuning experiments}
\subsection{Pretraining with additional variables}
The baseline ESFM\_s was pretrained for an additional 15\,k steps with a set of new surface variables $S_{new,1}$: total precipitation, potential evaporation, surface net solar radiation, surface solar radiation downwards, top net thermal radiation, top net solar radiation. 
This training approximates a model pretrained with variables that extend beyond the commonly used variables. 
Original variables are kept during this training phase. 

\begin{figure}[h]
    \centering
    \includegraphics[width=1\linewidth]{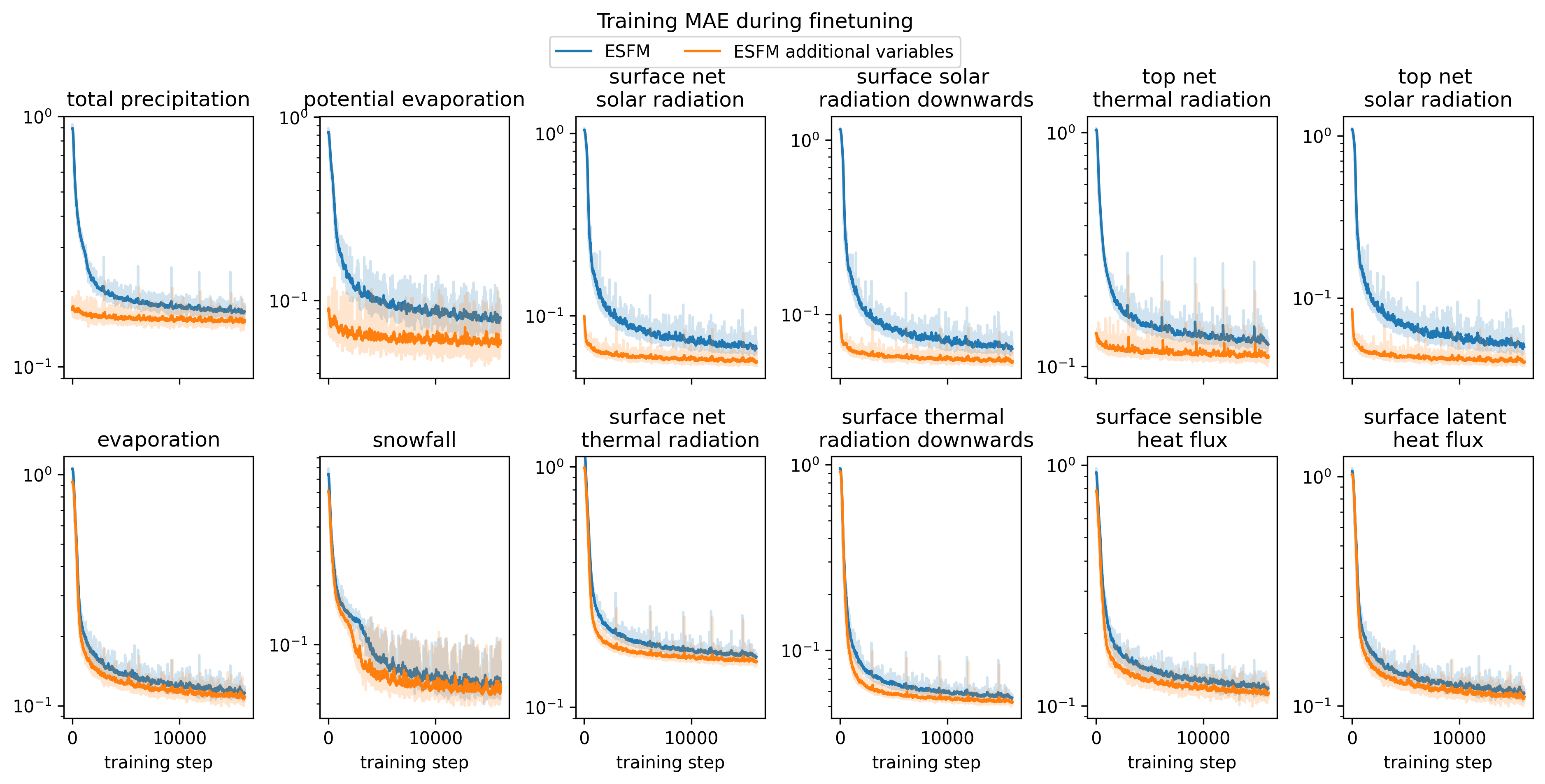}
    \caption{Training loss when finetuning ESFM\_s on a large set of new surface variables from two pretrained models: baseline pretraining (blue) and pretraining with a small set of new surface variables (orange).
    }
    \label{fig:loss_finetuning_newvars}
\end{figure}

Then, we finetune this extended ESFM\_s and the baseline ESFM\_s with the original variables, $S_{new,1}$, and a new set of surface variables $S_{new,2}$: evaporation, snowfall, surface net thermal radiation, surface thermal radiation downward, surface sensible heat flux, surface latent heat flux. 
Figure~\ref{fig:loss_finetuning_newvars} shows the loss curves for these two finetuning experiments.

\clearpage
\subsection{Doksuri typhoon in July 2023}
\begin{figure}[h]
    \centering
    \includegraphics[width=1\linewidth]{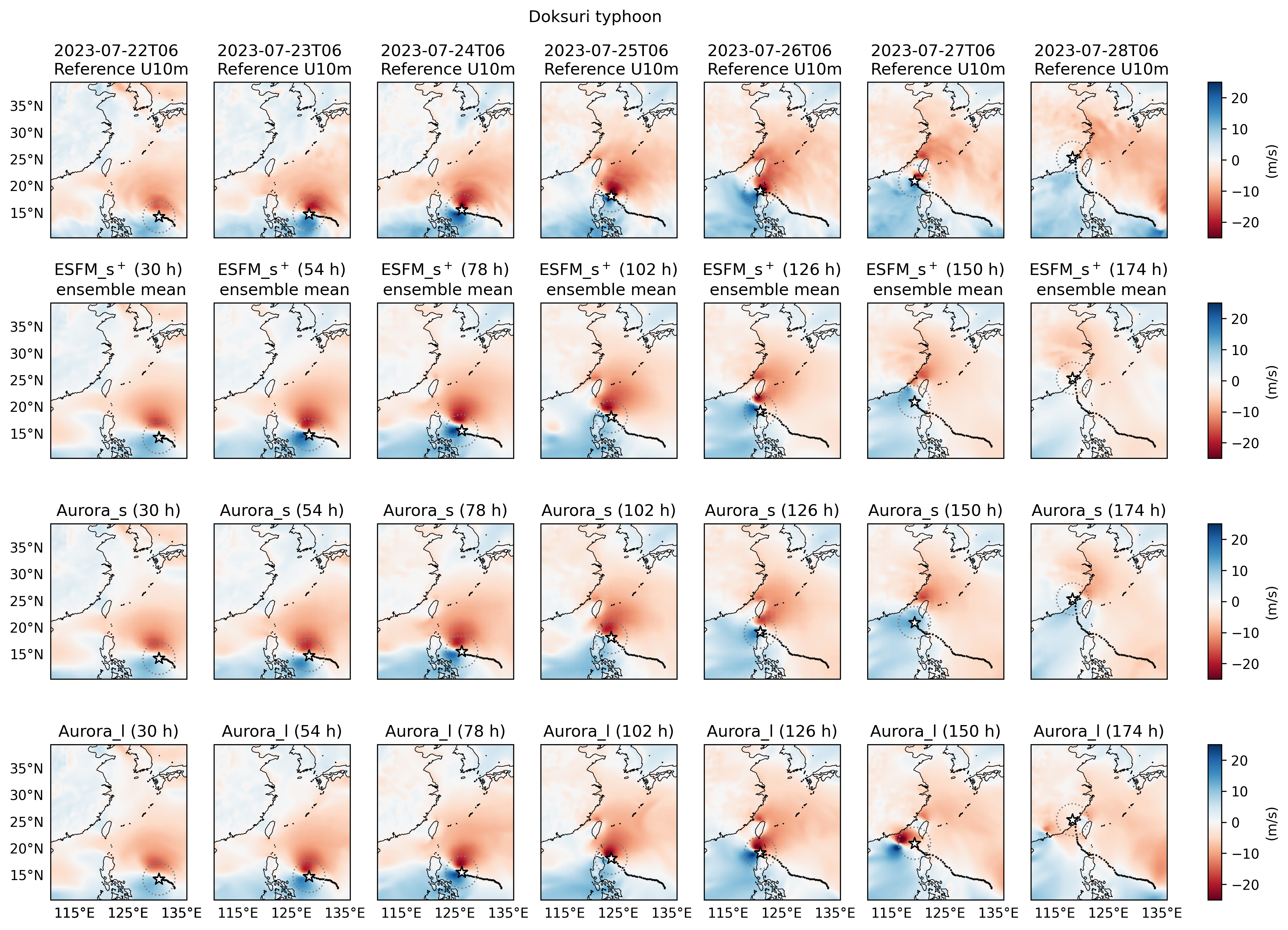}
    \caption{10-meter eastward component of wind during Doksuri typhoon (July 22 to July 28). 
    (First row) ERA5. 
    (Second row) mean of the 8 decoder heads of ESFM\_s. 
    (Third row) Aurora\_s. 
    (Fourth row) Aurora\_l. 
    ESFM\_s, Aurora\_s, and Aurora\_l are initialized on 21.07.2023 and rolled out from that date (from 30-hour to 174-hour lead time). 
    Each subpanel shows the cyclone tracking given by IBTrACS with the current position denoted with the star. 
    The circle indicates a 3-degree radius circle used to compute the maximum wind speed in Fig.~\ref{fig:finetuning_doksuri_timeseries}.}
    \label{fig:finetuning_doksuri_maps}
\end{figure}

Probabilistic ESFM (ESFM\_s+), Aurora small (Aurora\_s), and Aurora large (Aurora\_l) are initialized on 21.07.2023 and autoregressively rolled out from that date for 174 hour lead time forecasts. 
Figure~\ref{fig:finetuning_doksuri_maps} shows the maps of wind velocity in South-East Asia as the Doksuri typhoon progresses. 
Since the cyclone eye trajectory is extracted from the IBTrACS catalog and not computed from the models, it is placed at the same position across the compared models. 

\end{document}